\documentclass[fleqn,usenatbib]{mnras}

\usepackage{newtxtext,newtxmath}
\usepackage[T1]{fontenc}

\DeclareRobustCommand{\VAN}[3]{#2}
\let\VANthebibliography\thebibliography
\def\thebibliography{\DeclareRobustCommand{\VAN}[3]{##3}\VANthebibliography}

\usepackage{graphicx}	% Including figure files
\usepackage{amsmath}	% Advanced maths commands
\usepackage{subfigure}

\title[GSF instability general shears]{Linear and nonlinear properties of the Goldreich-Schubert-Fricke instability in stellar interiors with arbitrary local radial and latitudinal differential rotation}

\author[R. W. Dymott et al.]{
R.W. Dymott$^{1}$\thanks{E-mail: mmrwd@leeds.ac.uk},
A. J. Barker$^{1}$\thanks{E-mail: A.J.Barker@leeds.ac.uk},
C.A. Jones$^{1}$
and S.M. Tobias$^{1}$
\\
$^{1}$Department of Applied Mathematics, School of Mathematics, University of Leeds, Leeds, LS2 9JT, UK\\
}

\date{Accepted Nov 6 666. Received Nov 6 666; in original form Nov 6 666}
\pubyear{2022}

\begin{document}
\label{firstpage}
\pagerange{\pageref{firstpage}--\pageref{lastpage}}
\maketitle

\begin{abstract}
We investigate the linear and nonlinear properties of the Goldreich-Schubert-Fricke (GSF) instability in stellar radiative zones with arbitrary local (radial and latitudinal) differential rotation. This instability may lead to turbulence that contributes to redistribution of angular momentum and chemical composition in stars. In our local Boussinesq model, we investigate varying the orientation of the shear with respect to the `effective gravity', which we describe using the angle $\phi$. We first perform an axisymmetric linear analysis to explore the effects of varying $\phi$ on the local stability of arbitrary differential rotations. We then explore the nonlinear hydrodynamical evolution in three dimensions using a modified shearing box. The model exhibits both the diffusive GSF instability, and a non-diffusive instability that occurs when the Solberg-H\o iland criteria are violated. We observe the nonlinear development of strong zonal jets (“layering” in the angular momentum) with a preferred orientation in both cases, which can considerably enhance turbulent transport. By varying $\phi$ we find the instability with mixed radial and latitudinal shears transports angular momentum more efficiently (particularly if adiabatically unstable) than cases with purely radial shear ($\phi=0$). By exploring the dependence on box size, we find the transport properties of the GSF instability to be largely insensitive to this, implying we can meaningfully extrapolate our results to stars. However, there is no preferred length-scale for adiabatic instability, which therefore exhibits strong box-size dependence. These instabilities may contribute to the missing angular momentum transport required in red giant and subgiant stars and drive turbulence in the solar tachocline. 
\end{abstract} %AJB 250 words exactly! (250 limit for abstract in MNRAS!)

\begin{keywords}
Sun: rotation -- stars: rotation -- stars: interiors -- hydrodynamics -- waves -- instabilities
\end{keywords}

%%%%%%%%%%%%%%%%%%%%%%%%%%%%%%%%%%%%%%%%%%%%%%%%%%

%%%%%%%%%%%%%%%%% BODY OF PAPER %%%%%%%%%%%%%%%%%%

\section{Introduction}

The internal transportation of angular momentum (AM) and chemical composition throughout the life-cycle of a star can vastly affect its ultimate fate. Unfortunately, the dynamics of AM (and chemical) redistribution in stellar interiors is poorly understood, as can be seen particularly clearly for red and sub-giant stars, whose core-envelope differential rotations inferred from asteroseismology are not well explained by existing models \citep[e.g.][]{aerts2019}. 
In the following, we are interested in dynamics occurring in stably-stratified stellar radiative zones exhibiting differential rotation, particularly in regions of strong shear, such as the solar tachocline, which connects the radiative and convective regions and contains interesting wave, turbulence and magnetic field dynamics \cite[e.g.][]{GilmanFox1997,Wood2011,2017GApFD.111..282M,G2020}. 
Some of the main physical mechanisms that could enhance AM transport in stars involve (magneto-)hydrodynamic instabilities \cite[e.g.][]{2009Book,meynet2013,aerts2019}. One such instability that may occur within radiative zones is the Goldreich-Schubert-Fricke (GSF) instability \citep{GSF,Fricke1968}.  

The GSF instability is a doubly-diffusive instability of differential rotation, where the action of thermal diffusion on sufficiently small lengthscales reduces the stabilising effects of buoyancy, allowing for the development of a fingering-type instability \citep[analogous to the thermo-haline instability e.g.][]{Garaud2018}. In a rotating shear flow where the thermal gradient is stabilising (a radiative region) a reduction in thermal effects can allow AM fingers to develop and grow exponentially. Subsequently, these non-linearly saturate, e.g. by secondary parasitic shear instabilities, as they grow until turbulence develops. This configuration is visually analogous to salt fingering, and formally analogous for axisymmetric (2D 3-component) simulations performed with purely radial shear at the equator \citep[for a certain choice of diffusivity ratio,][]{barker2019}.   
Interestingly, the nonlinear development of the instability does not lead to a homogeneous turbulent state in general, and other interesting dynamics, such as the formation of layering in AM (often referred to as 'zonal-jets') has previously been observed, which can enhance turbulent transport \citep[particularly at non-equatorial latitudes, in the case with radial shear,][]{barker2020}.   

This paper builds directly upon \citet[][hereafter paper 1]{barker2019}, and \citet[][hereafter paper 2]{barker2020}. These papers consider a local Cartesian representation of a small patch of a stably-stratified, differentially-rotating, radiation zone, modelling a global `shellular' (or `vertical') differential rotation that varies only with spherical radius; first at the equator in paper 1, then at a general latitude in paper 2. Following these papers, we perform an axisymmetric linear stability analysis alongside complementary three-dimensional nonlinear numerical simulations, with the primary goal of understanding the nonlinear evolution of the GSF instability, and determining its potential role in AM transport and chemical mixing. 
Paper 1 found that 3D simulations at the equator with radial shear (primarily) exhibited homogeneous turbulence with sustained and enhanced AM transport. Significant differences between axisymmetric and 3D simulations were observed however, and where comparisons could be made their findings were in agreement with previous work \cite[e.g.][]{Korycansky1991}. A simple, easily-implementable theory to model AM transport in stars (motivated by \citealt{Denissenkov2010} and \citealt{Brown2013} for thermohaline convection), was also developed for possible use in stellar evolution codes.

Generalising the above study to an arbitrary latitude (but still with radial shear), paper 2 again found that the instability exhibited enhanced AM transport, with further increases typically seen away from the equator. Interestingly, the formation of zonal jets (or layering in AM) was observed in nonlinear simulations, which were tilted with respect to the local gravity vector by an angle that corresponded initially with the fastest growing linearly unstable modes, but later evolved with time. Paper 2 also analysed the linear stability and obtained the following simple criterion for onset of (diffusive) axisymmetric instability at a general latitude, for radial differential rotation: $ \mathrm{RiPr} < \frac{1}{4}$, where $\mathrm{Ri}=N^2/S^2$ is the local (gradient) Richardson number and $\mathrm{Pr}=\nu/\kappa$ is the (thermal) Prandtl number. Here $N^2$ is the squared buoyancy frequency, $S^2$ is the squared local shear rate, and $\nu$ and $\kappa$ are the kinematic viscosity and thermal diffusivity \citep[see also][]{Rashid2008}. For instability at the equator the flow only becomes unstable if the stricter Rayleigh criterion for non-diffusive centrifugal instability is violated.

The GSF instability belongs to a family of instabilities referred to as so-called 'secular' shear instabilities.
Standard shear instabilities, for which perturbations are usually assumed to be adiabatic, are not typically expected to develop in stellar radiation zones, thanks to the strong stabilising effects of the stratification. On the other hand, finite-amplitude `secular' (or diffusive) shear instabilities \citep[e.g.][]{Zahn1974,Zahn1992}, are believed to be important by producing thermally-diffusive shear-induced turbulence when the Richardson number Ri of the flow is large, provided the Peclet number Pe (which measures the ratio of thermal diffusion to advection timescales) is sufficiently small \citep[e.g.][]{Prat2013,Garaud2017,Gagnier2018,KulGaraud2018,Cope2020,Garaud2020}. The GSF instability is distinct from these in that it is a linear instability that only operates in the presence of rotation, but it is related in that it requires thermal diffusion to dominate over momentum diffusion. When both instabilities operate they can interact, leading to interesting nonlinear dynamics \citep{Chang2021}. The GSF instability and its co-existence with inflection-point instabilities has also been analysed in linear theory for horizontal shears with a $\tanh$ profile by \cite{Park2020,Park2021}. They referred to the GSF instability as the ``inertial instability" following its relation to this instability in the geophysical literature.

Our primary goal in this work is to develop an understanding of how the linear and nonlinear properties of the axisymmetric hydrodynamic (primarily GSF) instabilities are modified when the model is extended to allow arbitrary local differential rotation. Our model allows us to study the stability of horizontal/latitudinal shear, as well as mixed latitudinal and radial shears, thereby building considerably upon papers 1 and 2, which were restricted to purely radial shears. In our local model, this is accomplished by studying how variation in the orientation of the local effective gravity vector relative to the shear alters the linear and nonlinear dynamics of the GSF instability in a small-scale Cartesian box. This is done through studying its effects in the linear problem in section 3, where we derive criteria for the onset of linear adiabatic and diffusive instabilities. Here we will also present several figures that characterise the growth rates and wavevector magnitudes within the various possible regimes. Section 4 uses pseudo-spectral hydrodynamical simulations using an MPI-parallelised code to explore the fully nonlinear problem. In particular, we explore the effects of the orientation of the local effective gravity relative to the shear on both the kinetic energy and AM transport produced by the instability, and also present visual snapshots of the dynamics at various points throughout the evolution. In section 5 we present our conclusions and discuss motivations for further work, as well as a brief discussion of the application of our results to transport in stars.

\section{Local Cartesian Box model}

\subsection{The model and governing equations}
\label{sec:maths}

Our model represents a small-scale patch of a stably-stratified region of a differentially-rotating star, such as in the lower parts of the solar tachocline. We model this patch as a Cartesian box with coordinates $(x,y,z)$, where we define $y$ as the local azimuthal direction, and $x$ and $z$ as two directions in the meridional plane, which will be described in more detail below. We adopt the Bousinessq approximation \citep{Spiegel1960}, which is valid here since the scales on which the effects of thermal diffusion become strong enough to enable instability are typically far smaller than the pressure scale height \citep[e.g.~see the estimates in][]{barker2019}. 

\begin{figure}
\centering
\subfigure{\includegraphics[
    trim=0cm 0cm 0cm 0cm,clip=true,
    width=0.3\textwidth]{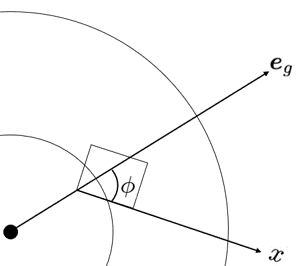}}
\subfigure{\includegraphics[width= 0.48\textwidth]{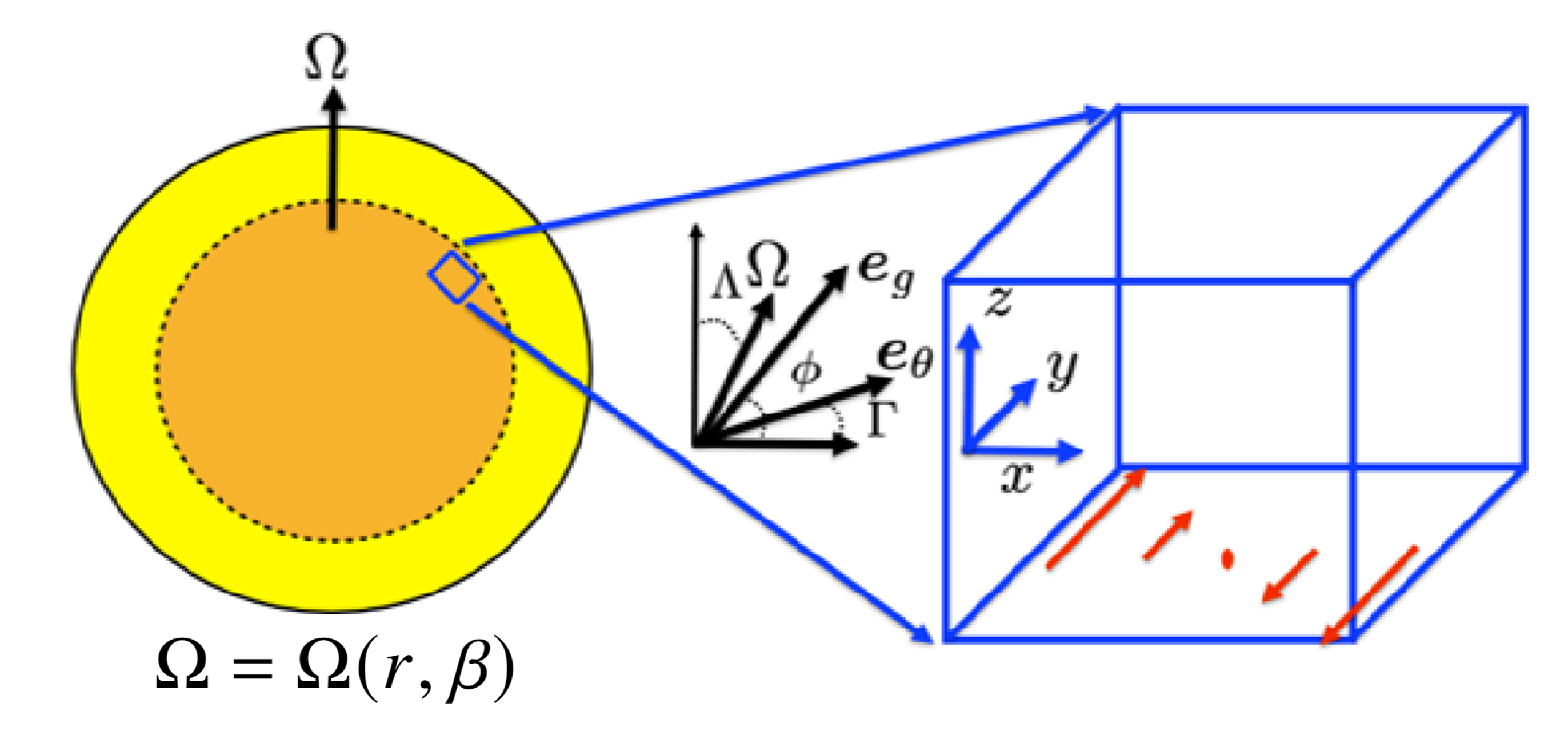}}
  \label{initgrowthfigs}
\caption{Two panels indicating the local Cartesian model with arbitrarily oriented local effective gravity $\boldsymbol{e}_g$ describing local differential rotation depending on both spherical radius and latitude in general (note, we refer to rotation profiles as $\Omega(r,\beta)$, where $\beta$ is the co-latitude). For illustration, the dark orange region may represent a radiation zone and the yellow region an overlying convection zone, if we consider application to the solar tachocline. We consider a general differentially-rotating star at a latitude $\Lambda+\phi$, with local shear along $x$, and normal to the stratification surfaces (i.e. along the temperature gradient) $\boldsymbol{e}_{\theta}$, which is inclined relative to the local radial direction (approximately along $\boldsymbol{e}_g$) by an angle $\Gamma-\phi$ that is determined by the thermal wind equation.}
    \label{fig:Boxmodel}
\end{figure}

The differential rotation in this model is locally decomposed into a rotation term ${\boldsymbol{\Omega}}(r,\beta)=\Omega(r,\beta)\hat{\boldsymbol{\Omega}}$ (which is locally constant with magnitude $\Omega$), along with a linear shear flow $\boldsymbol{U}_0 = -\mathcal{S}x\boldsymbol{e}_y$ which may in general vary with both spherical radius $r$ and co-latitude $\beta$ in the star. $\mathcal{S}$ is the constant value locally of $-\varpi|\nabla\Omega(r,\beta)|$, and $\varpi$ is the distance from the axis of rotation (cylindrical radius). We define $x$ to be aligned with the variation of the shear flow $\boldsymbol{U}_0$, and so it is in general misaligned with respect to the local effective gravity vector $\boldsymbol{e}_g=(\cos \phi,0,\sin\phi)$ (which is approximately along the spherical radial direction), where the components of this vector (and all subsequent ones) are expressed using our local Cartesian coordinates. We define the angle $\Lambda$ (as in paper 2), such that $\hat{\boldsymbol{\Omega}} = (\sin{\Lambda},0,\cos{\Lambda})$. Since $\Lambda$ is the angle from the equator (perpendicular to $\hat{\boldsymbol{\Omega}}$) to the $x$-axis and the latter is misaligned from the spherical radial direction ($\boldsymbol{e}_g$) by $\phi$, our latitude angle is $\Lambda$+$\phi$. Our model is illustrated in Fig.~\ref{fig:Boxmodel}, where the top panel shows the orientation of the box with respect to the local effective gravity $\boldsymbol{e}_g$ and the bottom panel illustrates the various angles and the shear flow considered.

Papers 1 and 2 adopted a similar model but with shear acting radially (i.e. co-linear with the effective gravity), so that $\phi=0$ and $x$ lies along $\boldsymbol{e}_g$. It is however known that shear flows can vary more generally in stellar interiors (for example, we know the Sun has both radial and latitudinal differential rotation, at least in the vicinity of the convection zone), and such mixed radial and latitudinal shears could possibly have enhanced mixing properties. Thus a natural extension for us is to investigate the effects of varying the angle between the direction of the local shear and the effective gravity, which we describe using the angle $\phi$.

The equations governing perturbations to the shear flow, $\boldsymbol{U}_0$, are
\begin{align}
\label{eq1}
   & D\boldsymbol{u}+2\boldsymbol{\Omega}\times\boldsymbol{u}+\boldsymbol{u}\cdot\nabla{\boldsymbol{U}_0} = -\nabla{p}+\theta\boldsymbol{e}_g+\nu\nabla^2{\boldsymbol{u}}, \\
\label{eq2}
    & D{\theta}+\mathcal{N}^2 \boldsymbol{u} \cdot \boldsymbol{e}_\theta = \kappa\nabla^2{\theta}, \\
\label{eq3}
    & \nabla\cdot{\boldsymbol{u}}=0, \\
\label{eq4}
    & D \equiv \partial_{t}+\boldsymbol{u}\cdot\nabla+ \boldsymbol{U}_0\cdot\nabla.
\end{align}
Here we have defined a new temperature perturbation as $\theta$, which has the units of an acceleration and is related to the standard temperature perturbation $\tilde{T}$ via $\theta = \alpha{g}\tilde{T}$, where $\alpha$ is the thermal expansion coefficient and $g$ is the local gravitational acceleration. We have set the background reference density to unity. A background temperature (entropy) profile $T(x,z)$ has also been adopted, with uniform gradient $\alpha{g}\nabla{T} = \mathcal{N}^{2}\boldsymbol{e}_{\theta}$, where $\boldsymbol{e}_\theta = (\cos{\Gamma},0,\sin{\Gamma})$, where we note that our buoyancy frequency $\mathcal{N}^2 > 0$ in the radiative zone of a star. The effective gravity vector $\boldsymbol{e}_g$ lies approximately in the spherical radial direction, and is inclined to $x$ by an angle $\phi$. For ease of presentation when referring to ``radial" and ``latitudinal", we consider sufficiently slowly rotating stars that $\boldsymbol{e}_g$ lies approximately along the spherical radial direction, though the model itself does not require this restriction. Throughout our system we also have constant kinematic viscosity $\nu$ and thermal diffusivity $\kappa$, both of which are vital ingredients to study the GSF instability.

\begin{figure}
    \centering
    \includegraphics[width = 8cm , height = 7cm]{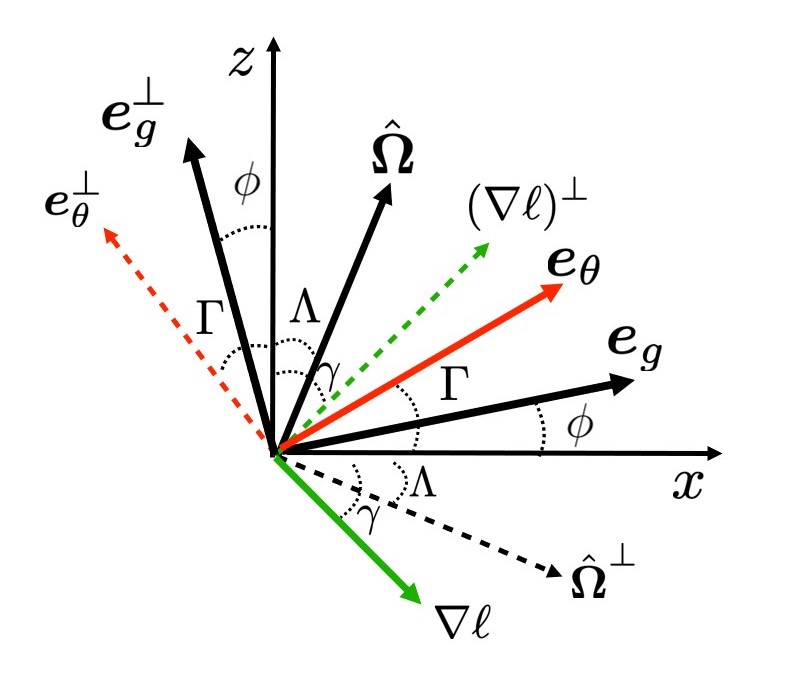}
    \caption{Illustration of the various vectors and corresponding angles in the $(x,z)$-plane as defined in the text. The cylindrical radial direction (along the equator) is along $\hat{\boldsymbol{\Omega}}^\perp$, and the rotation axis is along $\hat{\boldsymbol{\Omega}}$. The local radial direction is (approximately) along the effective gravity direction $\boldsymbol{e}_g$, with is misaligned with respect to the $x$ direction when $\phi$ is nonzero.}
    \label{Angles}
\end{figure}

We expect a realistic system would quickly adjust to be in thermal wind balance on a dynamical timescale. If this holds, we can then eliminate $\Gamma$ as a free parameter by assuming that a given basic flow $\boldsymbol{U}_0$ and its thermal state satisfies the thermal wind equation (TWE)
\begin{equation}
\label{TWE}
    2\Omega \mathcal{S}\sin{\Lambda}=\mathcal{N}^2\sin(\Gamma-\phi).
\end{equation}
It is left as a topic for future work to study how rapidly a stellar interior would adjust to satisfy thermal wind balance.
This equation is derived from the azimuthal component of the vorticity equation for the basic state and describes the degree of ``baroclinicity" in the system (i.e. the component of differential rotation along the rotation axis). When $\Lambda=0$, this indicates cylindrical differential rotation locally, where $\Omega(\varpi)$. This is a barotropic configuration, implying surfaces of constant density and pressure are aligned, i.e.~$\Gamma=\phi$. If $\sin\Lambda\ne 0$, this implies in general a misalignment between surfaces of constant density and pressure, such that $\Gamma$ and $\phi$ are unequal, which is referred to as a baroclinic configuration. In the latter, the rotation profile depends locally upon both spherical radius and co-latitude. The limit $\sin\Lambda=1$ implies $\Omega$ varies only with distance along its axis $\hat{\boldsymbol{\Omega}}$. The case $\phi=0$ refers (approximately) to spherical or shellular differential rotation, in which $\Omega$ depends locally upon spherical radius only. If $\phi=\pm90^\circ$, this approximately corresponds with purely latitudinal differential rotation, where $\Omega$ depends locally only upon co-latitude. These various cases are summarised in Table.~\ref{tableangles} below. We also illustrate the various angles in our problem in Fig.~\ref{Angles}.

\begin{table}
\begin{center}
\begin{tabular}{ |c|c|c|c| } 
 \hline
 $\Lambda$ & $\phi$ & Differential rotation profile & Baroclinic/barotropic? \\ 
 \hline
 $0$ & - & $\Omega(\varpi)$ (cylindrical) & barotropic \\ 
 $\pm 90^\circ$ & - & $\Omega(z)$ (axial variation only) & baroclinic \\ 
 - & 0 & $\Omega(r)$ (spherical/shellular) & baroclinic except at $\Lambda=0$ \\ 
 - & $\pm90^\circ$ & $\Omega(\beta)$ (horizontal/latitudinal) & baroclinic \\ 
  - & - & $\Omega(r,\beta)$ (arbitrary) & baroclinic in general \\ 
 \hline
\end{tabular}
\label{tableangles}
\caption{Table of differential rotation profiles as $\Lambda$ and $\phi$ are varied, where here $\beta$ is co-latitude, $z$ is distance along rotation axis, $r$ is spherical radius and $\varpi$ is cylindrical radius in this table.}
\end{center}
\end{table}

We take $\Omega^{-1}$ as our unit of time and the lengthscale $d$ to define our unit of length, where the latter is defined by 
\begin{equation}
    d = \left(\frac{\nu \kappa}{\mathcal{N}^2}\right)^{\frac{1}{4}}.
\end{equation}
The latter is chosen because the fastest growing mode typically has a wavelength $O(d)$. On this length-scale, the buoyancy timescale $\mathcal{N}^{-1}$ is equal to the geometric mean of the product of viscous ($d^2/\nu$) and thermal diffusion ($d^2/\kappa$) timescales \citep[see e.g.][for a discussion of its relevance for similar double-diffusive instabilities]{Radko2013}. This choice allows us to conveniently select a box size relative to the wavelengths of the fastest growing linear modes. We typically use a $(100d)^3$ box by default throughout our nonlinear simulations. We also define our dimensionless shear rate as $S= \mathcal{S}/\Omega$ and our dimensionless buoyancy frequency as $N=\mathcal{N}/\Omega$. 

In total, excluding the dimensions of the box (and numerical resolution), our problem has 5 independent parameters: $S$, Pr, $N^2$, $\Lambda$ and $\phi$ (since $\Gamma$ is constrained by Eq.~\ref{TWE}). Note that the Prandtl number, which is a crucial parameter in our system, is defined as
\begin{equation}
    \mathrm{Pr} = \frac{\nu}{\kappa}.
\end{equation}
The non-dimensional momentum and heat equations can be written,
\begin{align}
    & D\boldsymbol{u}+2\hat{\boldsymbol{\Omega}}\times \boldsymbol{u} - S u_{x} \boldsymbol{e}_{y} = -\nabla{p} + \theta {\boldsymbol{e}_{g}} + \mathrm{E} \nabla^{2}{\boldsymbol{u}}, \\
    & D \theta + N^2 \boldsymbol{u} \cdot \boldsymbol{e}_{\theta} = \frac{\mathrm{E}}{\mathrm{Pr}}{\nabla^2{\theta}},
\end{align}
along with incompressibility, where we have defined the local Ekman number $\mathrm{E}=\nu/(\Omega d^2)$ (which can be related to other parameters). Here lengths have been scaled using $d$, time has been scaled by $\Omega^{-1}$, velocities by $d\Omega^{-1}$ and temperatures by $d\Omega^{-2}$. Note that for simplicity hats have not been added to denote non-dimensional quantities.
 
\subsection{Numerical methods} 
 
Our nonlinear simulations are performed using a modified version of the Cartesian pseudo-spectral code SNOOPY \citep[e.g.][]{lesur2005}, as employed and tested in papers 1 and 2. SNOOPY uses a basis of shearing waves, meaning our Fourier modes have time-dependent wavevectors, which is equivalent to using shearing-periodic boundary conditions in a non-shearing frame. A periodic re-mapping procedure is applied for numerical reasons following \cite{Umurhan2004}. Each time-step is evaluated using a 3rd order explicit Runge-Kutta time-marching scheme and aliasing errors are eliminated using the 2/3-rule. Throughout our simulations we initialise the flow using solenoidal random noise with amplitude $10^{-3}$ for every wavenumber in the range $\hat{i},\hat{j},\hat{k}\in [1,21],$ where $k_{x}=\frac{2\pi}{L_x}\hat{i},k_{y}=\frac{2\pi}{L_y}\hat{j},$ and $k_{z}=\frac{2\pi}{L_y}\hat{k}$. Tables of our simulation parameters and of the linear properties of these cases are presented in Appendix~\ref{Tables}.

\section{Linear theory}\label{LinearTheory}
\subsection{Dispersion relation for axisymmetric modes}
We consider axisymmetric modes (with azimuthal wavenumbers $k_y = 0$) as these are likely to be the most unstable, and follow closely the methods in paper 2. \citet{knobloch.et.al.1982} argued that although non-axisymmetric baroclinic modes are unstable whenever surfaces of constant pressure and temperature are misaligned \citep[at least with a rigid boundary, as also found by][]{Rashid2008}, because the buoyancy frequency greatly exceeds the shear rate in most astrophysical situations, baroclinic modes will only be unstable for wavelengths larger than the stellar radius, which justifies our focus on axisymmetric modes. In this section we build upon \citet{knobloch.et.al.1982}, \citet{knobloch1982} and papers 1 and 2 by exploring in more detail the effects of varying $\phi$ on the linear stability problem, and we will derive several new results as well as reproducing some prior ones. We will explore graphically and in more detail the consequences of varying $\phi$ than in prior work, and we will for the first time compute the linear growth rates and wavenumbers of the most unstable modes in this problem as $\phi$ is varied.

We start by seeking solutions proportional to $\mathrm{exp}(\mathrm{i}k_x{x} + \mathrm{i}k_z{z} + st)$, where $k_x$ and
$k_z$ are the wavevector components in the $x$ and $z$ directions in the meridional plane. In our model $x$ is radial if $\phi=0$, but more generally it is aligned with the gradient in angular velocity, and $z$ is perpendicular to it. We define the complex growth rate $s=\sigma + \mathrm{i} \omega$, where the growth (decay) rate $\sigma\in\mathbb{R}$ and the oscillation frequency $\omega\in\mathbb{R}$. We manipulate the linearised versions of Eqs.~\ref{eq1}--\ref{eq4} for such perturbations, and define $s_{\nu}=s+\nu{k^2}$ and $s_{\kappa}=s+\kappa {k^2}$, to obtain
the cubic dispersion relation 
\begin{equation}
\label{DR}
    s^{2}_{\nu}s_{\kappa}+a s_{\kappa}+b s_{\nu}=0,
\end{equation}
where
\begin{align}
a &= \frac{2}{\varpi}(\hat{\boldsymbol{k}}\cdot\boldsymbol{\Omega})(\hat{\boldsymbol{k}}\cdot (\nabla\ell)^\perp), \\
&=\frac{2\Omega}{k^2}(s_{\Lambda}k_x+c_{\Lambda}k_z)(2\Omega k_x s_{\Lambda}+(2\Omega c_{\Lambda}-\mathcal{S})k_z),
\end{align}
and
\begin{align}
b&= \mathcal{N}^2(\hat{\boldsymbol{k}}\cdot\boldsymbol{e}_{\theta}^\perp)(\hat{\boldsymbol{k}}\cdot{\boldsymbol{e}_{g}^\perp}), \\
 &= \frac{\mathcal{N}^2}{k^2}(k_z{c_{\Gamma}}-k_x{s_{\Gamma}})(k_{z}c_{\phi}-k_x{s_{\phi}}).
\end{align}
In the above we have used $c_\Lambda$ and $s_\Lambda$ to refer to $\cos\Lambda$ and $\sin\Lambda$ for brevity, and similarly for trigonometric functions with other arguments. We also define the local angular momentum gradient
\begin{eqnarray}
 \nabla {\ell} =\nabla({{\varpi}^2{\Omega}})
          =\varpi(2\Omega{c_{\Lambda}}-\mathcal{S},0,-2\Omega{s_{\Lambda}}),
          =|\nabla{\ell}|(c_{\gamma},0,-s_{\gamma}),
\end{eqnarray}
which has magnitude
\begin{equation}
   |\nabla{\ell}|^2=\varpi^2\left({\mathcal{S}^2}+4{\Omega}(\Omega-\mathcal{S}c_{\Lambda})\right).
\end{equation}
The normal to the local angular momentum gradient is then
\begin{equation}
 (\nabla{\ell})^\perp = \varpi(2\Omega s_{\Lambda},0,2\Omega c_\Lambda-\mathcal{S})
 =|\nabla\ell|(s_\gamma,0,c_\gamma).
\end{equation}
We also define the vector perpendicular to the effective gravity
\begin{equation}
\boldsymbol{e}_{g}^\perp=(-s_{\phi},0,c_{\phi}),
\end{equation}
and the normal to stratification surfaces
\begin{equation}
\boldsymbol{e}_{\theta}^\perp=(-s_{\Gamma},0,c_{\Gamma}),
\end{equation}
as well as the cylindrical radial direction
\begin{equation}
\hat{\boldsymbol{\Omega}}^{\perp}=(c_{\Lambda},0,-s_{\Lambda}).
\end{equation}
The baroclinic shear (along the rotation axis) is 
\begin{align}
\label{baroshear}
\hat{\boldsymbol{\Omega}}\cdot(\nabla{\ell})&=\varpi((2\Omega{c_{\Lambda}}-\mathcal{S})s_{\Lambda}-2\Omega{s_{\Lambda}{c_{\Lambda}}})=-\mathcal{S}\varpi s_{\Lambda} \\
&=
|\nabla{\ell}|(s_{\Lambda}c_{\gamma}-s_{\gamma}c_{\Lambda})=-|\nabla{\ell}|s_{\gamma-\Lambda},
\end{align}
Hence, the angle between the rotation axis and local angular momentum gradient is $\cos^{-1}\left({-\frac{\mathcal{S}{\varpi}s_{\Lambda}}{|\nabla{\ell}|}}\right)$. We also define a modified Richardson number,
\begin{equation}
\mathrm{R}= \frac{\mathcal{N}^2\varpi}{2\Omega|\nabla{\ell}|},
\end{equation}
which is a potential measure of the stabilising effects of the stratification against the destabilising effects from the angular momentum gradient \citep{knobloch.et.al.1982}. We also have
\begin{equation}
\varpi\mathcal{S}s_\Lambda=-|\nabla{\ell}|s_{\gamma-\Lambda}.
\end{equation}
This means that, with some rearranging, the thermal wind equation (Eq.~\ref{TWE}) can be written as 
\begin{equation}
    \frac{2\Omega|\nabla{\ell}|}{\varpi}s_{\gamma-\Lambda}=\mathcal{N}^{2}s_{\Gamma-\phi},
\end{equation}
or in the form
\begin{equation}
\label{modTWE}
   s_{\gamma-\Lambda}=\mathrm{R}s_{\Gamma-\phi}.
\end{equation}

\subsection{Non-diffusive (in)stability}

For the non-diffusive case we take $\nu=\kappa=0$, giving a reduced dispersion relation
\begin{equation}
   s^2=-(a+b).
\end{equation}
A negative (or zero) real component of the growth rate is required for stability, thus we have stability when
\begin{equation}
  a+b\geq 0,
\end{equation}
To find a criterion independent of $\boldsymbol{k}$ we divide by $k_z$ and define $q = k_x/k_z$ to obtain
\begin{equation}
 ({q}s_{\Lambda}+c_{\Lambda})({q}s_{\gamma}+c_{\gamma})+\mathrm{R}({q}s_{\Gamma}-c_{\Gamma})({q}s_{\phi}-c_{\phi})\geq 0.
\end{equation}
This is a quadratic in $q$, hence for stability it 
must have no real roots, and a positive discriminant implies
\begin{equation}
(s_{{\Lambda}+\gamma}-\mathrm{R}s_{\Gamma+\phi})^2-4(s_{\Lambda}s_{\gamma}+\mathrm{R}s_{\Gamma}s_{\phi})(c_{\Lambda}c_{\gamma}+\mathrm{R}c_{\Gamma}c_{\phi})
< 0.
\end{equation}
We then use Eq.~\ref{modTWE} to eliminate R. After simplifying, we obtain the stability criterion
\begin{equation}
\frac{s_{\gamma+\Gamma}s_{\gamma-\Lambda}s_{\Lambda+\phi}}{s_{\Gamma-\phi}}> 0.
\label{adstab}
\end{equation}
Alternatively, this can be written $R s_{\gamma+\Gamma}s_{\Lambda+\phi}> 0$,
which is equivalent to $\nabla \ell \cdot \boldsymbol{e}_\theta^\perp<0$ when $\Lambda+\phi>0$ and $\nabla \ell \cdot \boldsymbol{e}_\theta^\perp>0$ when $\Lambda+\phi<0$. Together these are equivalent to the Solberg-H\o iland criteria, that the specific angular momentum should not decrease outwards (from the rotation axis) along an isentropic surface for adiabatic stability. When this condition is violated, we expect more violent dynamical instabilities to operate than the GSF instability that is our primary focus. Taking the limit $\phi\to 0$, we recover Eq.~30 in paper 2.

\subsection{GSF instability}

The presence of thermal diffusion ($\kappa\neq0$) offers a mechanism to relax the stabilising effects of gravitational buoyancy and allows operation of the GSF instability, as long as it overcomes viscous diffusion ($\nu\neq0$). When the constant term in our cubic dispersion relation becomes negative we have instability. Hence we may write the diffusive instability criterion as
 \begin{equation}
\nu^2\kappa k^6+a\kappa k^2 + b \nu k^2<0.
 \end{equation}
leading to the necessary instability condition ($\nu^2\kappa k^6$ is stabilising)
\begin{equation}
\label{instab}
    a+\mathrm{Pr} b <0 .
\end{equation}
For adiabatic stability $a+b\geq 0$, so if $\mathrm{Pr}<1$, it is possible for \ref{instab} to be satisfied, implying $(1-\mathrm{Pr})b>0$. Hence at small Pr, $b$ must be positive, so it is necessary that $a$ is negative for instability. In order to find a more stringent condition for instability, we follow a similar approach to the non-diffusive case involving the sign of the discriminant. The instability criterion is
\begin{equation}
\label{GSFcritfull}
(s_{\Lambda+\gamma}-\mathrm{RPr} s_{\Gamma+\phi})^2-4(s_\Lambda s_\gamma+\mathrm{RPr} s_\Gamma s_\phi)(c_{\Lambda}c_{\gamma}+\mathrm{RPr} c_\Gamma c_\phi)>0.
\end{equation}

In the limit of strong stratification, surfaces of constant pressure and density align and Eq.~\ref{TWE} implies $\Gamma \xrightarrow[]{} \phi $. This leads to 
 \begin{equation}
   \mathrm{R Pr} < \frac{ s_{\Lambda-\gamma}^2}{4 s_{\phi+\Lambda}s_{\phi+\gamma}},
\end{equation}
which \citep[cf Eq.2.30 of][]{knobloch1982} can also be written in terms of the usual Richardson number as
 \begin{equation}
 \label{GSFstab}
   \mathrm{Ri Pr} < \frac{ s_{\gamma}s_{\Lambda}}{4 s_{\phi+\Lambda}s_{\phi+\gamma}},
\end{equation}
for instability, by noting that $\mathrm{R}=\mathrm{Ri} (s_{\gamma-\Lambda}^2/(s_\gamma s_\Lambda))$ (see Appendix~\ref{Appendix1}).
This criterion reduces to $\mathrm{Ri Pr}<1/4$ for instability when $\phi\to 0$ (shellular differential rotation), as obtained in paper 2. However, instability is possible for weaker differential rotation for non-zero $\phi$.

We analyse the asymptotic limits as $\mathrm{Pr}\to 0$ in detail in Appendix~\ref{Appendix1}, both for the strongly driven case (where $\mathrm{RiPr}\to 0$) and the weakly driven case (where $\mathrm{RiPr}=O(1)$), where we derive several new results. The growth rate in the limit of small Pr where the instability is strongly driven (but adiabatically stable) can be determined by considering the limit $\mathrm{RiPr}\to 0$ as $\mathrm{Pr}\to 0$. In this regime, we find $s=\sqrt{-a}$, and hence both the maximum growth rate and the wedge angle of instability in the $(k_x,k_z)$-plane are independent of $\phi$ for fixed $\Lambda$. In reality though, we are usually interested in a fixed latitude $\Lambda+\phi$, in which case the growth rate and unstable wedge angles do depend on $\phi$, being maximised for mixed radial and horizontal shears rather than purely radial ones. In this regime, the maximum growth rate and wave-vector magnitude can be predicted from Eq.43-44 of \cite{barker2020}, which we reproduce here:
\begin{align}
\label{RiPr0}
    s^2&=\frac{2\Omega|\nabla \ell|}{\varpi}\sin^2\left(\frac{1}{2}\left(\gamma-\Lambda\right)\right), \\
    k^4&=\frac{1}{2d^4}\sin^2\left(\frac{\gamma+\Lambda}{2}\right).
\end{align}
These will be plotted later in Fig.~\ref{InitialGrowth} for $S=2$ as a function of $\phi$ for various latitudes $\Lambda+\phi$. On the other hand, if $\mathrm{RiPr}=O(1)$ as $\mathrm{Pr}\to 0$, the growth rate and unstable wedge may depend on $\phi$ for a fixed $\Lambda$ as well (as we show in Appendix~\ref{appendixsub1}).

\subsection{Oscillatory GSF instability}
\label{Osconset}

Our cubic dispersion relation Eq.~\ref{DR} also allows oscillatory instabilities, for which $s=\sigma+\mathrm{i}\omega$ and $\omega\ne 0$ at onset. These are essentially weakly destabilised inertia-gravity waves gaining energy from the differential rotation or baroclinicity. To derive a criterion for onset ($\sigma=0$) in this case, we substitute $s=\mathrm{i}\omega$ into Eq.~\ref{DR}, consider the limit of strong stratification for which $\Gamma\to \phi$, neglect terms with higher powers of $k$ \citep[following][]{knobloch1982}, and equate real and imaginary parts to obtain:
\begin{align}
\nonumber
    &-\omega^2(1+2\mathrm{Pr})(1+q^2)+2\Omega\frac{|\nabla \ell|}{\varpi} (c_\gamma +q s_\gamma)(c_\Lambda+q s_\Lambda) \\
    \label{osc1}
    &\hspace{4.5cm} +\mathcal{N}^2\mathrm{Pr}(c_\phi-q s_\phi)^2=0, \\
    \nonumber
    &\omega\left(-\omega^2 (1+q^2)+2\Omega\frac{|\nabla \ell|}{\varpi}(c_\gamma +q s_\gamma)(c_\Lambda+q s_\Lambda) \right.\\ 
    \label{osc2}
    &\hspace{4.5cm}\left.+\mathcal{N}^2(c_\phi-q s_\phi)^2 \right)=0.
\end{align}
Since we are looking for oscillatory instabilities, we omit the solution with $\omega=0$, so we can combine both of the above to eliminate $\omega^2$, giving the quadratic
\begin{align}
    \mathrm{Pr}(c_\gamma+q s_\gamma)(c_\Lambda+q s_\Lambda)+\mathrm{R}(1+\mathrm{Pr})(c_\phi-q s_\phi)^2=0.
\end{align}
We require a positive discriminant, so that
\begin{align}
\label{osconset1}
    \mathrm{R}<\frac{\mathrm{Pr}}{2(1+\mathrm{Pr})}\frac{s_{\Lambda-\gamma}^2}{s_{\phi+\Lambda}s_{\phi+\gamma}},
\end{align}
for oscillatory instability to onset \citep[cf Eq.2.32 of][]{knobloch1982}. Equivalently,
\begin{align}
\label{osconset2}
    \mathrm{Ri}<\frac{\mathrm{Pr}}{2(1+\mathrm{Pr})}\frac{s_\gamma s_\Lambda}{s_{\phi+\Lambda}s_{\phi+\gamma}}.
\end{align}
This can be contrasted with Eq.~\ref{GSFstab} for direct instability (steady modes with $\omega=0$). The ratio of the quantity $\mathrm{RiPr}$ predicted by Eq.~\ref{GSFstab} to that from Eq.~\ref{osconset2} is
\begin{align}
\label{oscomparison}
    \frac{(1+\mathrm{Pr})}{2\mathrm{Pr}^2}\to \infty \quad \mathrm{as} \;\; \mathrm{Pr}\to 0. 
\end{align}
Hence, GSF instability occurs first as a direct instability at onset for small Pr \citep[in agreement with][]{knobloch1982}, since oscillatory instability requires a much smaller value of RiPr.

To determine the properties of the modes at onset in the limit $\mathrm{Pr}\to 0$ (and RPr$\to 0$), we can solve Eqs.~\ref{osc1} and \ref{osc2} to obtain a preferred wavevector orientation and squared frequency 
\begin{align}
    q & = \cot \phi, \\
    \omega^2&=\frac{2\Omega |\nabla \ell|}{\varpi} s_{\gamma+\phi}s_{\Lambda+\phi}.
\end{align}
The first result implies that the waves have wavevectors $\boldsymbol{k}$ that lie approximately along $\boldsymbol{e}_g$. We have determined numerically for a range of parameters that these modes lie between $\boldsymbol{e}_g$ and $\boldsymbol{e}_\theta$ and that they always have smaller growth rates than the fastest growing direct GSF instability.

\subsection{Illustrative results from linear theory}

\begin{figure*}
  \subfigure[]{\includegraphics[
    trim=0cm 0cm 0cm 0cm,clip=true,
    width=0.49\textwidth]{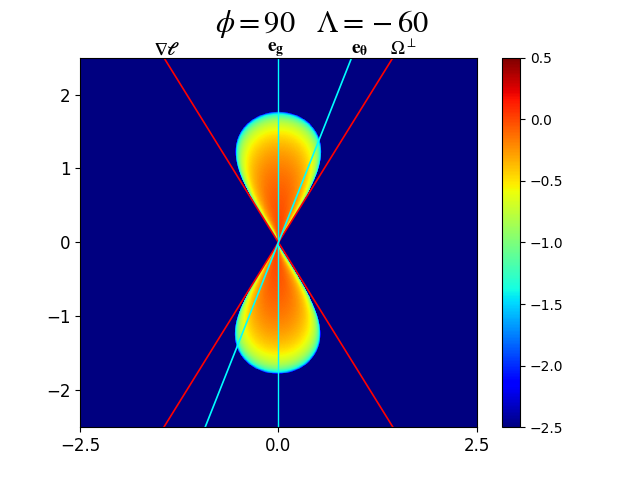}}
    \subfigure[]{\includegraphics[
    trim=0cm 0cm 0cm 0cm,clip=true,
    width=0.49\textwidth]{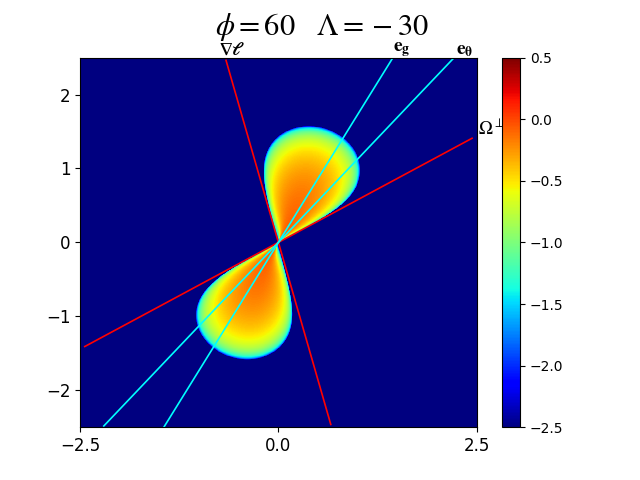}}
    \subfigure[]{\includegraphics[
    trim=0cm 0cm 0cm 0cm,clip=true,
    width=0.49\textwidth]{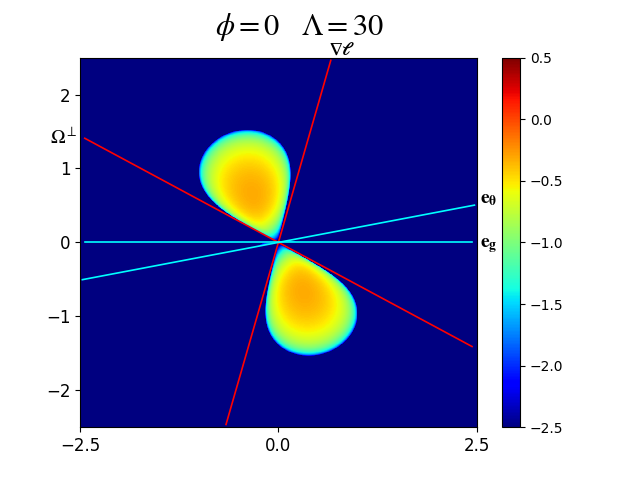}}
    \subfigure[]{\includegraphics[
    trim=0cm 0cm 0cm 0cm,clip=true,
    width=0.49\textwidth]{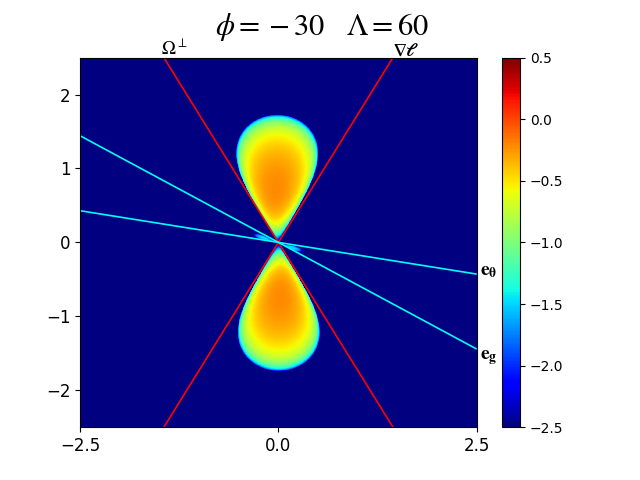}}
    \subfigure[]{\includegraphics[
    trim=0cm 0cm 0cm 0cm,clip=true,
    width=0.49\textwidth]{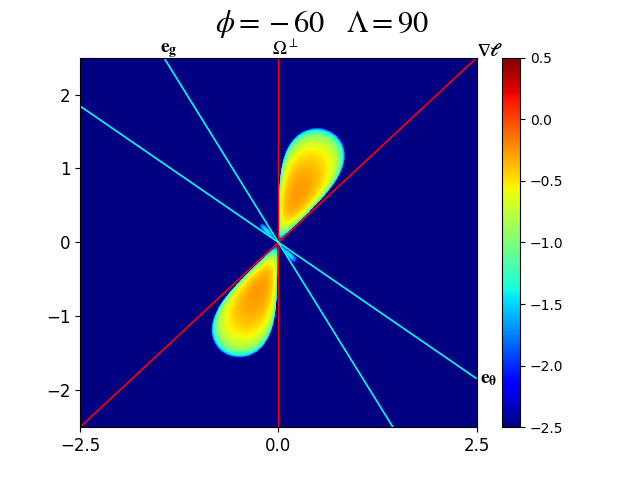}}
    \subfigure[]{\includegraphics[
    trim=0cm 0cm 0cm 0cm,clip=true,
    width=0.49\textwidth]{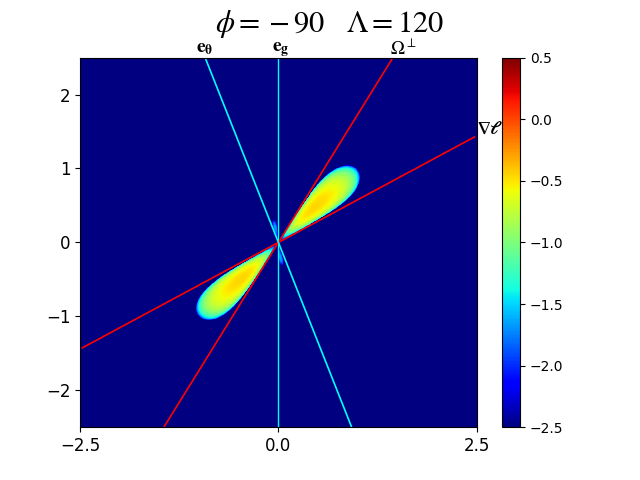}}
  \caption{Figures of linear growth rate $\log_{10} (\sigma/\Omega)$ for the axisymmetric GSF (or adiabatic) instability for various $\phi$ on the $(k_x, k_z)$-plane for $\mathcal{N}^2/\Omega^2 = 10$, Pr$=10^{-2}$, $\mathcal{S}/\Omega=2$, at a fixed latitude $\Lambda + \phi$ = $30^\circ$. Here we vary $\phi$ in multiples of $30^\circ$ from $-90^\circ$ to $90^\circ$. GSF (or adiabatically) unstable modes are contained within the wedge bounded by the two vectors $\hat{\boldsymbol{\Omega}}^\perp$ and $\nabla \ell$. Note that panels (a) and (b) are adiabatically unstable according to Eq.~\ref{adstab}. We also observe a smaller wedge outside this region in panels (d), (e) and (f) containing weakly growing oscillatory modes which are bounded by the light blue lines. The fastest growing modes (darkest red) in adiabatically unstable cases occur for $k \to 0$ suggesting that the presence of diffusion leads to the preference of the largest possible wavelengths in this regime. This is in comparison to the GSF cases where the darkest areas have a unique non-zero wavenumber and hence a preferred wavelength in real space.
}
  \label{Lobes}
\end{figure*}

 Fig.~\ref{Lobes} shows the base 10 logarithm of the growth rate $\sigma$ from solving Eq.~\ref{DR} on the $(k_{x},k_{z})$-plane for axisymmetric instabilities. In the majority of our investigations, we fix the latitude $\Lambda +\phi$ and choose $S=2$, $N^2=10$ and Pr $=10^{-2}$ to allow a direct comparison with paper 2. We then vary $\phi$ (and consequently $\Lambda$) to probe the effects of shear orientation on the linear instability. Additional cases, including fixing $\Lambda = 60^\circ$ (see Appendix \ref{LobesAppendix}) as well as probing the effects of Pr and $N^2$ were also considered. In these figures we also plot the vectors $\hat{\boldsymbol{\Omega}}^\perp$ and $(\nabla{\ell})$ as the solid red lines. These lines delineate the wedge within which $a<0$ and GSF-unstable modes are expected. We also plot the vectors {$\boldsymbol{e}_g$}  and {$\boldsymbol{e}_\theta$} as the light blue lines, and the wedge between them is where $b<0$, and oscillatory modes can be found. The angles of the red and blue lines can be found from Table~B1 and Figure~\ref{Angles}.

The main feature seen in all of these plots are ``primary lobes" corresponding to either the diffusive GSF instability or to the adiabatic instability when Eq.~\ref{adstab} is violated. These lobes contain (directly) unstable modes (with $\omega=0$) with a preferred wavevector orientation lying between the AM gradient $\nabla{\ell}$, and the line perpendicular to the rotation axis $\hat{\boldsymbol{\Omega}}^{\perp}$. Since $S \sim \Omega$ and given that our unit of time is $\Omega^{-1}$, the fastest growing modes have growth rates O(1) and are observed to lie along the line that is approximately half-way between these two vectors (as explained in Appendix \ref{Appendix1}). Note that this wedge is perpendicular to the physical wedge within which the GSF (or adiabatically) unstable mode displacements (and velocity perturbations) arise due to the incompressibility condition $\boldsymbol{k} \cdot \boldsymbol{u} = 0$.

\begin{figure*}
    \subfigure[Latitude $ \Lambda + \phi = 0^\circ$]{\includegraphics[
    trim=0cm 0cm 0cm 0cm,clip=true,
    width=0.49\textwidth]{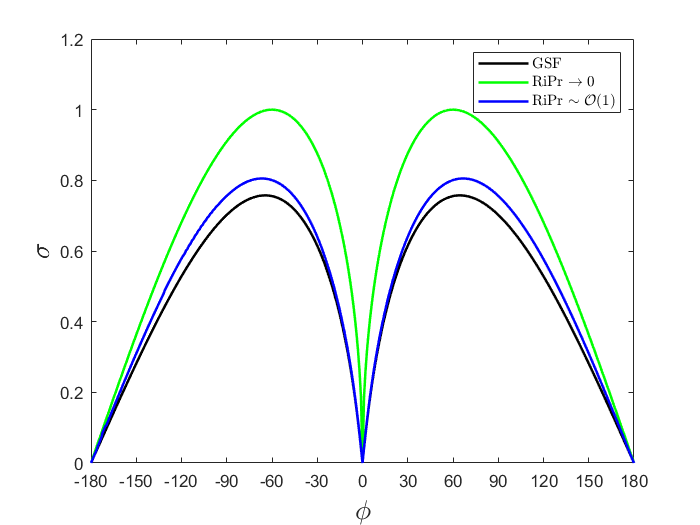}}
    \subfigure[Latitude $\Lambda +\phi = 30^\circ$]{\includegraphics[
    trim=0cm 0cm 0cm 0cm,clip=true,
    width=0.49\textwidth]{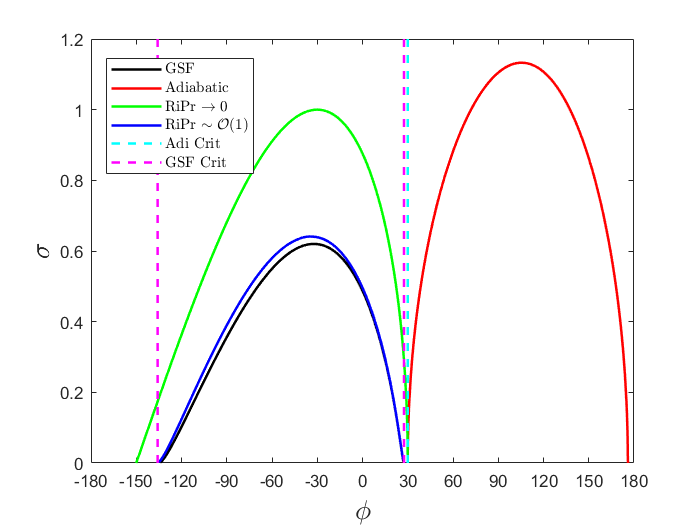}}
    \subfigure[Latitude $\Lambda +\phi = 60^\circ$]{\includegraphics[
    trim=0cm 0cm 0cm 0cm,clip=true,
    width=0.49\textwidth]{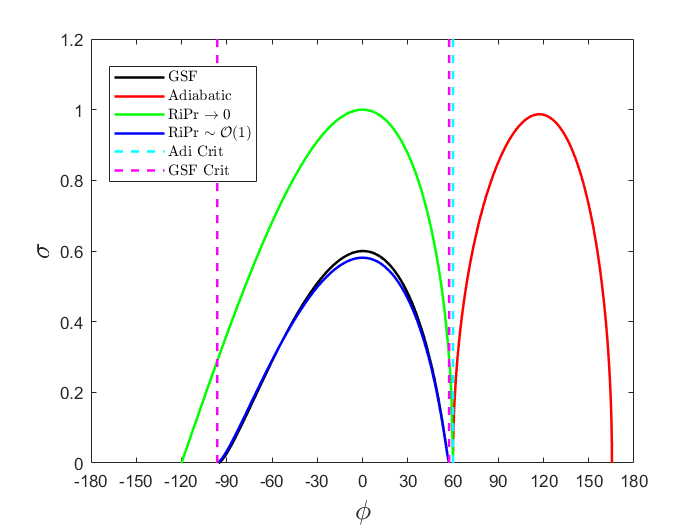}}
    \subfigure[Latitude $\Lambda + \phi = 90^\circ$]{\includegraphics[
    trim=0cm 0cm 0cm 0cm,clip=true,
    width=0.49\textwidth]{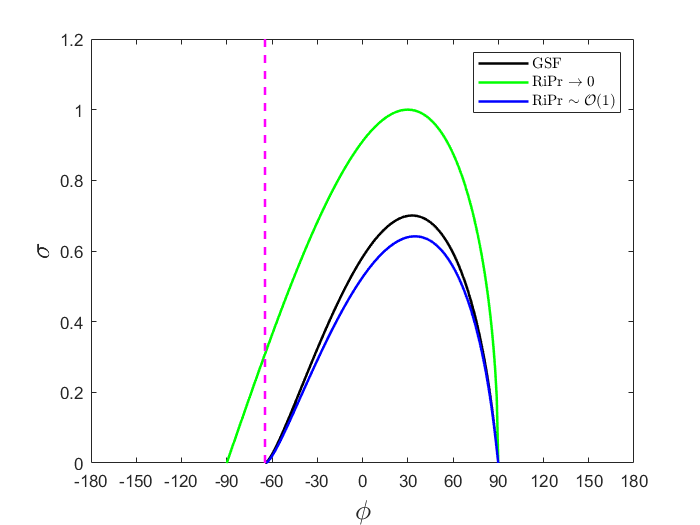}}
  \caption{A selection of figures comparing the maximum linear growth rates in both the adiabatic (red) and diffusive (black) regimes for $S=2, \mathrm{Pr}=10^{-2}, N^2=10$. The light blue dashed line corresponds to the critical $\phi$ for onset of adiabatic instability predicted by Eq.~\ref{adstab}. Onset of GSF instability as predicted by Eq.~\ref{GSFstab} is shown as magenta dashed lines. Predictions for the growth rate in the limits assuming $\mathrm{RiPr}\sim O(1)$ and $\mathrm{RiPr}\to 0$ are shown as the blue and green lines, respectively.
}
  \label{InitialGrowth}
\end{figure*}

We observe in Fig.~\ref{Lobes} that the orientation of the primary lobes and the maximum growth rates at a fixed latitude $\Lambda+\phi=30^\circ$ depends strongly on $\phi$. In particular, we observe the fastest growth (and the widest primary lobes) at this latitude for $\phi\in[60^\circ,90^\circ]$, which are also adiabatically unstable according to Eq.~\ref{adstab} (see also panel (b) in Fig.~\ref{InitialGrowth} that displays the maximum growth rate vs $\phi$). For adiabatically stable but GSF unstable cases, here for $\phi<30^\circ$, we observe somewhat slower growth (but still $O(1)$) and lobes that narrow as $\phi\to-90^\circ$. The fastest growing modes (darkest red) in adiabatically unstable cases occur for $|\boldsymbol{k}|=k \to 0$, suggesting that with the presence of diffusion the dominant modes grow on the largest possible wavelengths (without diffusion these modes do not have a preferred wavevector magnitude $k$, only a preferred wavevector orientation). This is in comparison to the GSF cases where the darkest areas have a unique non-zero wavenumber, and hence a preferred wavelength in real space.

On the other hand, we have observed that for a fixed $\Lambda$ (see Fig.~\ref{Lobes2} in Appendix \ref{LobesAppendix}), varying $\phi$ alone does not change the orientation or sizes of the primary lobes, but it does modify the maximum growth rates, with cases with horizontal shears for $\phi\sim 90^\circ$ exhibiting faster growth than radial shears with $\phi\sim 0^\circ$. This result might be expected because radial motions will be preferentially inhibited to a greater extent by the stable stratification. Decreasing Pr to a smaller, more realistic, value substantially increases the size of the primary lobes for a given RiPr. This result holds even in the presence of a more realistic and much larger buoyancy frequency (not shown) as the key parameter for diffusive instability is RiPr.

 \begin{figure*}
    \subfigure[$ \Lambda + \phi = 0^\circ$]{\includegraphics[
    trim=0cm 0cm 0cm 0cm,clip=true,
    width=0.49\textwidth]{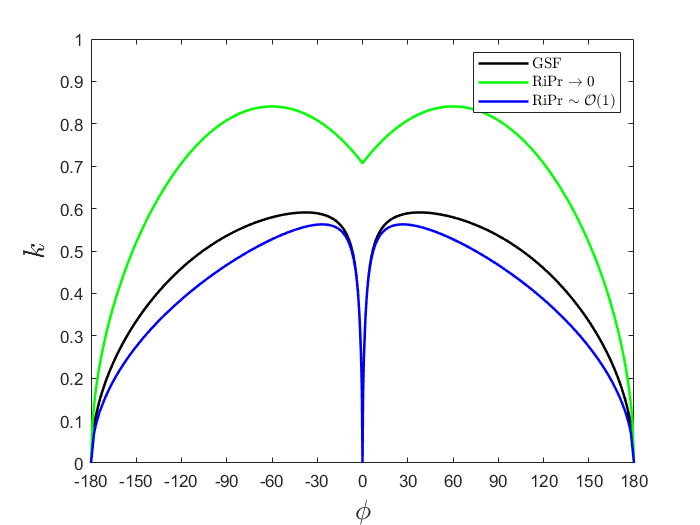}}
    \subfigure[$\Lambda +\phi = 30^\circ$]{\includegraphics[
    trim=0cm 0cm 0cm 0cm,clip=true,
    width=0.49\textwidth]{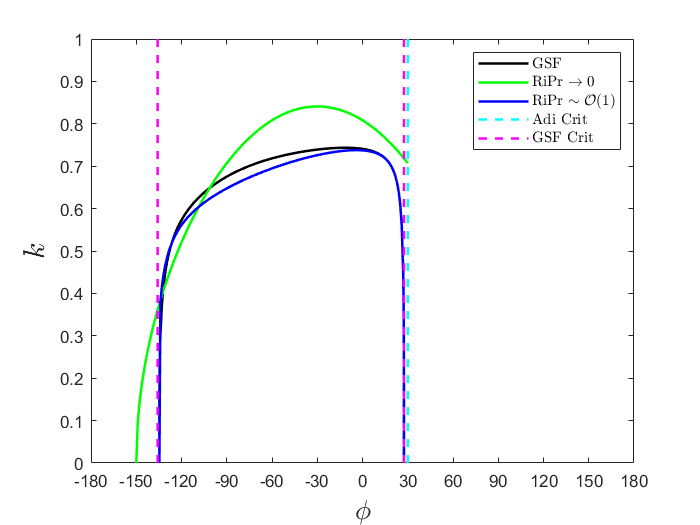}}
    \subfigure[$\Lambda +\phi = 60^\circ$]{\includegraphics[
    trim=0cm 0cm 0cm 0cm,clip=true,
    width=0.49\textwidth]{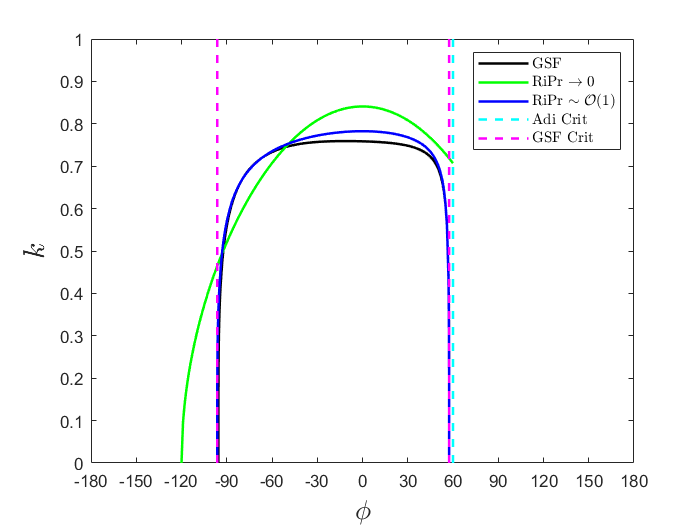}}
        \subfigure[$\Lambda + \phi = 90^\circ$]{\includegraphics[
    trim=0cm 0cm 0cm 0cm,clip=true,
    width=0.49\textwidth]{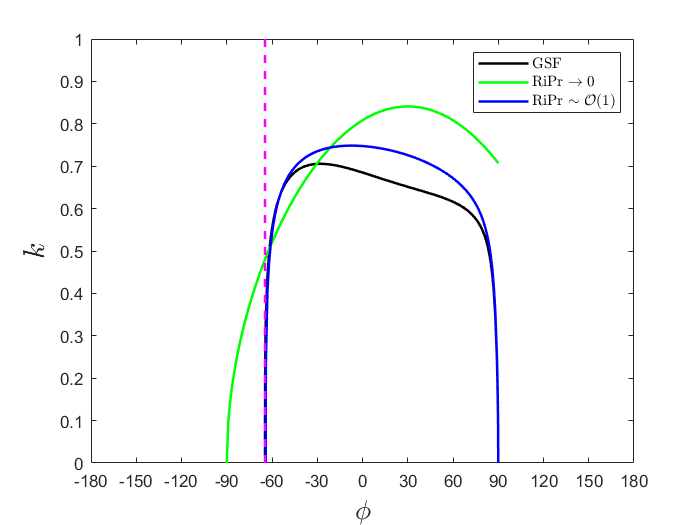}}
  \caption{A selection of figures comparing the fastest growing wavevector magnitudes against $\phi$ at the latitudes $0^\circ$, $30^\circ$, $60^\circ$, $90^\circ$ for $S=2, \mathrm{Pr}=10^{-2}, N^2=10$. Onset of adiabatic instability as predicted by Eq.~\ref{adstab} is shown as a blue dashed line. Onset of GSF instability predicted by Eq.~\ref{GSFstab} is shown as the magenta dashed lines, and onset of adiabatic instability predicted by Eq.~\ref{adstab} is indicated by the light blue dashed lines. We plot predictions from solving our cubic numerically (black) as well as the corresponding asymptotic predictions assuming $\mathrm{RiPr}\sim O(1)$ (blue) and $\mathrm{RiPr}\to 0$ (green).
}
  \label{MagK}
\end{figure*}

We additionally note the appearance in Fig.~\ref{Lobes} (and Fig.~\ref{Lobes2}) of two further, but much smaller ``secondary lobes", which are barely visible when $\phi=0$ and were not previously identified in paper 2 owing to the wavenumber resolution and colour-scale adopted for their figures. These lobes are most visible here for $\phi=-30^\circ$ and $\phi=-90^\circ$, and correspond to the oscillatory ($\omega\ne 0$) axisymmetric baroclinic instabilities that can develop in this system \citep{McIntyre1970,knobloch1982,LeBars2021,Labarbe2021}. These are oscillatory modes -- as noted in section~\ref{Osconset}  essentially weakly excited inertia-gravity waves -- in contrast to the usual GSF (or adiabatic) instability that onsets as a direct instability within the primary lobes. The smaller secondary lobes are likely to be overpowered by the GSF instability in stellar interiors (see \S~\ref{Osconset}), as their maximum growth rates are generally much smaller than the primary lobes for $\mathrm{Pr}\ll 1$, however they could potentially become important in the presence of strong chemical gradients where there are stricter criteria for instability \citep{knobloch.et.al.1982}.

Here we present a selection of figures showing how the maximum linear growth rates and wavenumber magnitudes vary with $\phi$. We show results for both the GSF instability (black) and the adiabatic instability (red) by solving the dispersion relations directly, as well as the growth rate in the asymptotic limits as $\mathrm{RiPr} \rightarrow 0$ (green; based on Eq.~\ref{RiPr0}) and $\mathrm{RiPr} \sim \mathcal{O}(1) $ (blue; based on Appendix~\ref{Appendix1}, Eqs.~\ref{eq:1.25} and \ref{eq:1.26}). 

In Fig.~\ref{InitialGrowth}, we plot the growth rate of a given instability against $\phi$ (in the full range between $\pm 180^\circ$), whilst keeping the latitude $\Lambda+\phi$ fixed, and setting $S=2$, $N^2=10$, Pr $=10^{-2}$. 

At the equator ($\phi+\Lambda=0$), $\phi=0$ is marginally stable with $S=2$, corresponding to Rayleigh stability. This is also true for any case where $\phi$ is such that $\Lambda=0$, which corresponds with cylindrical differential rotation ($\Omega(\varpi)$ only) which is neutrally stable for $\mathcal{S}=2$ (constant angular momentum as a function of cylindrical radius $\varpi$). Interestingly, we see that the effects of varying $\phi$ are symmetric about zero at the equator, and there is no adiabatic instability in this case. The fastest growing instability occurs for mixed radial/latitudinal shears with $\phi\sim 60^\circ$, rather than purely latitudinal shears with $\phi\sim 90^\circ$, which is intuitively surprising. We observe that the growth rate is in very good agreement with the prediction from the asymptotic limit as $\mathrm{RiPr}\sim O(1)$, but it is smaller than the ``upper bound" predicted by considering its evaluation in the limit as $\mathrm{RiPr}\to 0$.

Moving away from the equator, at $\phi+\Lambda=30^\circ$ latitude we again see the expected marginal stability when $\phi\sim 30^\circ$, but we also observe onset of adiabatic instability between $\phi =30^\circ$ and $\phi =90^\circ$. Only diffusive instability is observed for $\phi <30^\circ$, but adiabatic instability dominates instead when $\phi >30^\circ$, which typically has a larger growth rate than GSF unstable modes. The transition between diffusive and adiabatic instability is given by the dashed blue line, which shows the critical value of $\phi$ predicted by Eq.~\ref{adstab}, and the magenta dashed lines indicate the bounds for GSF instability given by Eq.~\ref{GSFstab}. These are in excellent agreement with our numerical results. Note that there is a tiny nonzero range of $\phi$ for which neither instability occurs near $\phi\sim 30^{\circ}$ between the magenta and light blue dashed lines. This is a finite Pr effect due to viscosity, which is not present in the $\mathrm{RiPr}\to 0$ prediction (in green; that matches the light blue dashed line). We observe that the numerically-computed growth rate of the GSF instability from directly solving the cubic (black) is again in very good agreement with the prediction from the asymptotic limit as $\mathrm{RiPr}\sim O(1)$, and is somewhat smaller than the prediction valid when $\mathrm{RiPr}\to 0$. The latter case also occurs for a wider range of $\phi$, occurring for $\phi>-150^\circ$ which is stable for the black and green curves.

The $60^\circ$ latitude case shows similar behaviour to the $30^\circ$ latitude case except that marginal stability for adiabatic instability occurs now at $\phi = 60^\circ$. At the pole ($\Lambda + \phi =90^\circ$), we firstly see that the system is stable to both diffusive and adiabatic instabilities for $-180^\circ <\phi< -90^{\circ}$ and onset for GSF occurs at $\phi =-64^\circ$ and lasts until $\phi=90^{\circ}$. Adiabatic instability is not observed for any $\phi$ at the poles. The transition to GSF instability is predicted by Eq.~\ref{GSFstab}, plotted as the dashed magenta line, which is in excellent agreement with our numerical results. This equation is singular when $\phi\to 90^\circ$ and so is not plotted there. Again, we see that the prediction in the asymptotic limit assuming $\mathrm{RiPr}\sim O(1)$ is in very good agreement with our results, whereas the $\mathrm{RiPr}\to 0$ prediction is an upper bound.

We thus observe from these figures that the maximum growth rates depend strongly on latitude and on the differential rotation angle $\phi$, but typically have similar maximum values $O(1)$ (when $S\sim O(1)$) when (either adiabatic or diffusive) instability occurs. In general, the fastest growing modes typically occur for mixed radial/latitudinal shears rather than purely radial or latitudinal shears, and the most unstable orientation of the shear depends on latitude.

Fig.~\ref{MagK} shows the corresponding wavevector magnitudes ($k=|\boldsymbol{k}|$) for the fastest growing modes as a function of $\phi$ between $\phi=-180^\circ$ and $180^\circ$ for each panel plotted in Fig.~\ref{InitialGrowth}. We observe that in the adiabatically unstable regime (to the right of the blue dashed lines showing the predictions of Eq.~\ref{adstab}) the preferred wavevector magnitude is not plotted. This is because the diffusion-free quadratic dispersion relation exhibits a preferred orientation but no preferred wavevector magnitude in this local model. There is a preference for $k\to 0$ however in this regime when diffusion is present, as we have observed by solving our cubic dispersion relation here, but we omit showing this.

Interestingly, we see that in the three cases that showed the largest growth rates (latitudes $0^\circ, 60^\circ$ and $90^\circ$) we also see that all of these cases have similar $|\boldsymbol{k}|$ values on average in the range 0.5-0.7 (in units of $d^{-1}$). We observe that the asymptotic predictions for $k$ assuming $\mathrm{RiPr}\sim O(1)$ are in very good agreement with the numerical results from solving our cubic (black), whereas the predictions assuming $\mathrm{RiPr}\to 0$ are typically slightly larger (indicating slightly smaller wavelength modes). The differences in both growth rates and wavevector magnitudes could be important for the nonlinear evolution (e.g.~as would be expected from simple parasitic mode saturation prescriptions like the one considered in paper 2). 

 \begin{figure}
    {\includegraphics[
    trim=0cm 0cm 0cm 0cm,clip=true,
    width=0.49\textwidth]{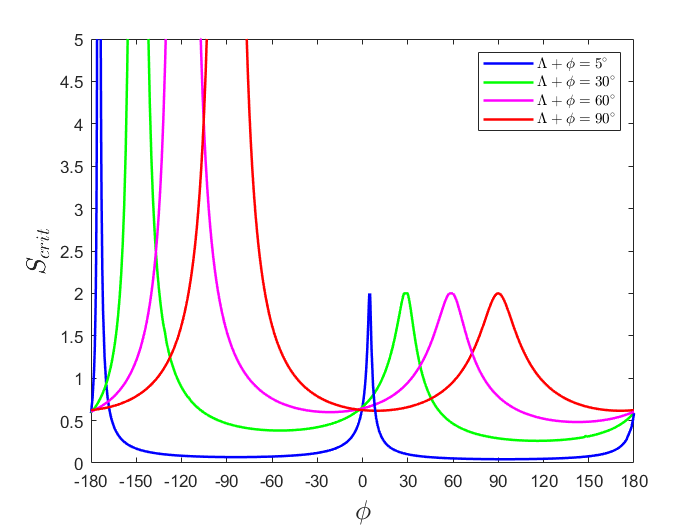}}
  \caption{Critical shear strength $S_{crit}$ required for onset of GSF instability for $\phi \in [-180^\circ,180^\circ]$ at the latitudes $\Lambda+\phi=5^{\circ}, 30^{\circ}, 60^{\circ}$ and $90^{\circ}$ (assuming $N^2=10, \mathrm{Pr}=10^{-2}$). Note: we choose $5^{\circ}$ to approximate the equatorial region for numerical reasons. For each latitude, $\Lambda \sim 0$, corresponding to an approximately cylindrical rotation profile, is a local maximum in $S_{crit}$ because GSF instability requires Rayleigh's stability criterion to be violated, whereas for other latitudes Eq.~\ref{GSFstab} is usually a less stringent condition except for large negative values of $\phi$.}
  \label{Scitvsphi}
\end{figure}

The critical shear strength $S_{crit}$ for onset of GSF instability (i.e. which occurs for $S>S_{crit}$) at a given latitude $\Lambda+\phi$ is highly dependent on $\phi$. To compute $S_{crit}$ for the GSF instability we solve numerically the equation given by setting the left hand side of Eq.~\ref{GSFcritfull} to zero for each $\phi$ (with all other parameters fixed). Results are shown in Fig.~\ref{Scitvsphi} for various latitudes (assuming $N^2=10, \mathrm{Pr}=10^{-2}$). Note that when $\Lambda=0$, corresponding with cylindrical rotation, and for values $\Lambda\sim 0$, there is a region with a local maximum constant value in $S_{crit}=2$. This is because cylindrical rotation profiles are only unstable if Rayleigh's stability criterion is violated, which is typically a more stringent condition than Eq.~\ref{GSFstab}.

On the other hand, we show that when $\phi$ is negative the instability is stabilised for sufficiently large values of $\-\phi$. Such $\phi$ values (e.g. $\phi\sim-180^\circ$ near the equator) correspond to outwardly varying angular momentum profiles when $S$ is positive, which are thus Rayleigh-stable. Hence for such negative values of $\phi$ (depending on latitude), large or even infinite values of $S_{crit}$ are required for instability.

Fig.~\ref{Scitvsphi} shows that a large reduction in $S_{crit}$ is possible when $\phi$ and $\Lambda$ are both nonzero, particularly near the equator. For shellular rotation ($\phi=0$), we note that the most readily destabilised cases are near the poles ($\Lambda+\phi\approx 90^\circ$), as identified in paper 2. On the other hand, the equatorial regions are most readily destabilised for primarily horizontal (or mixed radial/latitudinal) shears. In particular, note that $\Lambda+\phi\approx 5^\circ$ is unstable for very weak horizontal shears, and more generally for those with $|\phi-5^\circ|\gtrsim 5^\circ$. This figure illustrates the non-trivial behaviour of the GSF instability as a function of latitude and $\phi$.

In the next section, we will turn to analyse the results of a set of numerical simulations exploring the nonlinear evolution of these instabilities as $\phi$ is varied.

\section{Nonlinear simulations and results}
\label{nonlinearresults}

\subsection{Varying $\phi$ with $S=2$}
\label{overview}

\subsubsection{Nonlinear Evolution of GSF instability}
\label{GSFunstable}

\begin{figure*}
  \subfigure[Diagram]{\includegraphics[
    trim=0cm 0cm 0cm 0cm,clip=true,
    width=0.45\textwidth]{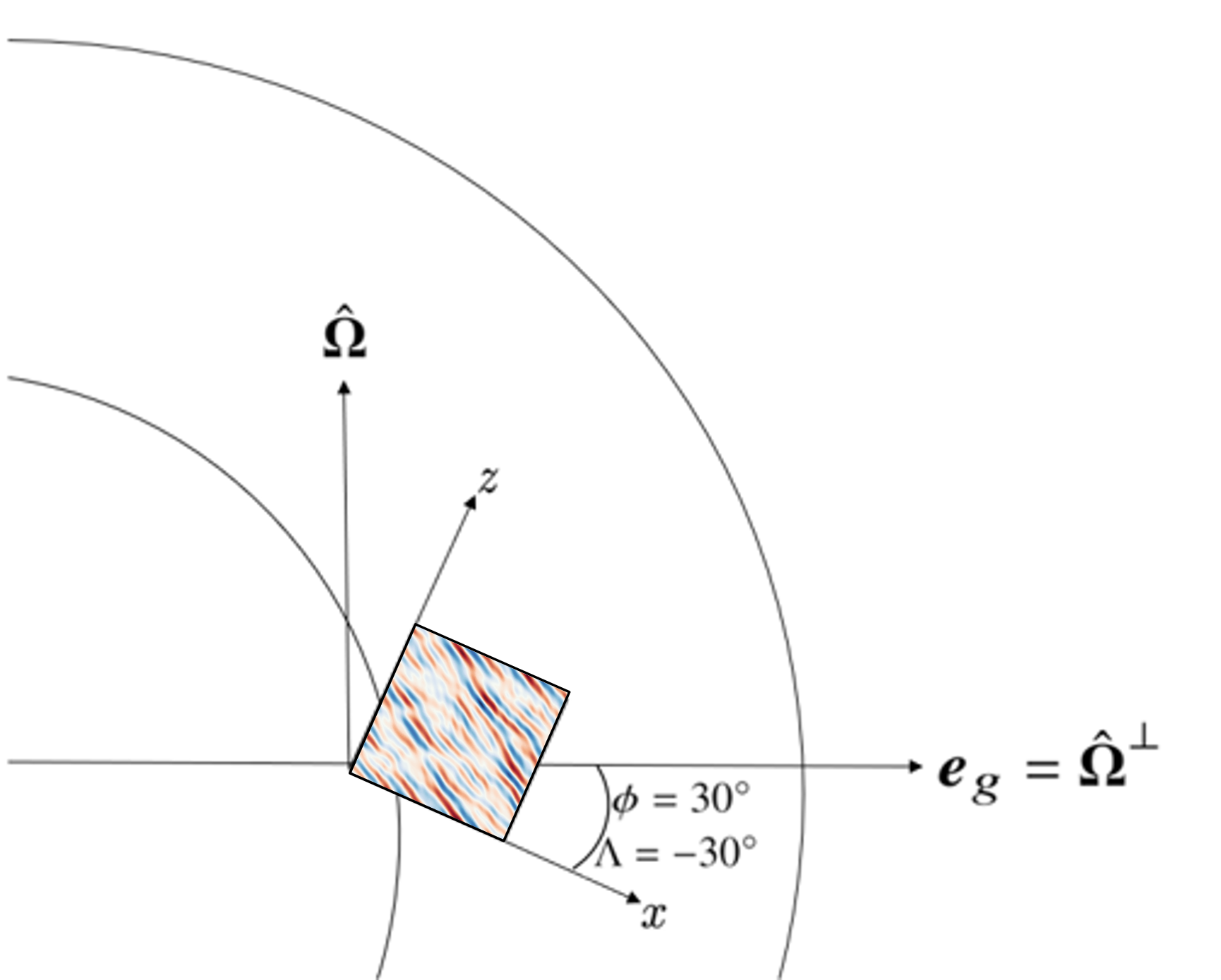}}
    \subfigure[KE spectrum at $t=250$]{\includegraphics[
    trim=0cm 0cm 0cm 0cm,clip=true,
    width=0.45\textwidth]{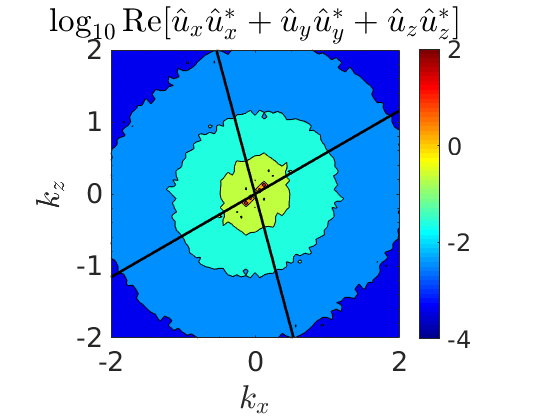}}
  \subfigure[$u_y$ at $t=10$]{\includegraphics[
    trim=0cm 0cm 0cm 0.7cm,clip=true,
    width=0.45\textwidth]{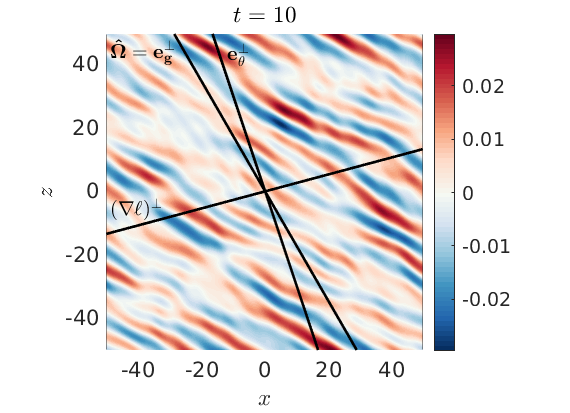}}
    \subfigure[$u_y$ at $t=50$]{\includegraphics[
    trim=0cm 0cm 0cm 0.7cm,clip=true,
    width=0.45\textwidth]{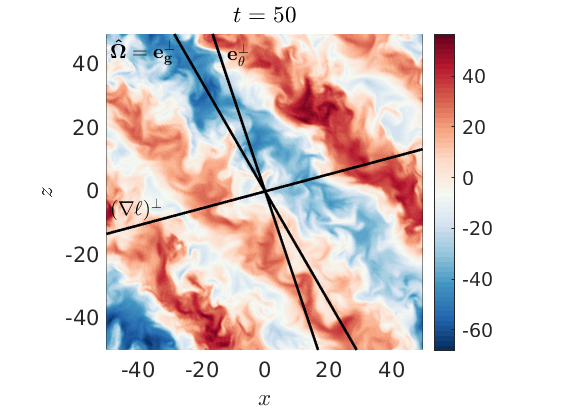}}
    \subfigure[$u_y$ at $t=100$]{\includegraphics[
    trim=0.cm 0.cm 0.cm 0.7cm,clip=true,
    width=0.45\textwidth]{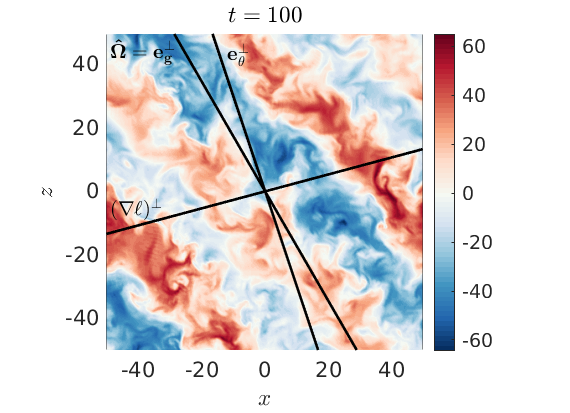}}
    \subfigure[$u_y$ at $t=250$]{\includegraphics[
    trim=0cm 0cm 0cm 0.7cm,clip=true,
    width=0.45\textwidth]{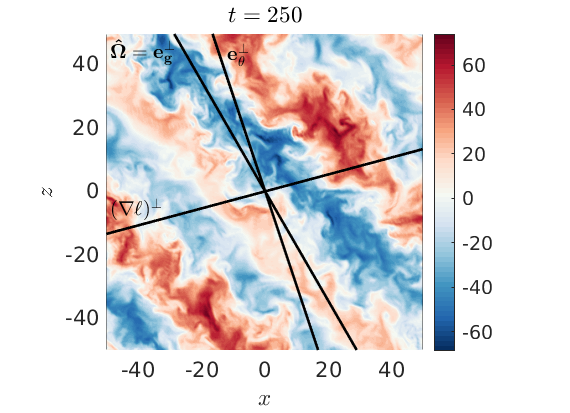}}
  \caption{Panel (a) depicts the local configuration of the box, filled with the snapshot of $u_y$ at $t=10$, within the global picture at the equator $\Lambda+\phi=0$ with $\phi=30^\circ, \Lambda = -30^{\circ}$, $S=2, \mathrm{Pr} = 10^{-2}$ and $N^2=10$. This is coupled with snapshots of the $y$-component of the velocity in $(x,z)$ slices at $y=0$ at various points throughout the evolution (at times $t=10,50,100$ and $250$) of the GSF instability in panels (c) to (f). By $t=10$ AM fingers have developed in the wedge of instability, in between lines of constant AM, $(\nabla{\ell})^{\perp}$, and the rotation axis, $\hat{\boldsymbol{\Omega}}$. By $t=50$ we already see clear layering, made up of oppositely-directed zonal jets. The jets are fully developed by $t=100$ and we see from its evolution at $t=250$, along with Figs.~\ref{KEplots} and \ref{RSplots}, that this is a statistically steady state transporting enhanced levels of AM. Panel (b) shows the kinetic energy spectrum in the ($k_x,k_z$)-plane at $t=250$ and shows the preferred orientation for the layered state and how it differs from that of the initial fingers.}
  \label{VTK1}
\end{figure*}

\begin{figure}
   
  \subfigure[Lat $= 0^\circ$]{\includegraphics[
    trim=3cm 9.5cm 4cm 10cm,clip=true,
    width=0.47\textwidth]{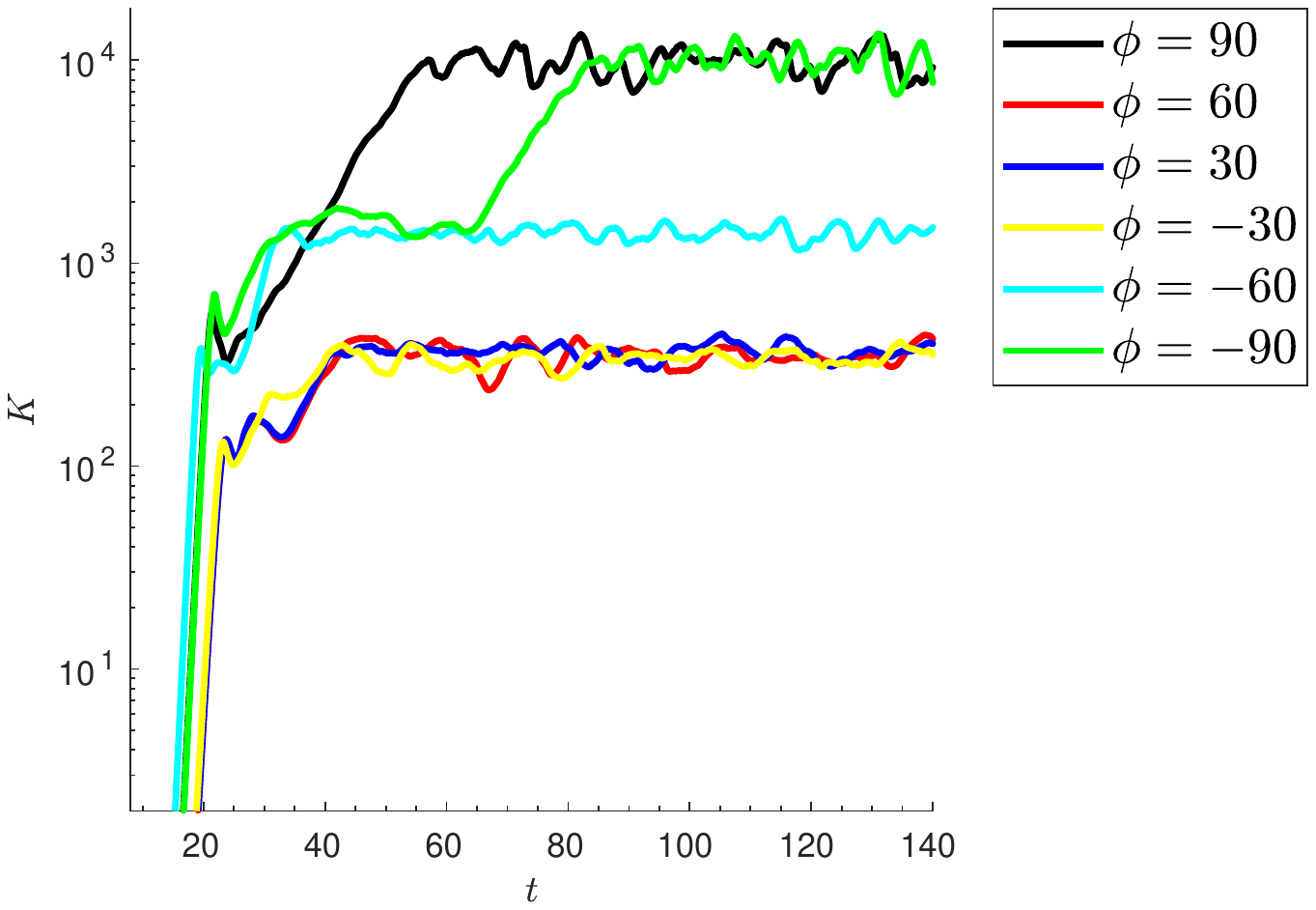}}
    \subfigure[Lat $= 30^\circ$]{\includegraphics[
    trim=3cm 9.5cm 4cm 9.5cm,clip=true,
    width=0.47\textwidth]{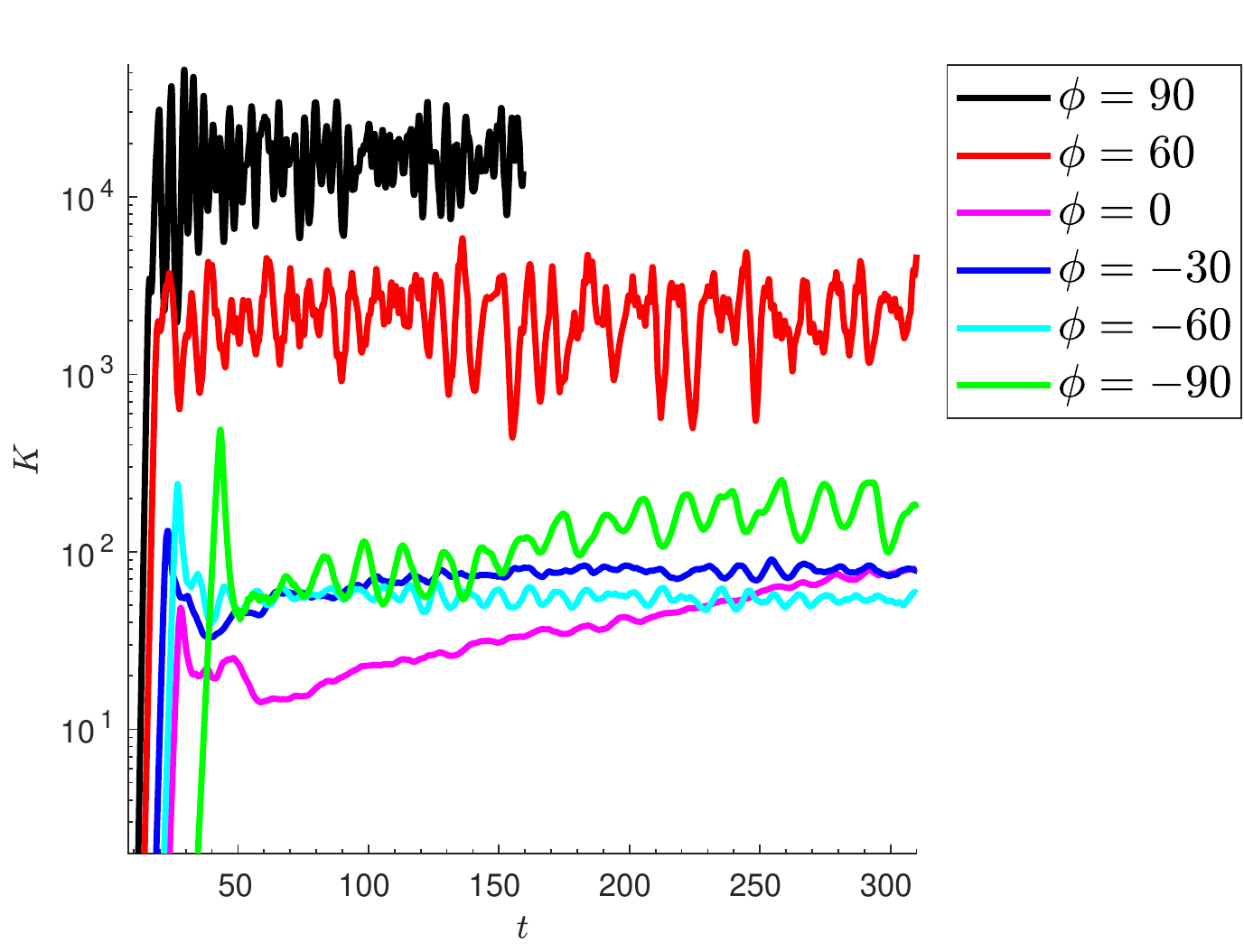}}
    \subfigure[Lat$ = 90^\circ$]{\includegraphics[
    trim=3cm 9.5cm 4cm 9.5cm,clip=true,
    width=0.47\textwidth]{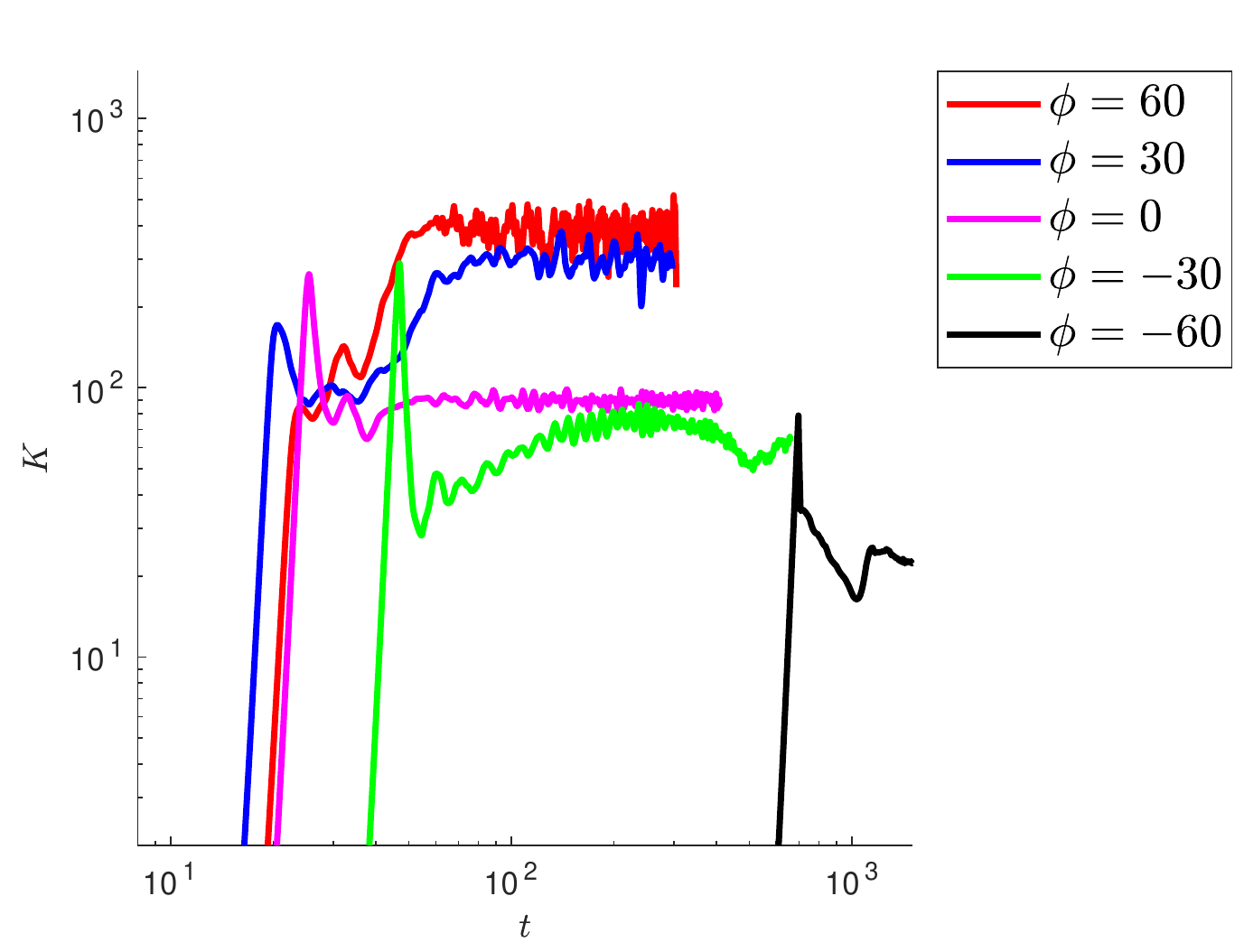}}
  \caption{Evolution of the kinetic energy $K$ ($K = \frac{1}{2}\langle|\boldsymbol{u}|^2\rangle$) at various latitudes for $S=2$, $N^2 = 10$, $\mathrm{Pr} = 10^{-2}$, varying $\phi$ at a fixed latitude $\Lambda+\phi$.  The orientation of the shear directly affects final $K$ levels within the GSF regime, as well as by leading to a stronger adiabatic instability for certain $\phi$. Panels (a) and (b) are plotted on a semi-log scale, whereas the slow evolution for $\phi = -60$ required a log-log scale for (c).}
  \label{KEplots}
\end{figure}

\begin{figure}
  \subfigure[Lat $= 0^\circ$]{\includegraphics[
    trim=3cm 9.5cm 4cm 9.5cm,clip=true,
    width=0.47\textwidth]{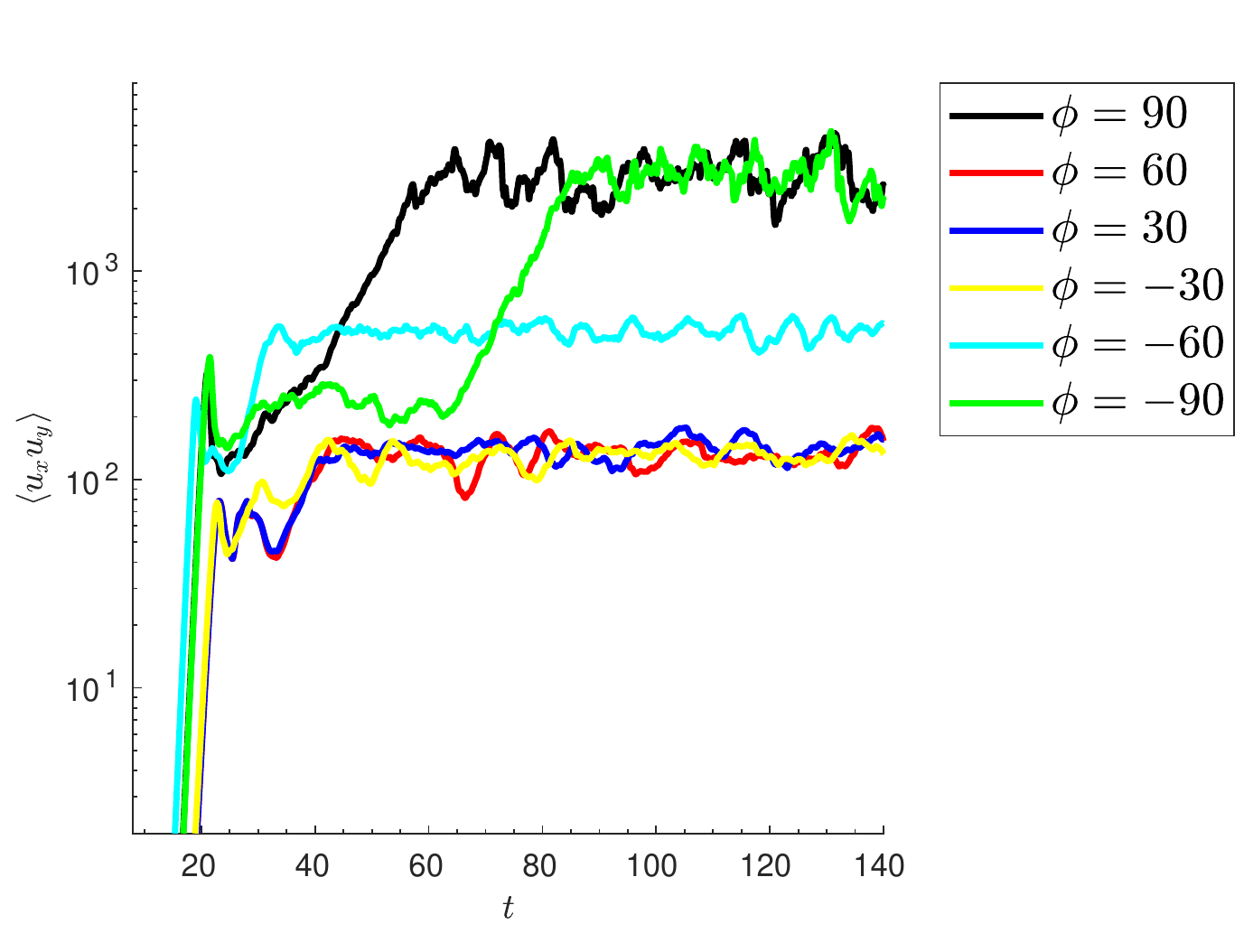}}
    \subfigure[Lat $= 30^\circ$]{\includegraphics[
    trim=3cm 9.5cm 4cm 9.5cm,clip=true,
    width=0.47\textwidth]{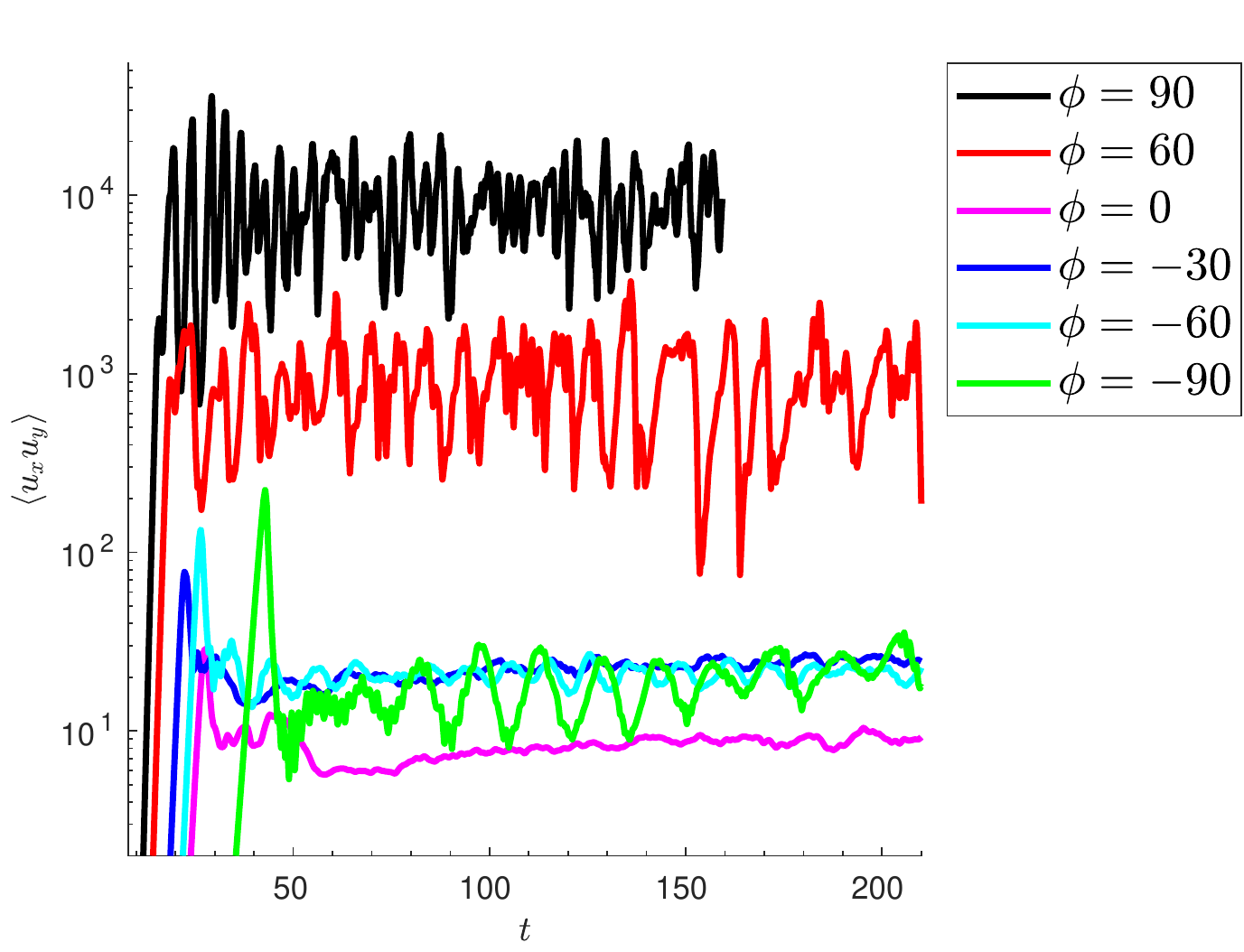}}
    \subfigure[Lat$ = 90^\circ$]{\includegraphics[
    trim=3cm 9.5cm 4cm 9.5cm,clip=true,
    width=0.47\textwidth]{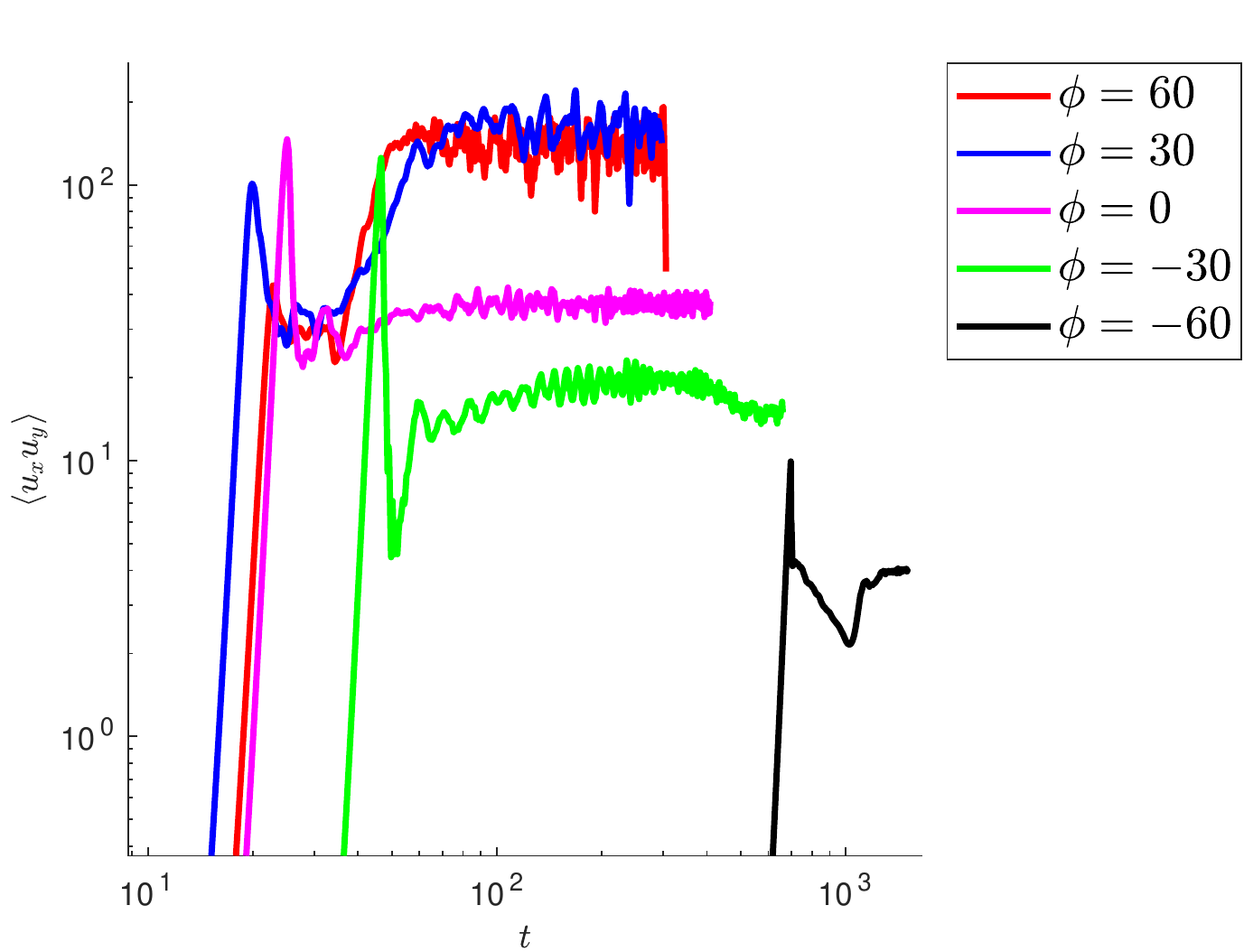}}
  \caption{Reynolds stress component $\langle u_{x}u_{y}\rangle$ illustrating AM transport for the same cases as Fig.~\ref{KEplots}.
}
  \label{RSplots}
\end{figure}

 \begin{figure*}
     \subfigure[Diagram]{\includegraphics[
    trim=0cm 0cm 0cm 0cm,clip=true,
    width=0.45\textwidth]{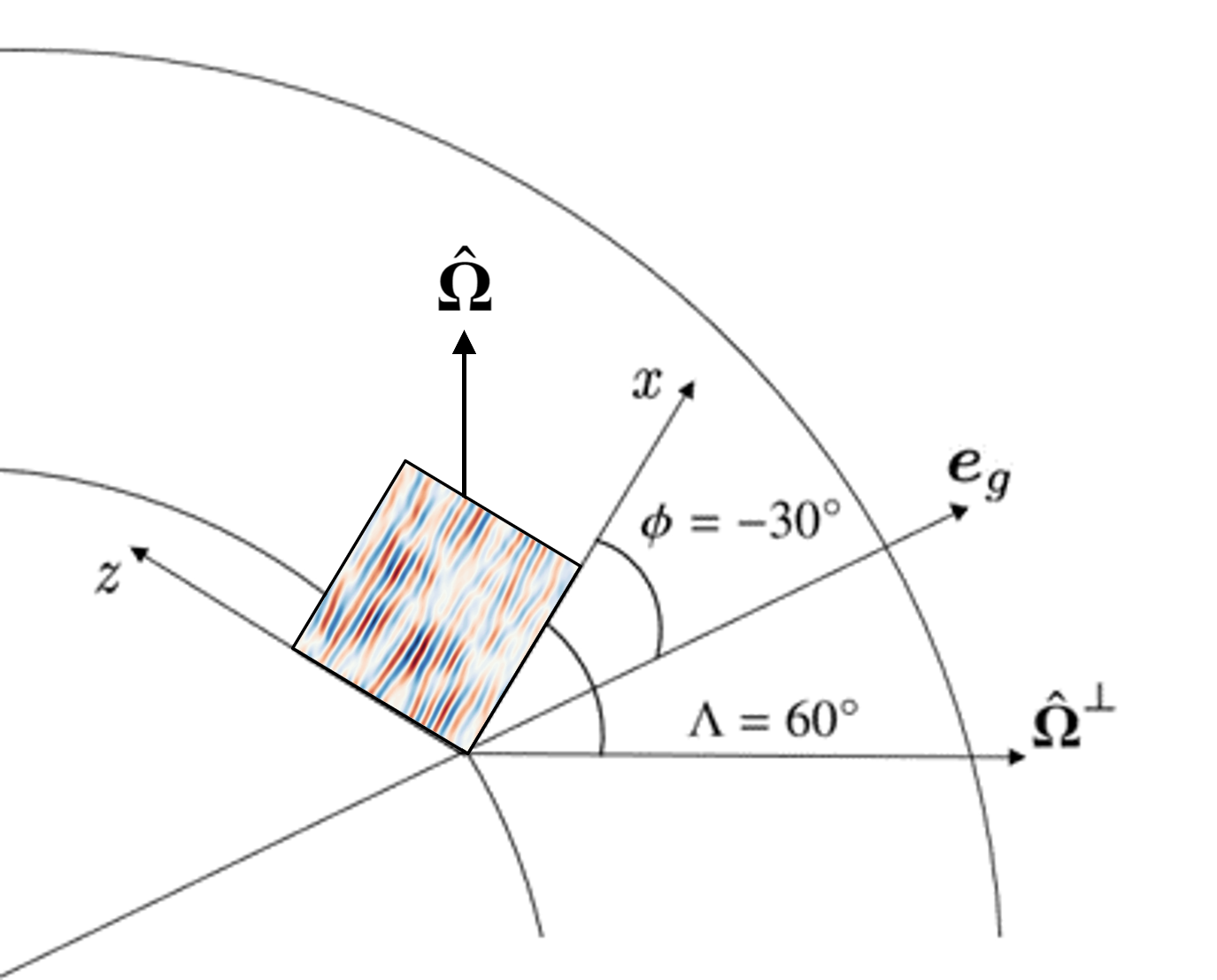}}
    \subfigure[KE spectrum at $t=10$]{\includegraphics[
    trim=0cm 0cm 0cm 0cm,clip=true,
    width=0.45\textwidth]{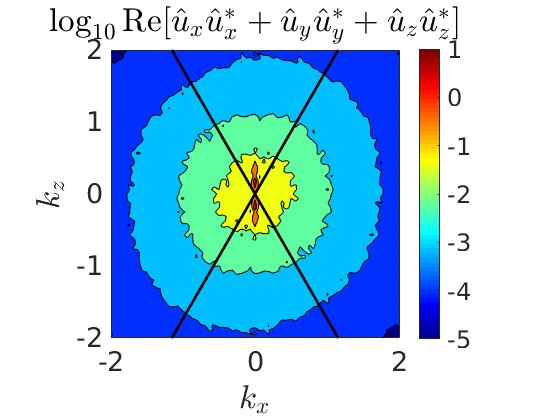}}
  \subfigure[$u_y$ at $t=10$]{\includegraphics[
    trim=0cm 0cm 0cm 0.7cm,clip=true,
    width=0.45\textwidth]{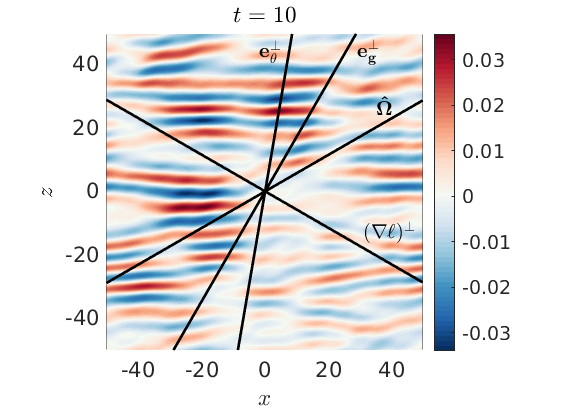}}
    \subfigure[$u_y$ at $t=50$]{\includegraphics[
    trim=0cm 0cm 0cm 0.7cm,clip=true,
    width=0.45\textwidth]{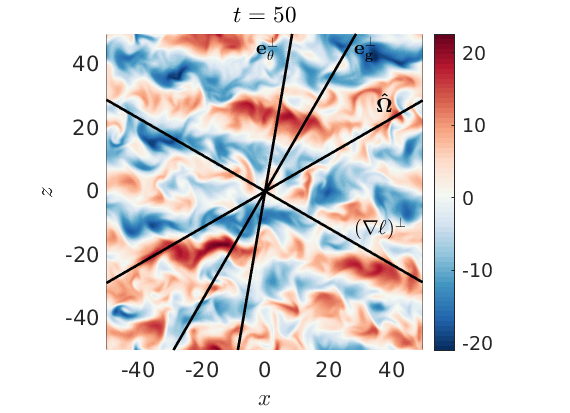}}
    \subfigure[$u_y$ at $t=100$]{\includegraphics[
    trim=0cm 0cm 0cm 0.7cm,clip=true,
    width=0.45\textwidth]{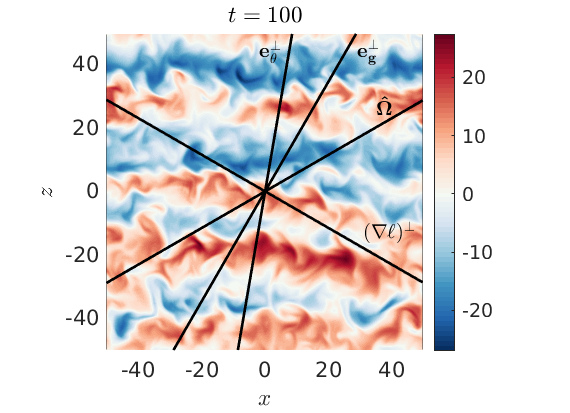}}
    \subfigure[$u_y$ at $t=250$]{\includegraphics[
    trim=0cm 0cm 0cm 0.7cm,clip=true,
    width=0.45\textwidth]{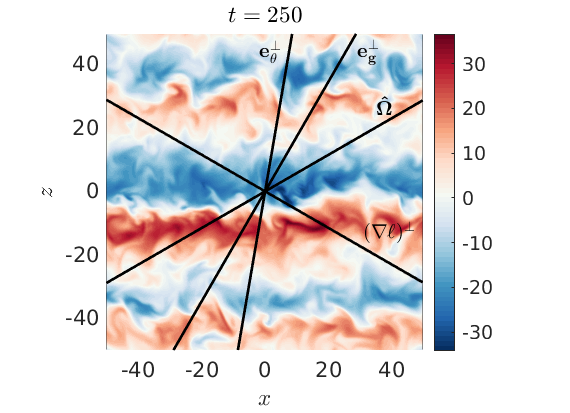}}
  \caption{Panel (a) depicts the local configuration of the box, filled with the snapshot of $u_y$ at $t=10$, within the global picture for a non-equatorial case at latitude $\Lambda+\phi=30^\circ$ with $\phi=-30^\circ, \Lambda = 60^{\circ}$, $S=2, \mathrm{Pr} = 10^{-2}$ and $N^2=10$. This is coupled with snapshots of the $y$-component of the velocity in $(x,z)$ slices at $y=0$ at various points throughout the evolution (at times $t=10,50,100$ and $250$) of the observed GSF instability in panels (c) to (f). At $t=10$ AM fingers are observed inside the wedge of instability, in between the lines $(\nabla{\ell})^{\perp}$ and $\hat{\boldsymbol{\Omega}}$. By $t=50$ AM fingers have saturated by parasitic instabilities and are now well within the non-linear regime and the onset of layer formation. The flow at $t=100$ and $t=250$ is weaker in magnitude than that in Fig.~\ref{VTK1}, and the layers are less distinct from one another here. Panel (b) shows the kinetic energy spectrum in the ($k_x,k_z$)-plane at $t=250$.}
  \label{VTK2}
\end{figure*}

 \begin{figure*}
    \subfigure[Diagram]{\includegraphics[
    trim=0cm 0cm 0cm 0cm,clip=true,
    width=0.45\textwidth]{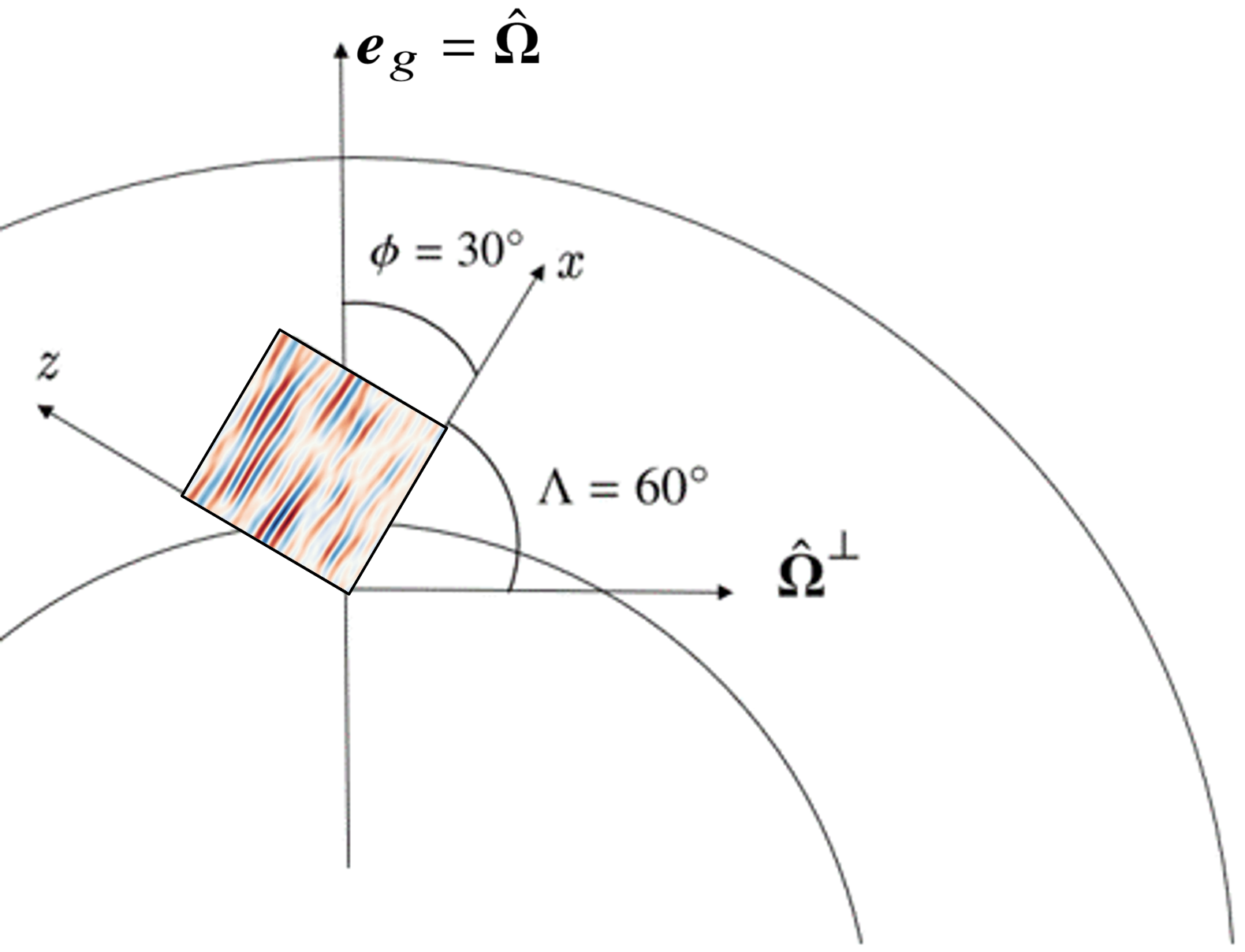}}
    \subfigure[KE spectrum at $t=250$]{\includegraphics[
    trim=0cm 0cm 0cm 0cm,clip=true,
    width=0.45\textwidth]{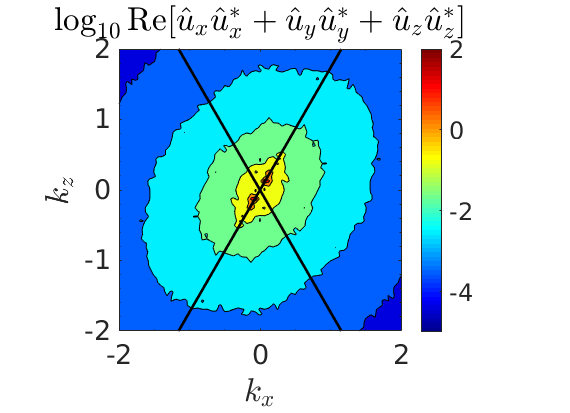}}
    \subfigure[$u_y$ at $t= 10$]{\includegraphics[
    trim=0cm 0cm 0cm 0.7cm,clip=true,
    width=0.45\textwidth]{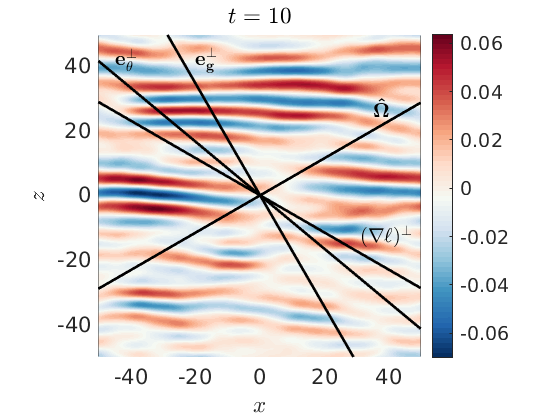}}
    \subfigure[$u_y$ at $t= 50$]{\includegraphics[
    trim=0cm 0cm 0cm 0.7cm,clip=true,
    width=0.45\textwidth]{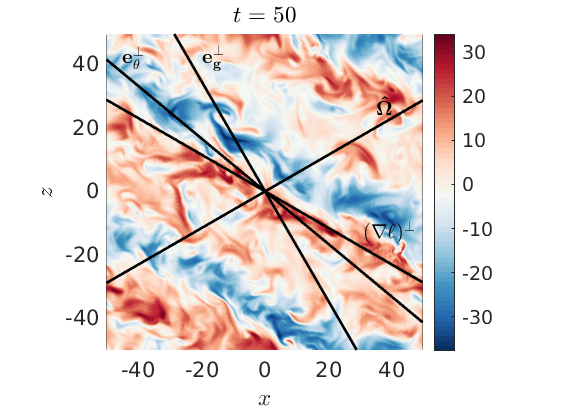}}
    \subfigure[$u_y$ at $t=100$]{\includegraphics[
    trim=0cm 0cm 0cm 0.7cm,clip=true,
    width=0.45\textwidth]{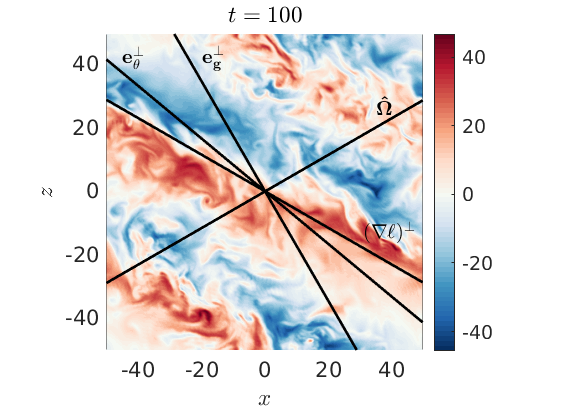}}
    \subfigure[$u_y$ at $t=250$]{\includegraphics[
    trim=0cm 0cm 0cm 0.7cm,clip=true,
    width=0.45\textwidth]{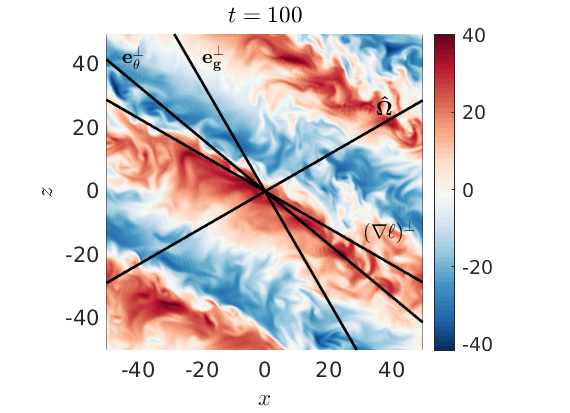}}
  \caption{Panel (a) depicts the local configuration of the box, filled with the snapshot of $u_y$ at $t=10$, within the global picture for a case at the north pole with latitude $\Lambda+\phi=90^\circ$ with $\phi=30^\circ, \Lambda = 60^{\circ}$, $S=2, \mathrm{Pr} = 10^{-2}$ and $N^2=10$. This is coupled with snapshots of the $y$-component of the velocity in $(x,z)$ slices at $y=0$ at several points in the evolution (at times $t=10,50,100$ and $250$) of the GSF instability in panels (c) to (f). At $t=10$ AM fingers develop within the unstable wedge, and by $t=50$ they have saturated and clear AM layers start to appear. Interestingly, we see throughout the $t=50, 100$ and $250$ snapshots that these layers rotate to orientate themselves along lines of constant AM. The lack of changes after $t=100$ indicates the system has attained a statistically steady state. Panel (b) shows the kinetic energy spectrum in the ($k_x,k_z$)-plane at $t=250$.}
  \label{VTK4}
\end{figure*}

Our primary aim is to understand the nonlinear effects of varying the orientation of the shear $\phi$ on the GSF instability, and to quantify the resulting turbulent angular momentum transport. This section presents results from our 3D simulations including snapshots of the flow at various stages throughout the evolution. We use non-dimensional parameters throughout and following Fig.~\ref{InitialGrowth} (and paper 2) we fix $S=2$, $N^2=10$, $\mathrm{Pr}=10^{-2}$, and use a domain with $L_x = L_y = L_z = 100$, unless otherwise stated. Note that $S=2$ is the critical value for instability to onset for a cylindrical differential rotation profile ($\Lambda=0$) or at the equator for a shellular profile ($\phi=0$). We will later investigate the effects of varying shear strength $S$ and box size $(L_x,L_y,L_z)$ in \S~\ref{shear} and \ref{BoxSection}. Table~\ref{Table1}.1 is a table of the linear properties of our simulations, and Table~\ref{Table2}.2 gives some of the resulting nonlinear properties.

Snapshots of the $y$ component of the velocity field $u_y$ (which is the variable that most clearly shows both the linear and nonlinear behaviour) in the $(x,z)-$plane are given in Figs.~\ref{VTK1}-\ref{VTK4}. The upper left image in each of these figures shows how the local model fits into the global picture for each choice of parameters. Panel (c) shows the linear growth phase, which is dominated by the fastest growing mode velocity perturbations (``AM fingers") that are orientated roughly half-way between $\hat{\boldsymbol{\Omega}}$ and $(\nabla{\ell})^\perp$ (indicated by solid black lines). Here, centrifugally-driven AM fingers develop. Panel (d) shows the initial nonlinear saturation of these fingers and the formation of zonal jets. Panels (e) and (f) show the evolution of the zonal jets, illustrating that these can grow in strength and tilt away from their initial orientation depending on the parameters. Given sufficient time all cases here with nonzero $\phi$ (and nonzero $\Lambda$) achieved a steady layered state, in which these zonal jets contribute to providing sustained AM transport.

Panel (b) in each of Figs.~\ref{VTK1}-\ref{VTK4} shows the kinetic energy spectrum on the $(k_x,k_z)$-plane (averaged over $y$) at $t=250$ in each case where all simulations had reached a statistically steady state exhibiting strong zonal jets. This shows the orientations of the modes as a function of their spatial scale, which strongly exhibit a preferred direction at small wavenumbers and become increasingly isotropic for larger wavenumbers. Note that the de-aliasing wavenumber in these simulations has magnitude $5.83$, so the decrease in spectral power by then is evidently more than a factor of $10^3$ from the peak, suggesting 
that our simulations are well resolved spatially. We have checked that this is also the case even in the more turbulent initial saturation phases at $t\sim 50$, in addition to verifying that our simulations are spatially resolved in $y$ by analysing the $k_y$ spectrum.

The first set of snapshots we present are given in Fig.~\ref{VTK1}. These illustrate the $y$-component of the velocity at various points throughout the evolution of the instability at the equator (at the times $t=10, 50, 100$ and $250$). This is an equatorial ($\Lambda+\phi=0$) case with $\phi=30^\circ$, $\Lambda = -30^{\circ}$, which we can see from Fig.~\ref{InitialGrowth} is within the GSF-unstable regime. By $t=10$ centrifugally-driven AM fingers have developed within the wedge of unstable directions. At $t=50$ the AM fingers have saturated nonlinearly and formed zonal jets or AM layers. Figs.~\ref{KEplots} and \ref{RSplots} show the corresponding volume-averaged kinetic energy ($K = \frac{1}{2}\langle|\boldsymbol{u}|^2\rangle$, where $\langle \cdot \rangle$ denotes a volume average), and AM transport (i.e. Reynolds stress component $\langle u_{x}u_{y}\rangle$), respectively. The jets are fully formed by $t=100$, and we see from the subsequent evolution at $t=250$, and Figs.~\ref{KEplots} and \ref{RSplots}, that this is a statistically steady state, which is transporting enhanced levels of AM.

Figs.~\ref{KEplots} and \ref{RSplots} indicate that the transport properties of the GSF (and adiabatic) instability depend heavily on shear flow orientation $\phi$ and latitude $\Lambda+\phi$. We notice that the magnitudes of turbulent transport in the final states are, on a whole, well ordered with respect to the predictions for the linear growth rate in Fig.~\ref{InitialGrowth}, in that $\langle u_x u_y\rangle$ is generally larger for cases with larger growth rates $\sigma$. However exceptions are observed, resulting from undetermined nonlinear factors such as the strengths of zonal jets in each case. We also notice that the case with $\phi = -60^\circ$ at the equator ($\Lambda+\phi=0$) doesn't behave in the same way as $\phi = 60^\circ$, despite the symmetrical nature (about zero) of the growth rate $\sigma$ predicted by linear theory in Fig.~\ref{InitialGrowth} and \ref{MagK}. Instead KE and AM transport properties are significantly  increased in comparison with $\phi = 60^\circ$, which Fig.~\ref{InitialGrowth} would suggest to be roughly equal.

Once the initial growth phase becomes nonlinearly saturated, jet migration and mergers dominate the dynamics. A merger can be seen particularly clearly in the equatorial case with $\phi = -90^{\circ}$ (the green line in Fig.~\ref{KEplots}), between times $t \approx 65$ and $t \approx 85$ the layers in the system merge to form larger scale jets that transport angular momentum more efficiently.

Fig.~\ref{VTK2} shows snapshots from a simulation at a non-equatorial latitude $\Lambda+\phi=30^\circ$ (cf. paper 2) within the GSF unstable regime, with $\phi=-30^\circ, \Lambda = 60^{\circ}$, with otherwise the same parameters. Instability onsets initially between lines $(\nabla{\ell})^{\perp}$ and $\hat{\Omega}$, so that here the preferred direction is along $x$. Parasitic instabilities acting on these fingers then lead into the non-linear regime, which quickly starts forming zonal-jets. Comparing these panels with Figs.~\ref{KEplots} and \ref{RSplots} clarifies that at $t \approx 50$ the initial exponential growth has subsided and the following growth in energy and turbulent transport results from strengthening or mergers of the jets. Potentially as a result of similarities in both the wavevector magnitudes and growth rates predicted by Figs.~\ref{MagK} and \ref{InitialGrowth} for $\phi < 30^{\circ}$, cases within $-90^{\circ} < \phi < 30^{\circ}$ have roughly the same velocities and hence mean kinetic energies. Interestingly a purely radial shear with $\phi=0$ produced the least AM transfer of these cases observed. 
 
The final case for which we will show snapshots in the GSF unstable regime in this section is a case at the (north) pole $\Lambda+\phi=90^\circ$ in Fig.~\ref{VTK4}. Here $\phi=30^\circ, \Lambda = 60^{\circ}$, $S=2, \mathrm{Pr} = 10^{-2}$ and $N^2=10$. Early phases of evolution have unstable mode flows excited between $(\nabla{\ell})^{\perp}$ and $\hat{\boldsymbol{\Omega}}$ but the nonlinear evolution orientates the subsequent zonal jets to become approximately parallel to lines of constant AM. This is consistent with what we might expect if the instability saturates by moving the system back towards marginal stability, though it is difficult to understand this quantitatively given the complexity of our shear flow at this time. Again, we conclude from the lack of changes between $t=100$ and $t=500$ that these layers have reached a statistically-steady state with enhanced transport properties.

We note that the small initial growth rate for $\phi=-60^\circ$ requires the use of a logarithmic time scale in \ref{KEplots} and \ref{RSplots} to see that this case does grow in strength and form layers; albeit over a much longer period of time. This case appears to saturate at a very low level, as might be expected.

\subsubsection{Adiabatically unstable cases}
\label{SHunstable}

 \begin{figure*}

    \subfigure[Diagram]{\includegraphics[
    trim=0cm 0cm 0cm 0cm,clip=true,
    width=0.45\textwidth]{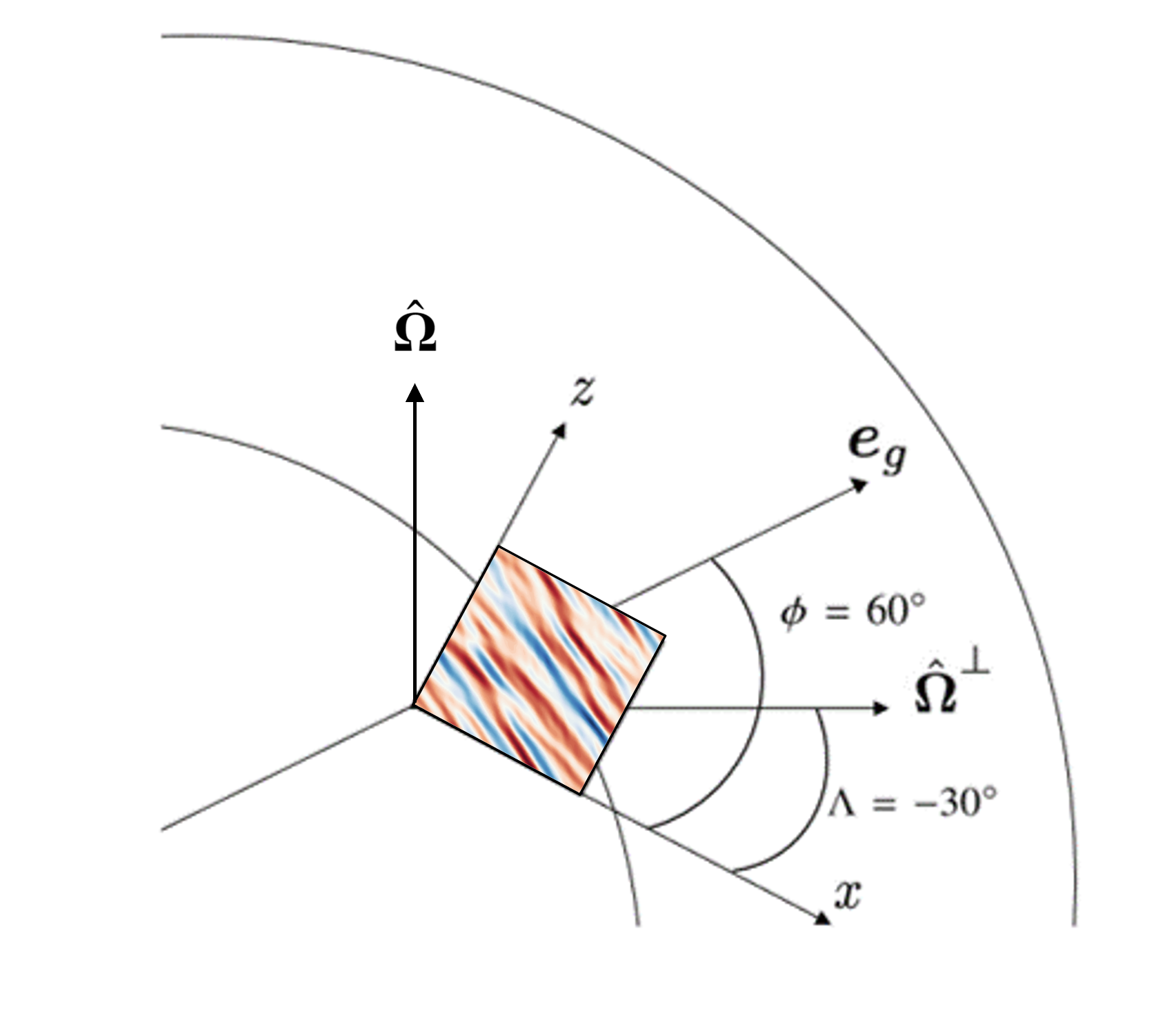}}
    \subfigure[KE spectrum at $t=250$]{\includegraphics[
    trim=0cm 0cm 0cm 0cm,clip=true,
    width=0.45\textwidth]{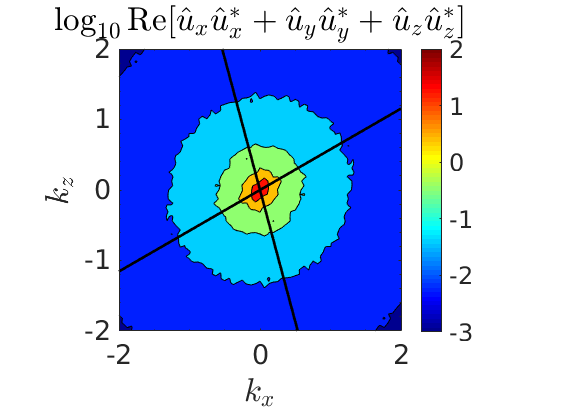}}
    \subfigure[$u_y$ at $t=10$]{\includegraphics[
    trim=0cm 0cm 0cm 0.7cm,clip=true,
    width=0.45\textwidth]{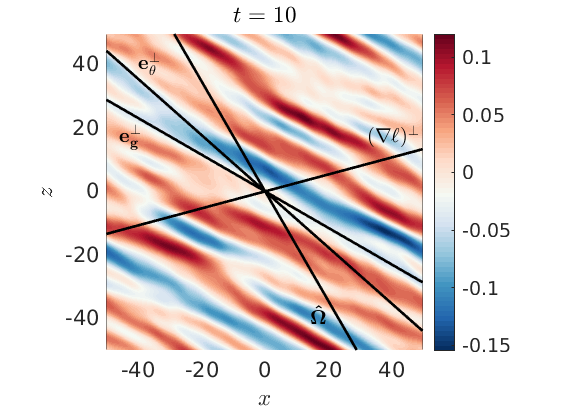}}
    \subfigure[$u_y$ at $t=50$]{\includegraphics[
    trim=0cm 0cm 0cm 0.7cm,clip=true,
    width=0.45\textwidth]{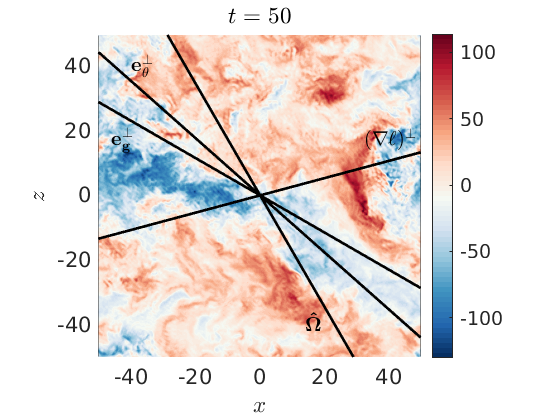}}
    \subfigure[$u_y$ at $t=100$]{\includegraphics[
    trim=0cm 0cm 0cm 0.7cm,clip=true,
    width=0.45\textwidth]{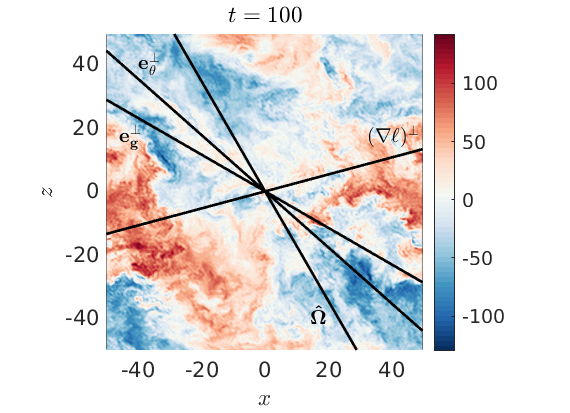}}
    \subfigure[$u_y$ at $t=250$]{\includegraphics[
    trim=0cm 0cm 0cm 0.7cm,clip=true,
    width=0.45\textwidth]{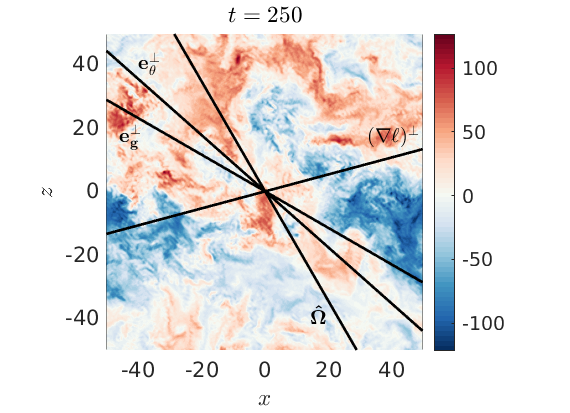}}
  \caption{Panel (a) depicts the local configuration of the box, filled with the snapshot of $u_y$ at $t=10$, within the global picture for a case in the adiabatically-unstable regime at $\Lambda+\phi=30^\circ$ latitude, with $\phi=60^\circ, \Lambda = -30^{\circ}$, $S=2, \mathrm{Pr}= 10^{-2}$ and $N^2=10$. This is coupled with snapshots of the $y$-component of the velocity in $(x,z)$ slices at $y=0$ at various points throughout the evolution (at the times $t=10, 50, 100$ and $250$) in panels (c) to (f).  This adiabatically unstable case exhibits rapidly growing AM fingers, which saturate shortly after $t=10$. By $t=50$, the system is highly turbulent, and $|u_y|$ far larger here than observed in any of the adiabatically stable cases. By $t=100$ and $250$ we see clear evidence of AM layers orientated along the same line as the initial fingers, despite the strong flows that develop. The zonal jets formed are much stronger than any in the GSF-unstable regime, in agreement with Figs.~\ref{KEplots} and \ref{RSplots}. Panel (b) shows the kinetic energy spectrum in the ($k_x,k_z$)-plane at $t=250$.}
  \label{VTK3}
\end{figure*}

As we have identified in \S~\ref{LinearTheory} and shown in Fig.~\ref{InitialGrowth}, the system can be adiabatically unstable for certain $\phi$ and latitudes $\Lambda+\phi$. When the adiabatic stability criterion Eq.~\ref{adstab} is violated, we expect much more violent instabilities that do not require diffusion to operate. These are essentially adiabatic centrifugal instabilities. We have shown in Fig.~\ref{MagK} that in this regime the unstable modes do not have a finite preferred wavevector magnitude in the absence of diffusion, with all modes having the same orientation growing at the same rate, but that  the presence of diffusion prefers modes to have arbitrarily large length-scales, with $k\to 0$.

We show the flow for one case at latitude $\Lambda+\phi=30^\circ$ with  $\phi=60^\circ, \Lambda = -30^{\circ}$ in Fig.~\ref{VTK3}. This is the case in Figs.~\ref{KEplots} and \ref{RSplots} with the highest levels of turbulent transport (and one of the highest for energy, only below the other adiabatically unstable case with $\phi=90^\circ$) for this latitude. The growth rate in this regime from Fig.~\ref{InitialGrowth} is only marginally higher than that with $\phi=0$ (as shown in paper 2). However, the lack of a finite preferred wavevector magnitude permits large wavelength modes on the scale of the box to grow, resulting in a dependence on the size of our Cartesian box we will analyse in \S\ref{BoxSection}. These then saturate leading to flows with much larger amplitudes than any of the GSF unstable cases in Figs.~\ref{VTK1}-\ref{VTK4}. The zonal jets in these cases are correspondingly stronger, and these adiabatically unstable cases lead to the highest values of turbulent transport. Note that Fig.~\ref{RSplots} shows up to three orders of magnitude stronger AM transport in these adiabatically-unstable cases when compared to the GSF-unstable ones.

Ostensibly the dynamics are similar to the GSF-unstable cases, however the timescales for the different phases to occur, and strengths of the flows, vary largely between these regimes. We notice that the AM fingers at $t=10$ are larger than those in the GSF regime, and by $t=50$, the system is in a highly turbulent state, with $|u_y|$ being far larger than observed in any of the adiabatically-stable cases. Whilst fluctuations tend to be large in the adiabatic regime, by $t=100$ we reach a statistically steady state in which large-scale layers have formed, orientated along the same line as the initial fingers by $t=250$, despite their large amplitudes.

Adiabatically-unstable differential rotations, which here primarily involve horizontal rather than radial shears if $\phi\sim 90^\circ$, can be expected to evolve much more rapidly than the diffusive instability analysed in \S~\ref{GSFunstable}.

 We note that an increase in angular momentum transport always accompanies an increase in lengthscale of the layered flow. For cases where the initial instability occurs at finite small scale, this increase in lengthscale arises owing to mergers of the zonal jets; these mergers can take significant time to complete. For the adiabatically unstable cases where the initial instability occurs on a large spatial scale, the increase in angular momentum transport is significant even at early times.

\subsection{Variation of shear strength $S$}
\label{shear}

 \begin{figure*}
  \subfigure[$\Lambda=60^\circ, \phi=-30^\circ$]{\includegraphics[
    trim=3cm 9.5cm 4cm 9cm,clip=true,
    width=0.47\textwidth]{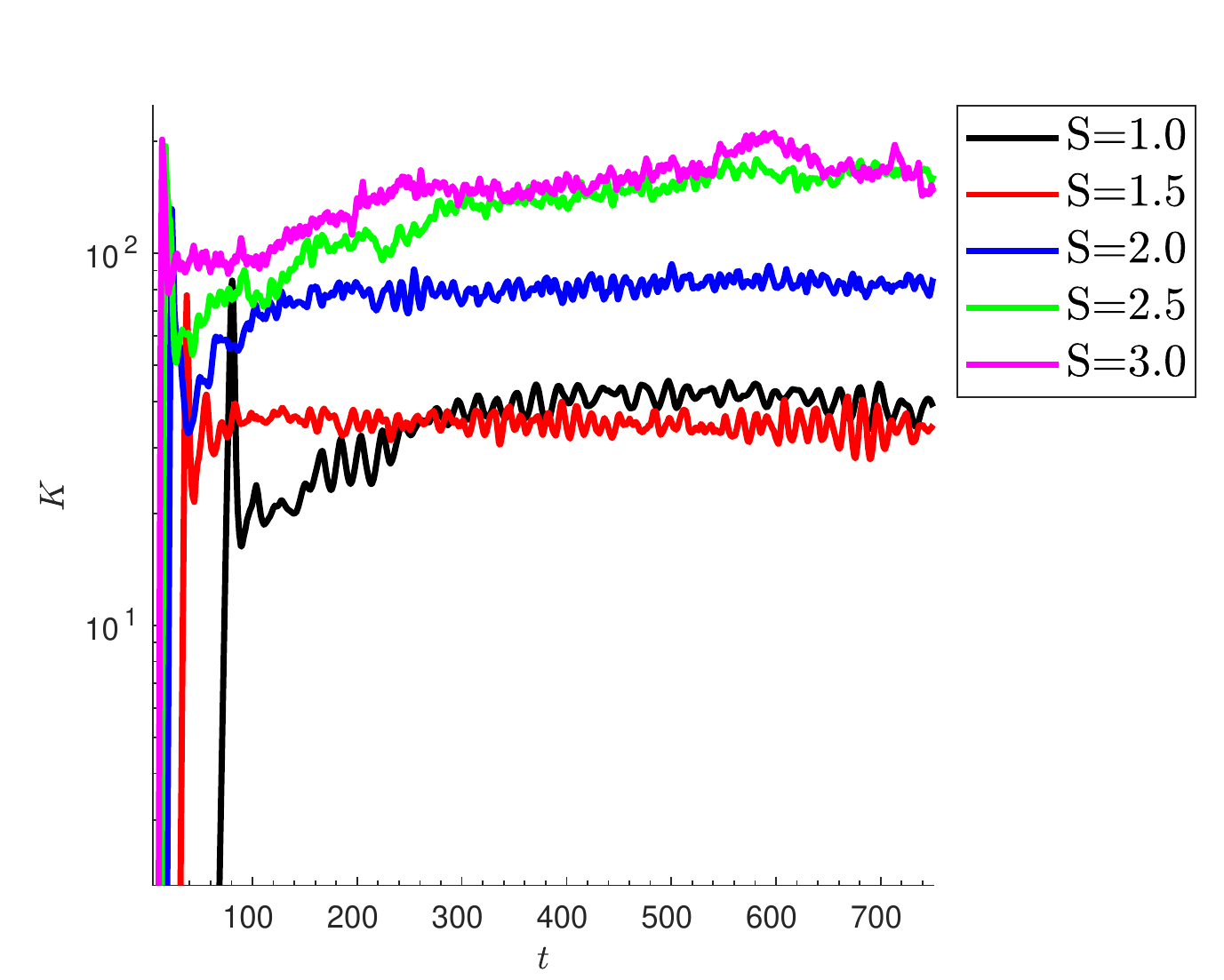}}
    \subfigure[$\Lambda=60^\circ, \phi=-30^\circ$]{\includegraphics[
    trim=3cm 9.5cm 4cm 9cm,clip=true,
    width=0.47\textwidth]{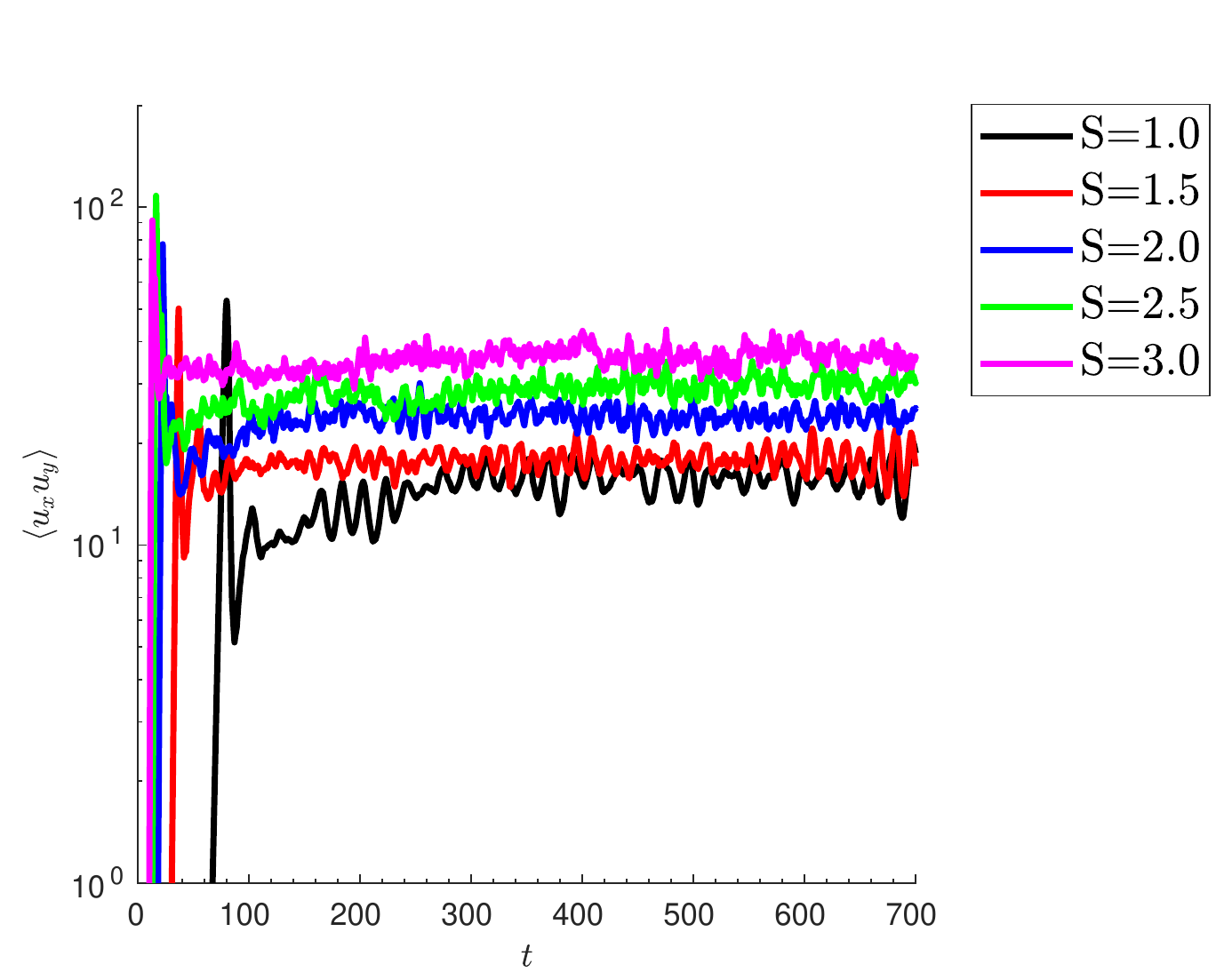}}
    \subfigure[$\Lambda=-60^\circ, \phi=90^\circ$]{\includegraphics[
    trim=3cm 9.5cm 4cm 9cm,clip=true,
    width=0.47\textwidth]{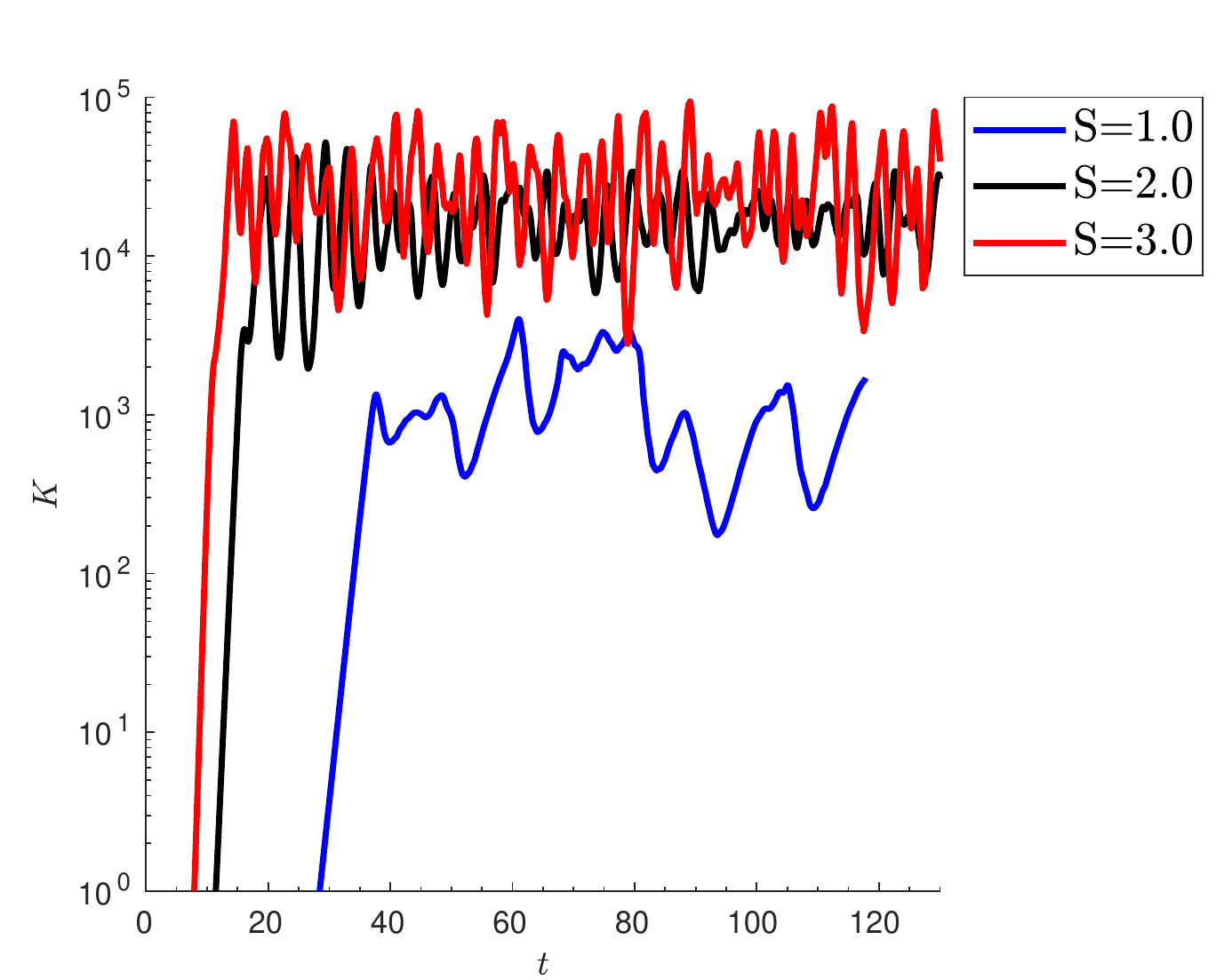}}
    \subfigure[$\Lambda=-60^\circ, \phi=90^\circ$]{\includegraphics[
    trim=3cm 9.5cm 4cm 9cm,clip=true,
    width=0.47\textwidth]{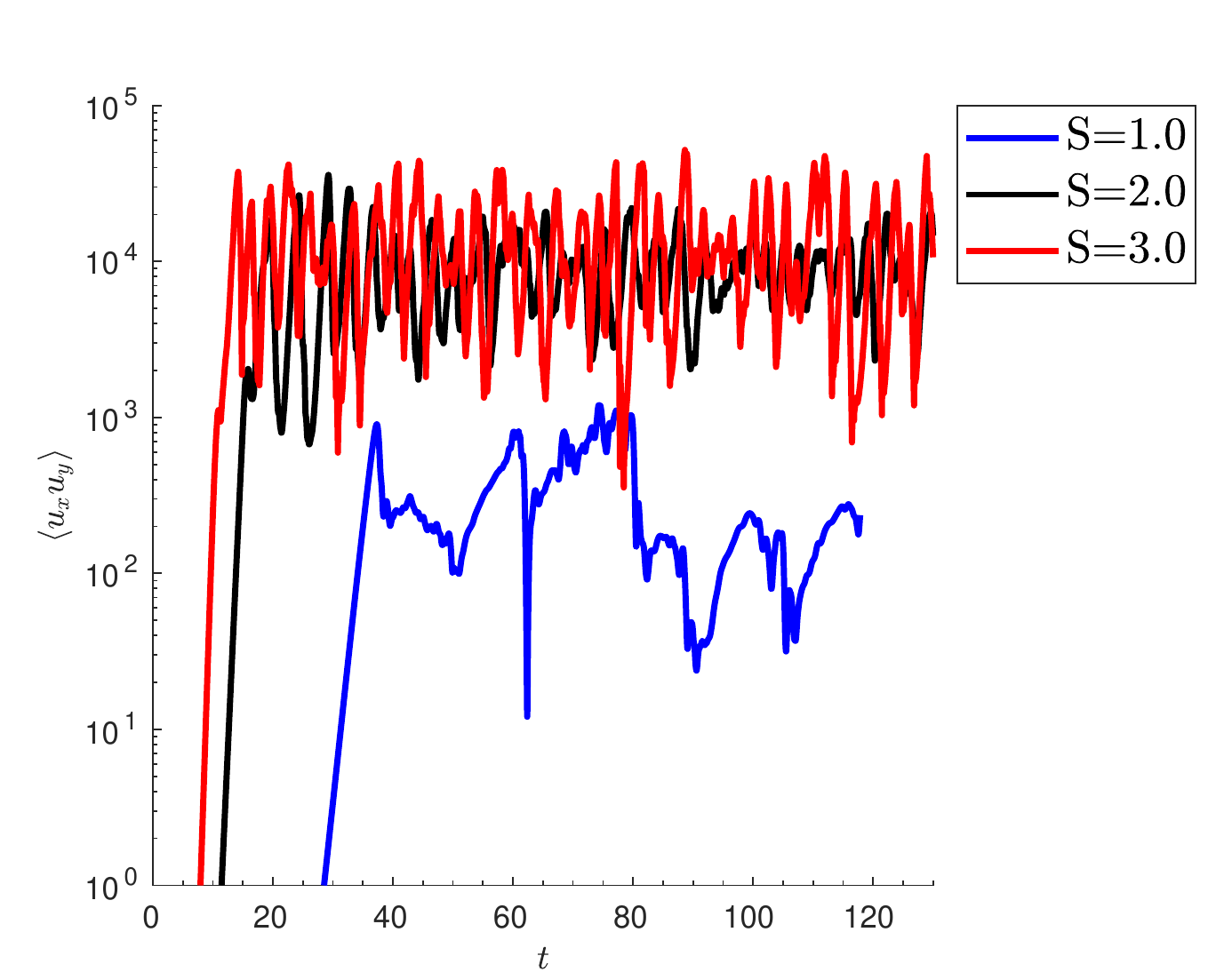}}
  \caption{Dependence of mean kinetic energy and AM transport (Reynolds stress) on shear flow strength $S$ for two different shear orientations. Panels (a) and (b) show a case with $\Lambda = 60^\circ$ and $\phi = -30^\circ$ that is GSF-unstable when $S=2$, and panels (c) and (d) show a case with $\Lambda = -60^\circ$ and $\phi=90^\circ$ that is an adiabatically unstable when $S=2$. Both of these are at $\Lambda+\phi=30^\circ$ latitude with $N^2=10$ and $\mathrm{Pr} = 10^{-2}$. AM transport in the final steady states are well ordered with respect to initial linear growth rates. $K$ is not however, perhaps due to the relatively stronger and larger wavelength zonal jets for $S=1$. Note that every shear strength we tested for $\Lambda = 60^\circ$ and $\phi = -30^\circ$ was GSF unstable and similarly every $\Lambda = -60^\circ$ and $\phi = 90^\circ$ case was adiabatically unstable.}
  \label{ShearComp}
\end{figure*}

We now examine the effect of varying the shear strength $S$ for two different shear orientations at a latitude $\Lambda+\phi=30^\circ$. The first has $\Lambda = 60^\circ$ and $\phi = -30^\circ$ (``mixed radial/horizontal shear") and is GSF-unstable (but adiabatically stable) when $S=2$, and the second has $\Lambda = -60^\circ$ and $\phi=90^\circ$ (``horizontal shear") and is adiabatically unstable when $S=2$. However, note that whether these cases are diffusively or adiabatically unstable depends on $S$. We have observed the qualitative behaviour of the flow in these simulations to be very similar to the cases presented in \S~\ref{overview}, so we restrict our presentation to the volume-averaged quantities. The mean kinetic energy and AM transport are shown in Fig.~\ref{ShearComp}. Cases that are adiabatically unstable in Fig.~\ref{ShearComp} are indicated by dashed lines.

The criterion for diffusive instability given by Eq.~\ref{GSFstab}, and for adiabatic instability given by Eq.~\ref{adstab}, both indicate that the shear strength $S$ directly affects the onset of each type of instability. In addition, larger shears lead to larger growth rates. Interestingly though, whilst larger shears lead to larger initial growth rates, the results of Fig.~\ref{ShearComp} suggest that this doesn't necessarily translate into a higher kinetic energy $K$ in the final steady state. Counter-intuitively, if we look carefully at panel (a) we can see that after the initial growth phase, the energy for $S=1$ grows to slightly overtake that for $S=1.5$. This was also observed in some cases in paper 2, and is potentially related to the relatively stronger, larger wavelength, and potentially more stable zonal jets for lower shears.

However the AM transport, shown via $\langle u_x u_y \rangle$, in the final steady state is in all cases ordered in the way predicted by their initial linear growth rates, with larger $S$ cases providing larger $\langle u_x u_y \rangle$. Adiabatically unstable cases have more energetic flows and provide higher levels of AM transport than GSF unstable cases, indicating that when the Solberg-H\o iland stability criteria (i.e.~Eq.~\ref{adstab}) are violated in stars we would expect much more rapid dynamical evolution. To be able to extrapolate our nonlinear results to stars we must verify whether the AM transport or other turbulent properties vary as the box size is varied. We turn to such a study in the next section.

\subsection{Dependence on box size}\label{BoxSection}

 \begin{figure}
  \subfigure[$\Lambda=60^\circ, \phi=-30^\circ$]{\includegraphics[
    trim=3cm 9.5cm 4cm 9cm,clip=true,
    width=0.47\textwidth]{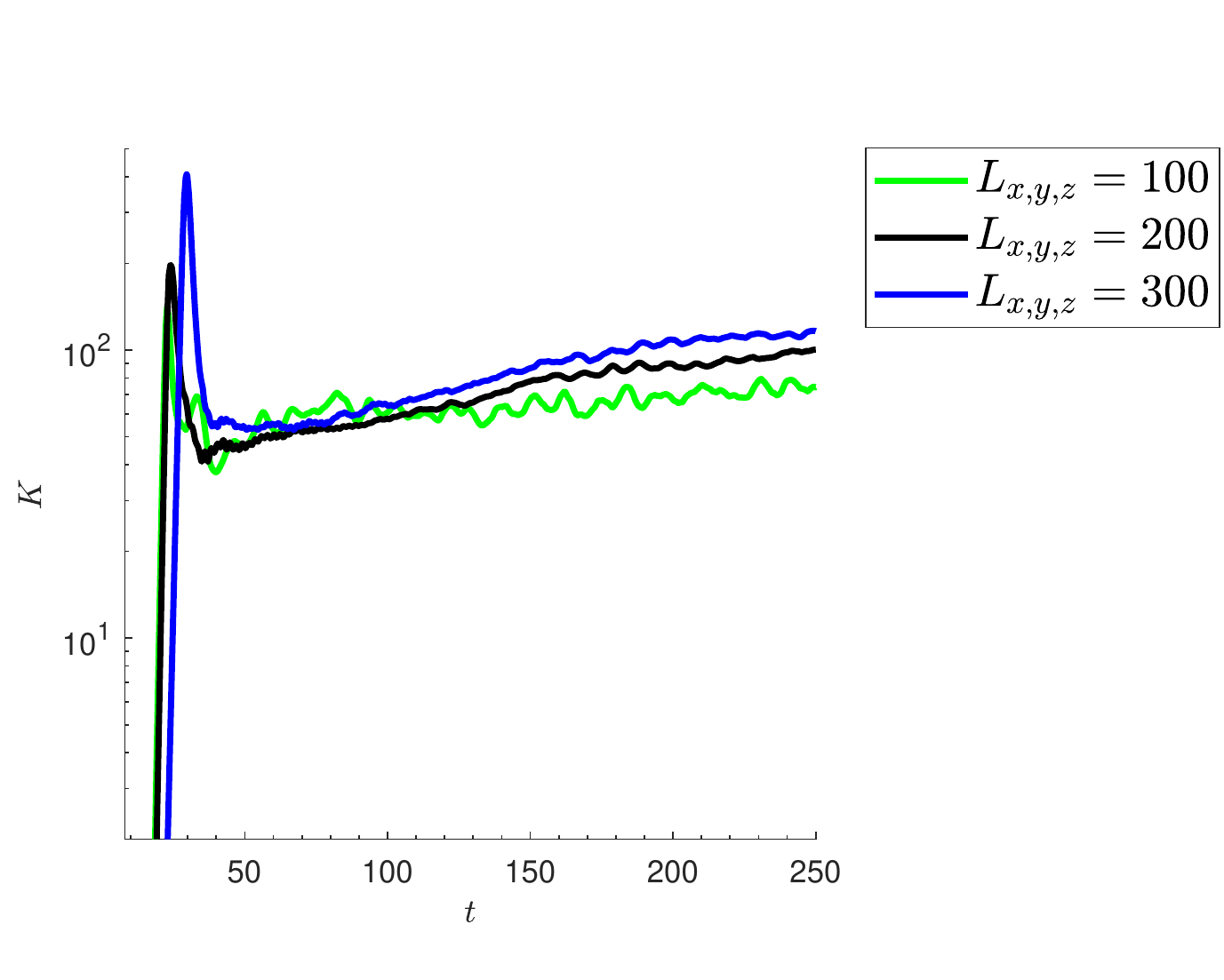}}
    \subfigure[$\Lambda=60^\circ, \phi=-30^\circ$]{\includegraphics[
    trim=3cm 9.5cm 4cm 9cm,clip=true,
    width=0.47\textwidth]{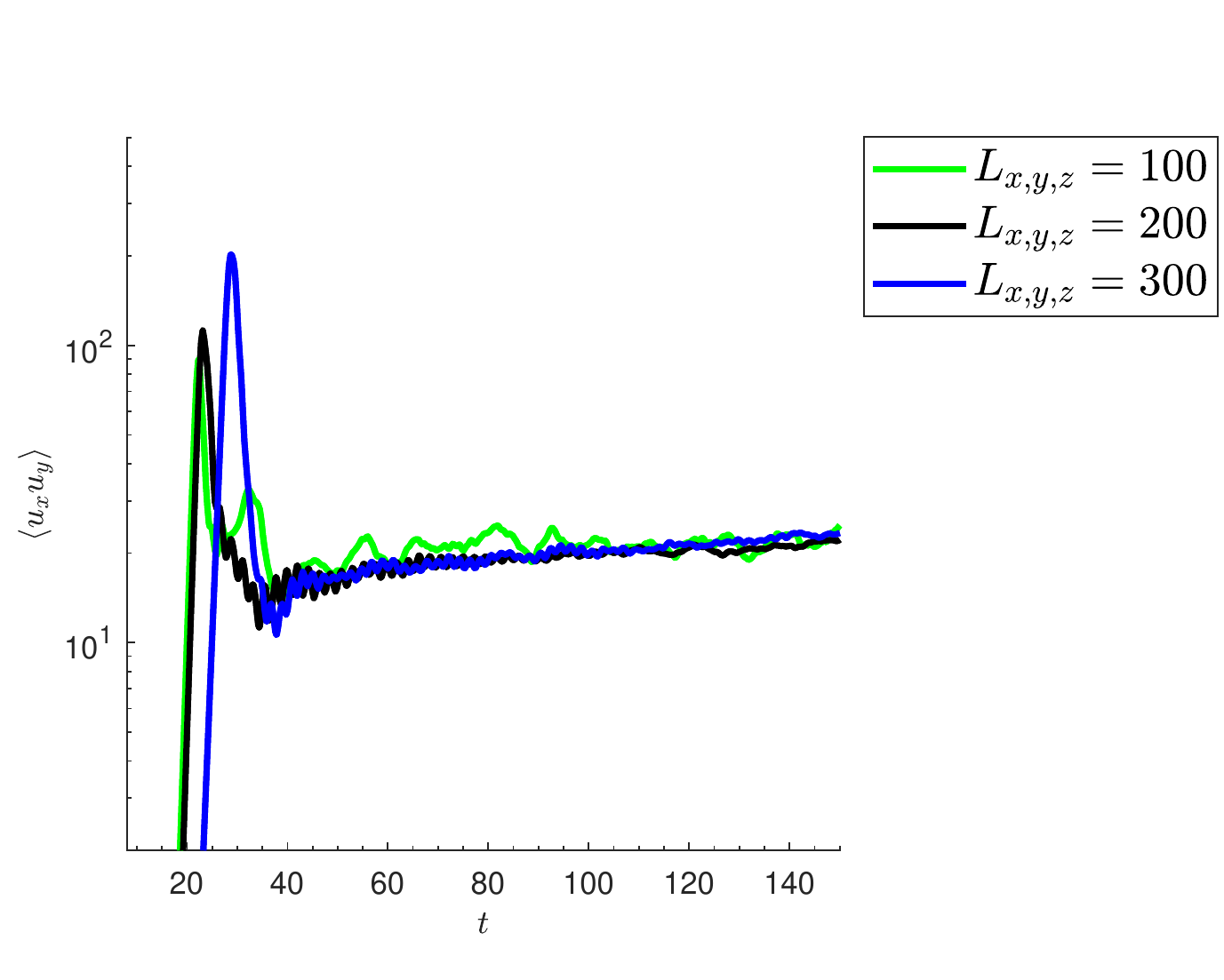}}
  \caption{Dependence of $K$ and $\langle u_x u_y \rangle$ on box size for the GSF instability at a latitude $\Lambda+\phi=30^\circ$, where we vary each of $(L_x,L_y,L_z)$. We fix $S=2$, $N^2 = 10$, $\mathrm{Pr}= 10^{-2}$, $\Lambda=60^\circ$ and $\phi=-30^\circ$. The AM transport is approximately independent of box size, and $K$ only depends weakly on it, implying that we can extrapolate results in this GSF-unstable regime to stars.}
  \label{GSFBox}
\end{figure}

 \begin{figure}
  \subfigure[$\Lambda=-60^\circ, \phi=90^\circ$]{\includegraphics[
    trim=3cm 9.5cm 3.95cm 9cm,clip=true,
    width=0.47\textwidth]{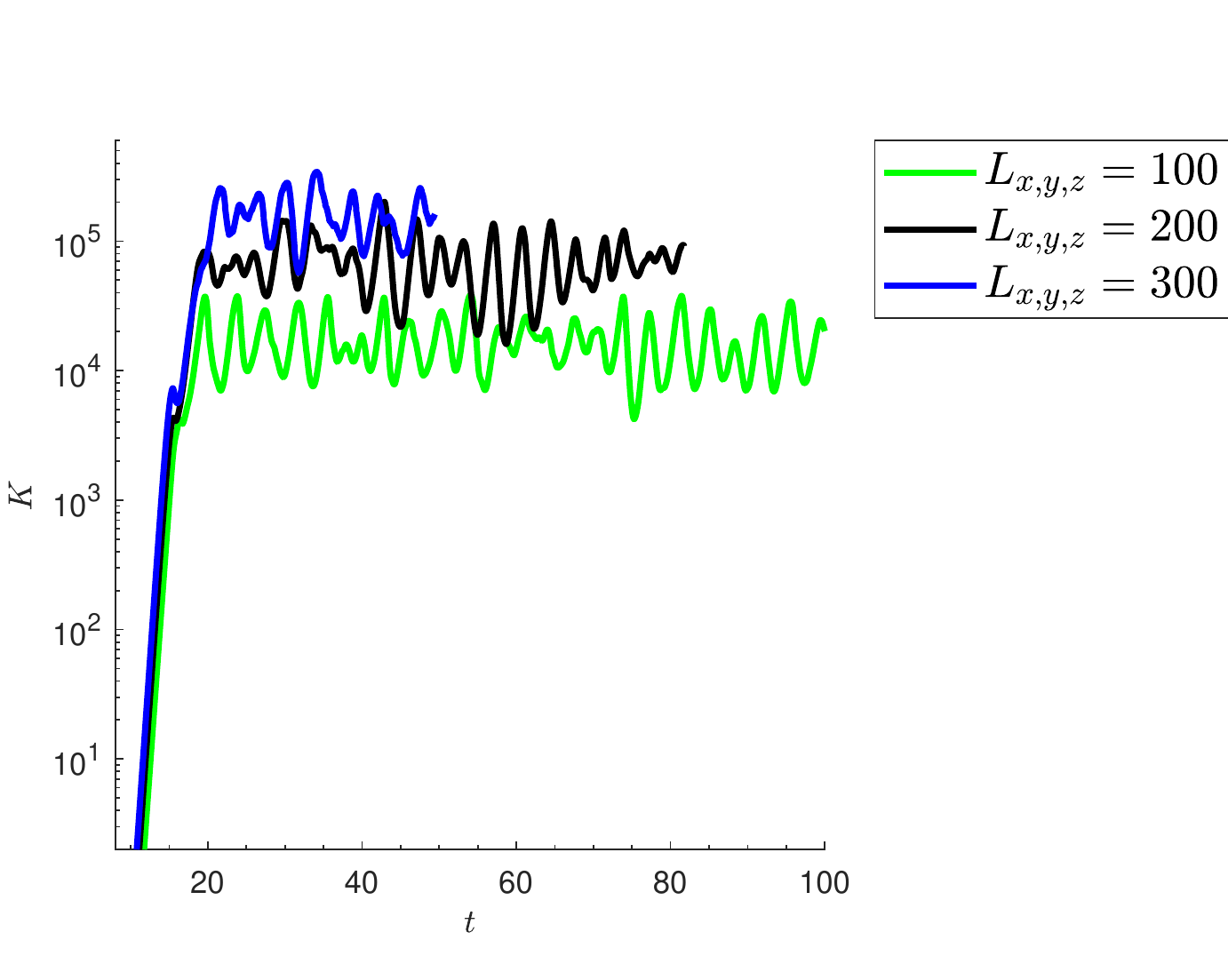}}
    \subfigure[$\Lambda=-60^\circ, \phi=90^\circ$]{\includegraphics[
    trim=3cm 9.5cm 4cm 9cm,clip=true,
    width=0.47\textwidth]{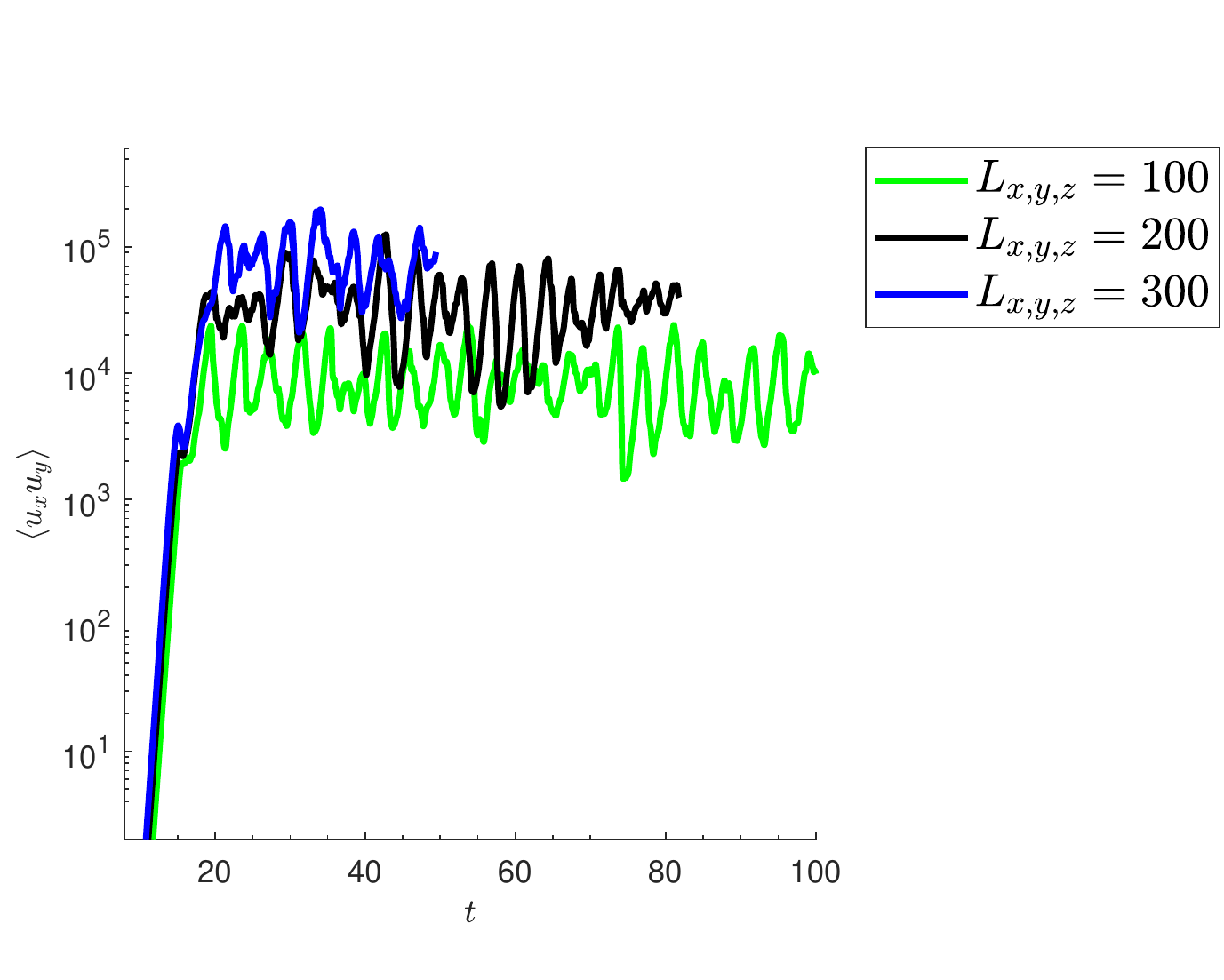}}
  \caption{Dependence of $K$ and $\langle u_x u_y \rangle$ on box size for the adiabatic instability at a latitude $\Lambda+\phi=30^\circ$, where we vary each of $(L_x,L_y,L_z)$. We fix $S=2$, $N^2 = 10$, $\mathrm{Pr}= 10^{-2}$, $\Lambda=-60^\circ$ and $\phi=90^\circ$. The strong dependence on box size implies that results obtained in this regime in our local model cannot reliably be extrapolated to stars.}
  \label{AdiBox}
\end{figure}

\begin{table}
\begin{center}
\begin{tabular}{ |c|c|c|c|c| } 
 \hline
 $ $ & $N_x$ & $N_y$ & $N_z$ \\ 
 \hline
\multicolumn{4}{c}{$ \text{GSF unstable} \quad (\phi = -30^{\circ} \ \Lambda = 60^{\circ}) $} \\

\hline
 $L_{x,y,z} = 100$ & 256 & $256$ & 256 \\ 
 
 $L_{x,y,z} = 200$ & 512 & 256 & 512 \\ 
 
 $L_{x,y,z} = 300$ & 512 & $512$ & 512 \\ 

 \hline
\multicolumn{4}{c}{$ \text{Adiabatically unstable} \quad (\phi = 90^{\circ} \ \Lambda = -60^{\circ}) $} \\

\hline
 
 $L_{x,y,z} = 100$ & $256$ & $256$ & 256 \\ 

  $L_{x,y,z} = 200$ & $512$ & 256 & 512 \\ 

  $L_{x,y,z} = 300$ & 512 & 512 & 512 \\ 
  
 \hline
\end{tabular}
\label{tableRES}
\caption{Table of resolutions (number of grid points in each dimension before de-aliasing) used when testing the effects of boxsize on the nonlinear properties of both types of instability. Cases with $L_{x,y,z} =100$ were also re-run with higher resolution but no differences were found over the original resolution.}
\end{center}
\end{table}

In order to verify whether our nonlinear results in a small domain might be applicable to astrophysical objects we wish to check whether the nonlinear saturation properties of each instability depend on the box size. To do this, we performed additional simulations with $S=2$ at a latitude of $\Lambda+\phi=30^\circ$ with $L_x = L_y = L_z = 200$ and $300$, and with appropriate spatial resolutions (as indicated in \ref{tableRES}), for both a GSF unstable case with $\Lambda=60^\circ$ and $\phi=-30^\circ$, and an adiabatically unstable case with $\Lambda=-60^\circ$ and $\phi = 90^\circ$. Results are shown in Figs.~\ref{GSFBox} and \ref{AdiBox} for the mean kinetic energy and AM transport. In order to ensure the smallest scales remain well resolved even in larger boxes we increase our resolutions as specified in Table~\ref{tableRES}.

Fig.~\ref{GSFBox} demonstrates that AM transport ($\langle u_x u_y \rangle$) in GSF-unstable cases is approximately independent of box size. This is a very promising and important result, since it indicates that the turbulent transport predicted by our simulations is robust, and can potentially be applied to model AM evolution in stars. The kinetic energy attained in the final state is also very similar, though this varies slightly more as the box is enlarged. These results are consistent, but not obvious, from the fact that the GSF instability has preferred wave-vector magnitude in linear theory, as predicted by Fig.~\ref{MagK}. This useful result means that the results from this paper, within the GSF unstable regime, can be applied to astrophysical problems with confidence.

On the other hand, Fig.~\ref{AdiBox} shows that both the kinetic energy and AM transport in adiabatically unstable cases that violate Eq.~\ref{adstab} exhibit a strong dependence on box size. This might be predicted from linear theory, because in this regime there is a preferred orientation but not a preferred wavevector magnitude for adiabatic instability, and Fig.~\ref{MagK} indicates that in this regime the fastest growing mode in the presence of diffusion has $k\to 0$. As a result, the wavelengths of the fastest growing modes grow without bound to fit within the box. Fig.~\ref{AdiBox} verifies that in this regime, where the mechanism limiting $k$ is the box size, then the nonlinear properties of the instability also depend on it. While this leads to a violent instability that transports AM very efficiently, our results in this regime cannot therefore be reliably extrapolated to stars and planets due to this clear box size dependence. To simulate this regime reliably would require models with shear profiles that are not linear, spherical geometry or other effects that could introduce a preferred scale or limit the wavelengths of the modes (e.g.~inclusion of the $\beta$ effect or compressibility).

\subsection{Momentum transport as a function of $\phi$}
\label{RSvsphisec}

\begin{figure}
  \subfigure{\includegraphics[
    trim=3cm 9cm 4cm 9cm,clip=true,
    width=0.45\textwidth]{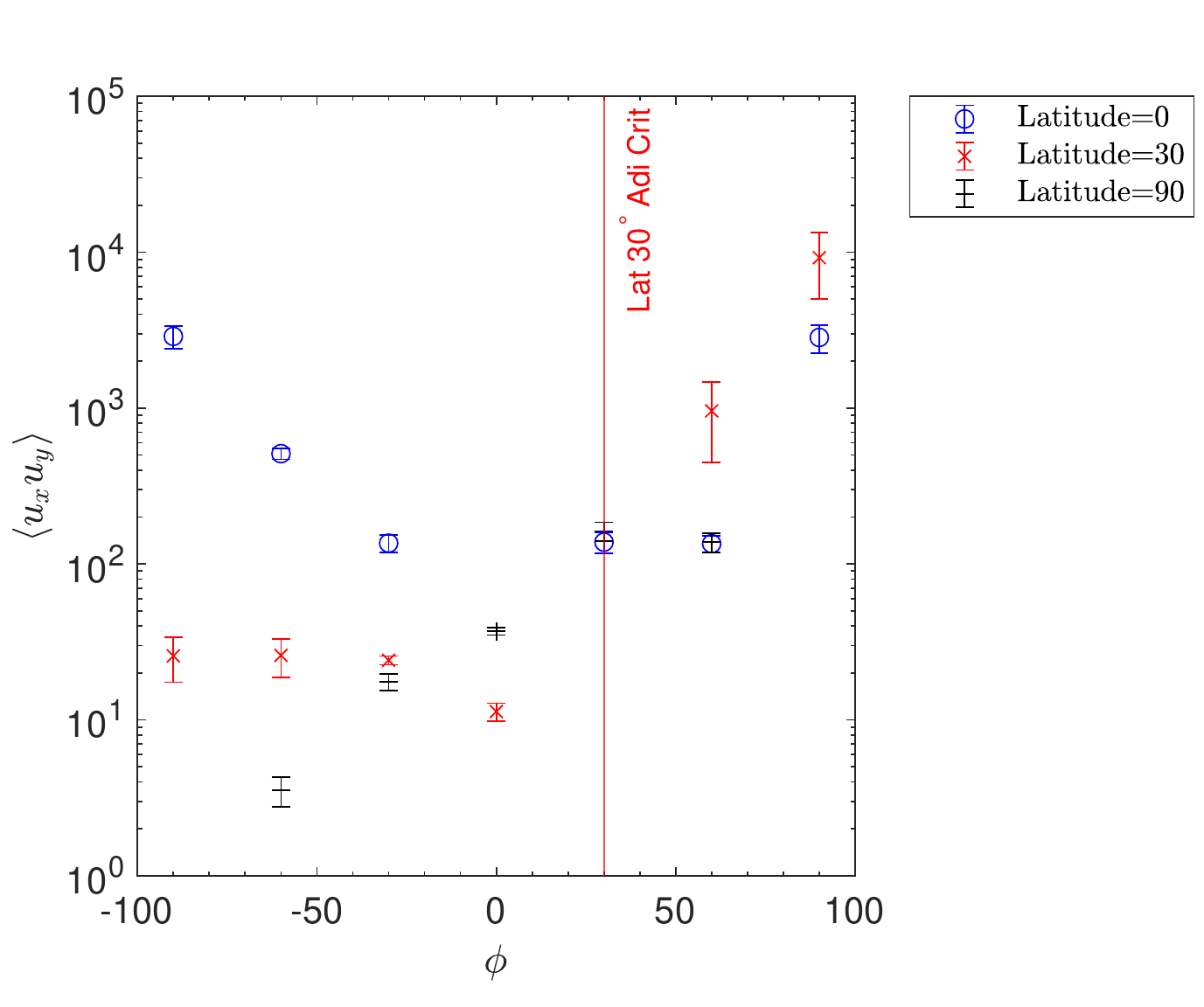}}
    \subfigure{\includegraphics[
    trim=3cm 9cm 4cm 9cm,clip=true,
    width=0.45\textwidth]{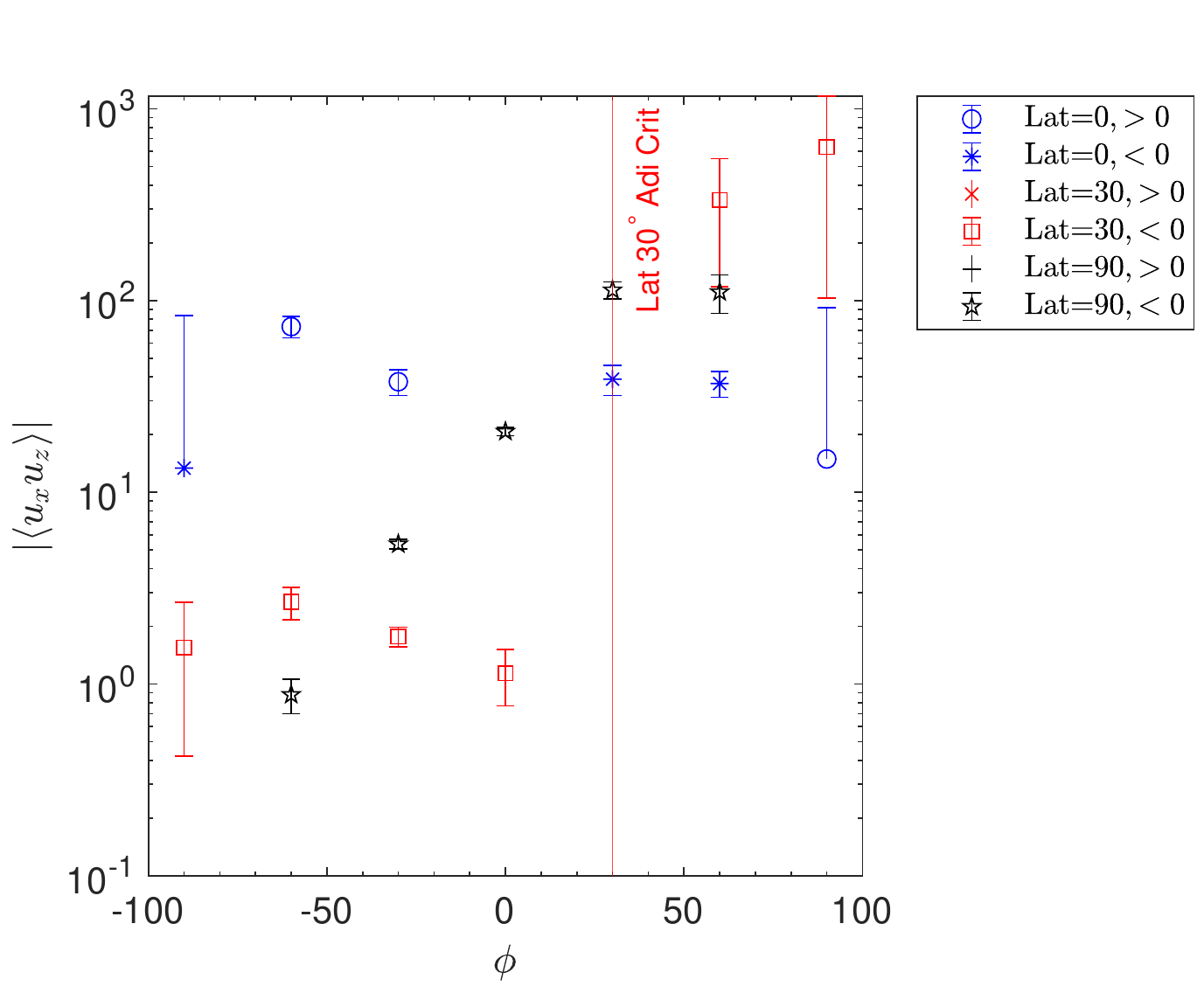}}
    \subfigure{\includegraphics[
    trim=3cm 9cm 4cm 9cm,clip=true,
    width=0.45\textwidth]{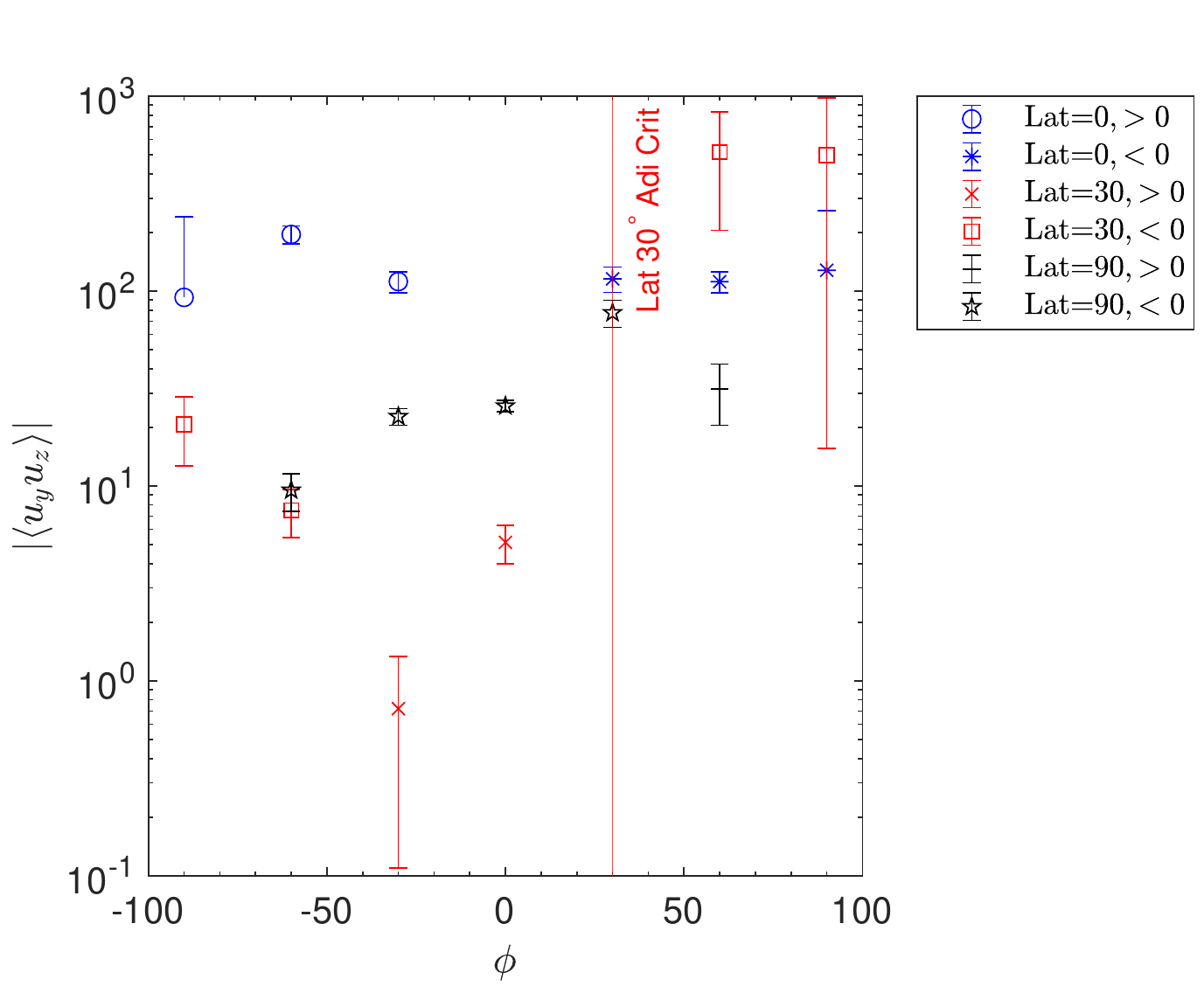}}
  \caption{Figure showing the Reynolds stresses (and hence momentum  transport) as a function of $\phi$ for various latitudes. The mean value in each case is calculated once the instability has reached its final turbulent/layered state after mergers, and the error bars indicating fluctuations are the standard deviations from the mean. All the Latitude $= 0^{\circ}$ and $90^{\circ}$ cases are GSF unstable (adiabatically stable) and adiabatically unstable cases occur for latitude $= 30^{\circ}$ after the vertical red lines at $\phi =30^\circ$.}
  \label{RSvsphi}
\end{figure}

In Fig.~\ref{RSvsphi}, we summarise the mean Reynolds stress components $\langle u_xu_y\rangle$, $\langle u_xu_z\rangle$ and $\langle u_xu_z\rangle$ as a function of $\phi$, after performing both spatial and temporal averaging in the final turbulent state (after layer mergers). The angular momentum transport is quantified by $\langle u_x u_y\rangle$, whereas the other two would correspond with turbulent driving of mean flows/circulations in the meridional plane.

Out of all the GSF-unstable (but adiabatically stable) cases studied, we found mixed radial/latitudinal shears ($\phi\ne 0$) and particularly latitudinal shears ($\phi\sim \pm90^\circ$) at the equator ($\Lambda+\phi=0$) to lead to the most transport. As shown in Fig.~\ref{RSvsphi}, purely latitudinal shears are the most unstable and produce AM transfer over three orders of magnitude greater than we previously found in paper 2 for the case of radial differential rotation ($\phi=0$). The increased transport properties when ($|\phi|\sim 90^\circ$) are at least in part due to the nearly perpendicular directions of the buoyancy and shear, such that buoyancy restoring forces are expected to be weaker. This nonlinear finding is consistent with the linear results shown in Fig.~\ref{Scitvsphi}, which suggests that the configuration is least stable near the equator for $|\phi| \sim 90^\circ$ compared both with $\phi=0$ and with other latitudes in the GSF-unstable regime. On the other hand, adiabatically-unstable cases generally have much larger transport  than the GSF-unstable ones for a latitude of $30^\circ$. It is interesting that in the GSF-unstable regime, $\langle u_xu_y\rangle$ is only weakly dependent on $\phi$ for latitude $30^\circ$.

In summary, we have found that the GSF instability is typically much more efficient at transporting momentum in stars with mixed radial/latitudinal or purely latitudinal differential rotations vs the shellular (radial) case, particularly near the equatorial regions. When adiabatic instability occurs, it also significantly enhances the transport. The most efficient transport is found near the equator for primarily latitudinal differential rotation profiles. A configuration with purely latitudinal shear at the equator would be unusual, but this tendency for predominantly horizontal shears to be more unstable and to  transport momentum more efficiently than vertical/radial shears, and for the growth rates and transport rates for primarily horizontal shears to be maximised near the equator are the general trends we have observed. Note that $\langle u_xu_y\rangle$ for a purely latitudinal shear would correspond with latitudinal transport of angular momentum, which we have shown is generally much more efficient than radial transport. When $\phi\ne 0$, $\langle u_xu_y\rangle$ does not correspond with radial momentum transport, as would be most commonly parameterised in 1D stellar models. Indeed, it is unclear how relevant 1D stellar models with rotation -- even with a suitable parameterisation for turbulent transport -- would be at capturing the long-term consequences of angular momentum transport due to these (and other) fluid and MHD instabilities.

\section{Conclusions}
\label{Conclusions}

We have presented a detailed study into the local hydrodynamic instabilities of differentially-rotating stably-stratified flows in stellar and planetary interiors. Our primary focus has been the GSF instability, an axisymmetric double-diffusive instability of differentially-rotating flows that requires thermal diffusion to operate, but we have also analysed the adiabatic instability occurring when the Solberg-H\o iland stability criteria are violated. We have built upon prior work \citep{barker2019,barker2020} by studying the linear and nonlinear properties of these instabilities for arbitrary orientations of the local shear with respect to the local effective gravity (by defining the angle $\phi$). Our model allows us to study radial ($\phi=0$), latitudinal/horizontal ($\phi=\pm 90^\circ$) and mixed radial and latitudinal shears, so it is more generally applicable to stellar interiors than prior studies that were restricted to considering radial shear. 

We first revisited the linear stability problem \citep[building upon][]{knobloch.et.al.1982,barker2019,barker2020}, discussed its properties in detail and derived several new results. An in-depth linear analysis of the most important regimes of the GSF instability is presented in Appendix \ref{Appendix1}. In particular, we derived a criterion Eq.~\ref{GSFstab} for the critical value of RiPr for onset of (diffusive) axisymmetric instability, where Ri is the local gradient Richardson number and Pr is the Prandtl number. We solved the cubic linear dispersion relation numerically on the ($k_{x}$,$k_{z}$) plane for axisymmetric instabilities and we discovered lobes of oscillatory instability previously predicted by \cite{knobloch1982} but never analysed in detail before. These grow more weakly in stellar interiors than the directly growing GSF modes that are our primary focus however. 

We solved for the linear growth rates and wavevectors for both the diffusive GSF and adiabatically-unstable regimes. The GSF instability is found to have broadly similar linear properties for radial, horizontal and mixed radial/horizontal shears, though there are important dependencies on the local orientation of the shear as a function of latitude for both the growth rate and dominant wavenumber. On the other hand, we found the adiabatic instability excited when the Solberg-H\o iland stability criteria are violated typically has a larger growth rate than the diffusive GSF instability. More importantly though, it has a preferred orientation but no preferred wavevector magnitude in our local model in the absence of diffusion. With diffusion, we find there is a preference for arbitrarily long length-scales.

Using a modified version of the pseudo-spectral code SNOOPY we performed a suite of non-linear simulations to explore the outcome of both types of instability as the properties of the local differential rotation are varied (both the orientation $\phi$ and shear rate $S$).
Our simulations have confirmed the predictions of linear theory for the linear growth phase and identified two distinct regimes (as $\phi$ is varied) in the nonlinear evolution corresponding to the GSF-unstable and adiabatically-unstable regimes. 

The GSF-unstable cases typically exhibit the formation of strong zonal jets which merge, with a preferred orientation that is consistent with that of the fastest-growing modes, but later evolves depending on the parameters of the simulation. The mean AM transport and turbulent kinetic energy in this regime are demonstrated to converge as the box size is increased. This is consistent with what we might have predicted based on the linear modes possessing a preferred length-scale. This key result means that our local simulations of the GSF instability can potentially be used to fruitfully study the turbulent transport and its relevance for stellar interiors (with suitable extrapolation to astrophysical parameter values). The zonal jets we have observed could play an important role in AM evolution in stars.

The adiabatically-unstable cases, on the other hand, lead to much more efficient AM transport and energetically stronger flows, in some cases leading to sustained AM transfer that is several orders of magnitude larger than the GSF-unstable cases. However, consistently with the properties of the linear modes in this regime preferring the largest length-scales, we have found that the AM transport continues to increase and does not converge as our box size is increased. This suggests that turbulent transport in stars within the adiabatically-unstable regime cannot be reliably studied using similar local Boussinesq models with linear shear.

Our linear analysis and simulations which probe the effects of shear strength $S$ suggest that it more likely for the instability to occur in earlier stages in the life of a star where it rotates more rapidly and potentially has stronger differential rotation. These stages in the life of a star have much more efficient AM transport. However, this instability could potentially operate (e.g.~at mid-latitudes) in the solar tachocline, for example \citep[e.g.][]{barker2020}. In addition, this instability is likely to operate on the equatorial atmospheric jets of hot Jupiters that advect heat from day-side to night-side \citep[e.g.][]{Goodman2009,LiGoodman2010,barker2020}. It  would be difficult or impossible to resolve in existing global simulations \citep[e.g.][]{Showman2009,Mayne2017} but could potentially significantly modify their atmospheric flows and should be studied further.

In the future we plan to continue our investigation of this system with the inclusion of magnetic fields, as stellar radiative interiors are also highly likely to be magnetised. We are also interested to see how these instabilities alter the chemical composition within stellar interiors, for which local fluid instabilities could play an important role.

\section*{Acknowledgements}
RWD was supported by an STFC studentship (2443617). AJB was supported by STFC grants ST/S000275/1 and ST/W000873/1. Simulations were undertaken on ARC4, part of the High Performance Computing facilities at the University of Leeds, and using the DiRAC Data Intensive service at Leicester, operated by the University of Leicester IT Services, which forms part of the STFC DiRAC HPC Facility (\href{www.dirac.ac.uk}{www.dirac.ac.uk}). The equipment was funded by BEIS capital funding via STFC capital grants ST/K000373/1 and ST/R002363/1 and STFC DiRAC Operations grant ST/R001014/1. DiRAC is part of the National e-Infrastructure. SMT would like to acknowledge support of funding from
the European Union Horizon 2020 research and innovation programme (grant agreement no. D5SDLV-786780). Additionally, we would like to thank the Isaac Newton Institute for Mathematical Sciences, Cambridge, for support and hospitality during the programme DYT2. This work was supported by EPSRC grant no EP/R014604/1.

%%%%%%%%%%%%%%%%%%%%%%%%%%%%%%%%%%%%%%%%%%%%%%%%%%
\section*{Data Availability}
The data underlying this article will be shared on reasonable request to the corresponding author.

%%%%%%%%%%%%%%%%%%%% REFERENCES %%%%%%%%%%%%%%%%%%

\bibliographystyle{mnras}
\bibliography{Main}

%%%%%%%%%%%%%%%%% APPENDICES %%%%%%%%%%%%%%%%%%%%%
\appendix
\section{Instability in the asymptotic limit of small Prandtl number}
\label{Appendix1}
The dispersion relation Eq.~\ref{DR} can be written
\begin{equation}
s^3  + s^2(2 \nu k^2 + \kappa k^2) + s (a+b+\nu^2 k^4 + 2 \nu \kappa k^4) + \nu^2 \kappa k^6 + a \kappa k^2 + b \nu k^2 =0.
\label{eq:1.2}
\end{equation}
Considering the orders of the terms here when $\mathrm{Pr}\ll 1$ (implying $\nu \ll \kappa$) we have 
\begin{equation}
a \sim O(\Omega^2), \;\; b \sim O(\Omega^2 / \mathrm{Pr}),  \;\; s \sim O(\Omega), \;\; k^2 \sim O(\Omega / \nu).
\label{eq:1.3}
\end{equation}
Since $\nu k^2$ is small compared to $\kappa k^2$ it can thus be neglected, reducing Eq.~\ref{eq:1.2} to
\begin{equation}
s^3  + s^2 \kappa k^2 + (a+b+2 \nu \kappa k^4)s + \nu^2 \kappa k^6 + a \kappa k^2 + b \nu k^2 =0.
\label{eq:1.4}
\end{equation}
We have $s^3=O(\Omega^3)$ and $s^2 \kappa k^2=O(\Omega^3 \kappa / \nu)$, therefore since $\kappa \gg \nu$ this second term is much larger than the first, so we may ignore the first one. 
Similarly $a s=\Omega^3$ but $b s=O(\Omega^3 \kappa / \nu)$, which is much larger, so we may also neglect $a s$ by comparison with $b s$. This means that we can reduce Eq.~\ref{eq:1.4} to 
\begin{equation}
 s^2 \kappa k^2 + 2 s  \nu \kappa k^4 + \nu^2 \kappa k^6 + bs + a \kappa k^2 + b \nu k^2 =0,
\label{eq:1.5}
\end{equation}
where all terms in Eq.~\ref{eq:1.5} are $O(\Omega^3  \kappa / \nu)$. 
Note both Ri and R are $O(1/\mathrm{Pr})$ in this scaling, and since Pr is small, Ri and R are large and Ri Pr (and R Pr) are assumed to be $O(1)$. We express
\begin{equation}
a = \frac{2 \Omega | \nabla \ell |}{\varpi} \sin ( \Lambda - \theta_k) \sin (\gamma - \theta_k), \quad b =  {\cal N}^2 \sin^2 ( \theta_k  + \phi).
%\eqno{(\theequation \textit{a,b})}
\label{eq:1.6}
\end{equation}
by defining the wavevector as $\boldsymbol{k}=k(\cos\theta_k,0,-\sin\theta_k)=(k_x,0,k_z)$.
To maximise $s$ over all possible wavenumbers $k_x$ and $k_z$, we may first maximise over $k^2$, and then maximise over the angle $\theta_k$. Note $a$ and $b$ only depend on $\theta_k$ and not on the magnitude $k$, so $\partial a / \partial k^2$ and $\partial b / \partial k^2$ are both zero. Differentiating Eq.~\ref{eq:1.5} with respect to $k^2$ and setting $ \partial s / \partial k^2 =0$, it follows that
\begin{equation}
\kappa s^2 + 4 s \nu \kappa k^2 + 3 \nu^2 \kappa k^4 + a \kappa + b \nu = 0.
\label{eq:1.7}
\end{equation}
Multiplying by $k^2$ and subtracting Eq.~\ref{eq:1.5} from this equation (to eliminate the $s^2$ term), gives 
\begin{equation}
2 s \nu \kappa k^4 + 2 \nu^2 \kappa k^6 - b s = 0.
\label{eq:1.8}
\end{equation}
The only difference between the non-zero $\phi$ case investigated here and the case analysed in paper 2 is that our expression for $b$ is different. If we define
\begin{equation}
\lambda = \frac {\kappa \nu k^4}{b},
\label{eq:1.9}
\end{equation}
we obtain
\begin{equation}
s = \frac{2 \nu k^2 \lambda}{1 - 2 \lambda}.
\label{eq:1.10}
\end{equation}
To get positive $s$, i.e. growing modes, we require $0 < \lambda < 1/2$ in Eq.~\ref{eq:1.10}. Now we eliminate $s$ and $k^4$ from Eq.~\ref{eq:1.5} and divide by $\kappa k^2$ to get
\begin{equation}
 s^2  + 2 s  \nu k^2 + \nu^2  k^4 + \frac{bs}{\kappa k^2}  + a  + b \mathrm{Pr} =0.
\label{eq:1.11}
\end{equation} 
Eliminating $s$ using Eq.~\ref{eq:1.10} we get
\begin{equation}
\frac{4 \nu^2 k^4  \lambda^2 }{(1-2 \lambda)^2 }  + \frac{4 \nu^2 k^4 \lambda}{(1-2 \lambda)  }+ \nu^2 k^4+ \frac{2 \nu k^2  \lambda b}{(1 - 2 \lambda) \kappa k^2}  + a + b \mathrm{Pr} = 0.
\label{eq:1.12}
\end{equation}
Now we use Eq.~\ref{eq:1.9} to eliminate $k^4$ 
and multiply up by $(1-2 \lambda)^2$ to get 
\begin{equation}
(1-2\lambda)^2 a = (\lambda -1) \mathrm{Pr} b.
\label{eq:1.14}
\end{equation}
We now have the two key equations \ref{eq:1.10}, which gives $s$ in terms of $\lambda$, and \ref{eq:1.14}, which relates $\lambda$ to $a$ and $b$. Maximising $s$ over $\theta_k$ requires us to differentiate \ref{eq:1.10} with respect to $\theta_k$ and set
$\partial s / \partial \theta_k = 0$ to obtain the maximum growth rate. First we eliminate $k$ between \ref{eq:1.9} and \ref{eq:1.10} to get 
\begin{equation} 
 (1-2 \lambda) s = 2 \mathrm{Pr}^{1/2} \lambda^{3/2} b^{1/2}.
\label{eq:1.15}
\end{equation} 
Now we differentiate \ref{eq:1.15} and \ref{eq:1.14} partially with respect to $\theta_k$. Since we require $s$ to be a maximum,  we set $\partial s / \partial \theta_k = 0$. These two equations allow us to eliminate
$\partial \lambda / \partial \theta_k$, giving an equation between  $\partial a / \theta_k$ and $\partial b / \theta_k$. Since we have both $a$ and $b$ in terms of $\theta_k$ in \ref{eq:1.6}, this is the equation
that determines the critical value of $\theta_k$ that corresponds to $\partial s / \partial \theta_k=0$, i.e. the maximum growth rate. Differentiating \ref{eq:1.15} with respect to $\theta_k$ and then multiplying by $2 \lambda (1- 2 \lambda)$ gives
\begin{equation}
(2 \lambda -3) b \frac{\partial \lambda}{\partial \theta_k} = \lambda (1 - 2\lambda) \frac{\partial b}{\partial \theta_k}.
\label{eq:1.18}
\end{equation}
Differentiating \ref{eq:1.14} with respect to $\theta_k$ and 
then multiplying by $(1-2 \lambda)(\lambda-1)$ gives
\begin{equation}
\frac{(1-2\lambda)(\lambda-1)}{a} \frac{\partial a}{\partial \theta_k} = (2 \lambda -3)  \frac{\partial \lambda}{\partial \theta_k}  +  \frac{(1-2\lambda)(\lambda-1)}{b} \frac{\partial b}{\partial \theta_k}.
\label{eq:1.21}
\end{equation}
Now \ref{eq:1.18} can be used to eliminate $(2\lambda -3) \partial \lambda / \partial \theta_k$ and \ref{eq:1.14} can be used to eliminate $a/b$, to obtain 
\begin{equation}
(2 \lambda -1) \frac{\partial a}{\partial \theta_k} = \mathrm{Pr} \frac{\partial b}{\partial \theta_k}.
\label{eq:1.23}
\end{equation}
Differentiating our expressions for $a$ and $b$ in \ref{eq:1.6} and substituting them into \ref{eq:1.23}, we obtain 
\begin{equation} (1- 2\lambda) \frac{2 \Omega |\nabla \ell | }{\varpi} \sin( \Lambda + \gamma - 2 \theta_k) = \mathrm{Pr} {\cal N}^2 \sin 2(\theta_k+\phi),
\label{eq:1.24}
\end{equation} 
or using the definition of R,
\begin{equation}
 (1- 2\lambda)  \sin( \Lambda + \gamma - 2 \theta_k) =  \mathrm{R Pr} \sin 2(\theta_k+\phi).
\label{eq:1.25}
\end{equation}
If the parameters RPr, $\Lambda$, $\gamma$ and $\phi$ are given,
\ref{eq:1.14} is
\begin{equation}
(1-2 \lambda)^2 \sin(\Lambda - \theta_k ) \sin(\gamma - \theta_k) = (\lambda -1) \mathrm{RPr} \sin^2 (\theta_k+\phi).
\label{eq:1.26}
\end{equation}
\ref{eq:1.25} and \ref{eq:1.26} are a pair of simultaneous equations for $\lambda$ and $\theta_k$ which were solved numerically (see Figs.~\ref{InitialGrowth} and \ref{MagK}). Note that once $\theta_k$ is found, $b$ can be found from \ref{eq:1.6}, then \ref{eq:1.9} determines $k^4$ from $\lambda$, such that
knowing $\lambda$ determines the magnitude of the critical $k$ for maximum growth rate, and $\theta_k$ gives the direction of the vector $\boldsymbol{k}$. Once $\theta_k$ is found, $a$ and $b$ can be constructed, and so $s$, the maximum growth rate, can be found from \ref{eq:1.10}. 

\subsection{ Limit $\lambda \to 0$}
The limit $\lambda \to 1/2$ (in which RiPr$\to 0$ as Pr$\to 0$) is not affected by $\phi$ because only $a$, and not $b$, matters in this limit, and $a$ is independent of $\phi$. However in the limit $\lambda \to 0$ (in which RiPr$=O(1)$ as Pr$\to 0$), $\phi$ does matter, and
\ref{eq:1.14} becomes
\begin{equation}
a+ \mathrm{Pr} b = 0,
\label{eq:2.1}
\end{equation}
and \ref{eq:1.25} becomes 
\begin{equation}
  \sin( \Lambda + \gamma - 2 \theta_k) =  \mathrm{RPr} \sin 2(\theta_k+\phi).
\label{eq:2.2}
\end{equation}
Putting in the expressions \ref{eq:1.6} for $a$ and $b$ into \ref{eq:2.1}, which are valid in the $\mathrm{Pr} \to 0$ limit, it follows
\begin{equation}
 \sin ( \Lambda - \theta_k) \sin ( \gamma - \theta_k) + \mathrm{RPr} \sin^2 ( \theta_k + \phi) = 0.
\label{eq:2.3}
\end{equation} 
Eliminating RPr between \ref{eq:2.2} and \ref{eq:2.3} we obtain an equation for the optimum $\theta_k$. To do this, we let
\begin{equation}
{\tilde \theta}_k = \theta_k +\phi , \quad {\tilde \Lambda} = \Lambda + \phi,  \quad {\tilde \gamma} = \gamma + \phi.
\label{eq:2.4}
\end{equation}   
Then \ref{eq:2.2} and \ref{eq:2.3} become 
\begin{equation}
  \sin( {\tilde \Lambda} + {\tilde \gamma} - 2 {\tilde \theta_k}) =  \mathrm{RPr} \sin 2{\tilde \theta_k}
\label{eq:2.5}
\end{equation}
and 
\begin{equation}
 \sin ( {\tilde \Lambda} - {\tilde \theta_k}) \sin ( {\tilde \gamma} - {\tilde \theta}_k) + \mathrm{RPr} \sin^2  {\tilde \theta}_k = 0.
\label{eq:2.6}
\end{equation}
We expand the sines in \ref{eq:2.6} and divide by $\sin^2 {\tilde \theta_k}$, giving
\begin{equation}
\sin{\tilde \Lambda}\sin{\tilde \gamma} \cot^2 {\tilde \theta}_k - \sin ( {\tilde \Lambda} + {\tilde \gamma}) \cot {\tilde \theta}_k + \cos {\tilde \Lambda} \cos {\tilde \gamma} = -\mathrm{RPr}.
\label{eq:2.7}
\end{equation}
Expanding \ref{eq:2.5} using
\begin{equation}
\sin({\tilde \Lambda + {\tilde \gamma} - 2 {\tilde \theta}_k }) = \sin({\tilde \Lambda + {\tilde \gamma}) \cos 2 {\tilde \theta}_k } -  \cos({\tilde \Lambda + {\tilde \gamma}) \sin 2 {\tilde \theta}_k }
\label{eq:2.8}
\end{equation}
and dividing by $\sin 2 {\tilde \theta}_k$ we obtain
\begin{equation}
 \sin({\tilde \Lambda} + {\tilde \gamma}) \frac{ \cos^2 {\tilde \theta}_k - \sin^2 {\tilde \theta}_k}{\sin 2 {\tilde \theta}_k } -  \cos({\tilde \Lambda} + {\tilde \gamma}) = R Pr,
\label{eq:2.9}
\end{equation}
or noting that $\sin 2 {\tilde \theta}_k = 2 \sin {\tilde \theta}_k \cos {\tilde \theta}_k$ 
\begin{equation}
 \frac{1}{2} \sin({\tilde \Lambda} + {\tilde \gamma})  \cot {\tilde \theta}_k  - \frac{1}{2}   \sin({\tilde \Lambda} + {\tilde \gamma})  \tan {\tilde \theta}_k  -  \cos{\tilde \Lambda} \cos {\tilde \gamma} + \sin{\tilde \Lambda} \sin {\tilde \gamma}  = \mathrm{RPr}.
\label{eq:2.10}
\end{equation}
Adding \ref{eq:2.7} and \ref{eq:2.10} in order to eliminate R Pr along with some helpful cancellations,
\begin{align}
\nonumber
&\sin{\tilde \Lambda}\sin{\tilde \gamma} \cot^2 {\tilde \theta}_k - \frac{1}{2} \sin({\tilde \Lambda} + {\tilde \gamma})  \cot {\tilde \theta}_k  \\ & \hspace{2cm} - \frac{1}{2}   \sin({\tilde \Lambda} + {\tilde \gamma})  \tan {\tilde \theta}_k  + \sin{\tilde \Lambda} \sin {\tilde \gamma}  = 0,
\label{eq:2.11}
\end{align}
 this can be written
\begin{equation}
\sin{\tilde \Lambda}\sin{\tilde \gamma} (\cot^2 {\tilde \theta}_k  + 1)  - \frac{1}{2} \sin({\tilde \Lambda} + {\tilde \gamma})  \frac{(\cot^2 {\tilde \theta}_k  +1)}{\cot {\tilde \theta}_k} = 0.
\label{eq:2.12}
\end{equation}
Here a factor $\cot^2 {\tilde \theta}_k +1$, which must be nonzero, cancels out, and expanding $\sin ( {\tilde\Lambda} + {\tilde \gamma})$ gives
\begin{equation}
\cot {\tilde \theta_k} = \frac{1}{2} ( \cot {\tilde \gamma}  + \cot {\tilde \Lambda} )
\label{eq:2.13}
\end{equation}
which may be rewritten as
\begin{equation}
\cot {(\theta_k}  + \phi   ) = \frac{1}{2} [ \cot {(\gamma + \phi)}  + \cot {(\Lambda + \phi)} ].
\label{eq:2.14}
\end{equation}
in the original variables. In the $\lambda \to 0$ limit, this simple equation gives $\theta_k$, the angle of $\boldsymbol{k}$ for the fastest growing mode. 

\subsection{Shellular $\phi =0$ case as in paper 2}

If $\Lambda$ is positive, ${\nabla} \ell$ has $z$-component $-2 \varpi  \Omega \sin \Lambda$, so $\gamma$ is also positive. If $\Lambda$ is negative, $\gamma$ is also negative. Hence whatever the sign of $\Lambda$, in the $\phi=0$ case there is always just one root $\theta_k$ of \ref{eq:2.14} and it lies between $\Lambda$ and $\gamma$. The wedge of instability
between $\Lambda$ and $\gamma$ is the range of angles for $\theta_k$ where $a$ is negative. At large ${\cal N}^2$, which means large R and Ri, $b$ is large and positive (from $O(\Omega^2 / \mathrm{Pr})$  where Pr is small). Inside the wedge, $a$ is negative, but it has smaller magnitude than $b$, only $O(\Omega^2)$.  This means that $a+b$ is positive, so it is adiabatically stable, but $a+b \mathrm{Pr}$ can be negative, implying GSF/diffusive instability. Diffusion at low Pr reduces the stabilising effects of the $b$ term, allowing the shear instability corresponding to $a$ to overcome it, leading to the GSF instability. 

There is a small wedge angle in which $b$ is negative for $0 < \theta_k < \Gamma$, since $b={\cal N}^2 \sin \theta_k \sin ( \Gamma+ \theta_k)$.
However, for large R and small Pr the thermal wind equation implies $\Gamma$ is small, only $O(\Omega^2 / {\cal N}^2)$, so although $b$ is negative it has a very small magnitude, which will normally be
wiped out by $a$ in this tiny wedge of $b$ instability. It might be possible for the angles $\Lambda$ and $\gamma$ to be very small also, so that both $a$ and $b$ are both very small, and negative $b$ might be bigger than positive $a$, but this unusual limit has yet not been explored.

 \subsection{The non-shellular case  $\phi \neq 0$ case}

Since $\phi$ is unrestricted, we have more possibilities than in the shellular case with $\phi=0$. If $\Lambda + \phi$ and $\gamma+\phi$ both have the same sign, and both lie between $0$ and $\pi$, then the previous argument for the shellular case still holds, and $\theta_k + \phi$ lies in the wedge between $\Lambda + \phi$ and $\gamma+\phi$, meaning $\theta_k$ lies between $\Lambda$ and $\gamma$, i.e. in the unstable wedge of negative $a$. Example: for $\Lambda=30^{\circ}$, $\gamma=60^{\circ}$, $\phi=15^{\circ}$ the solution of \ref{eq:2.14} is $\theta_k= 42.626^{\circ}$, in the required wedge giving negative $a$, positive $b$ and positive $s$, so its a local maximum of $s$. 
This case is very similar to the shellular case, and we complete the analysis of this case below in subsection \ref{appendixsub}. 

However, we could ask, what happens if the vector $\boldsymbol{e}_g$ lies between $\boldsymbol{\Omega}^{\perp}$ and ${\nabla} \ell$? If $\Lambda$ and $\gamma$ are both positive, this would mean $\phi$ is negative, and
$ \Lambda < -\phi < \gamma$. Now $\Lambda + \phi$ is negative and  $\gamma + \phi$ is positive. This means that $\cot x$ is no longer continuous as $x$ increases from  $\Lambda + \phi$ to  $\gamma + \phi$ because it goes to infinity at $x=0$. Example:  $\Lambda=30^{\circ}$, $\gamma=60^{\circ}$, $\phi=-45^{\circ}$. Looking at \ref{eq:2.14}, $\cot (\theta_k + \phi) = 0.5(\cot 15^{\circ} - \cot 15^{\circ}) = 0$,
so $\theta_k + \phi = 90^{\circ}$, so $\theta_k = 135^{\circ}$ which is not in the unstable wedge. We have a solution for \ref{eq:2.14} here, but it has both $a$ and $b$ positive, so from \ref{eq:2.6} RPr is negative. 
This is not what we want physically in the radiation zone, because we want the stratification to be stable, with $R >0$ as it is in the tachocline. 
If $\boldsymbol{e}_g$ lies between $\boldsymbol{\Omega}^{\perp}$ and $\nabla \ell$, then there is a value of $\theta_k = -\phi$ which lies in the unstable wedge and has $b$ 
zero. If $b$ is zero, our original scaling breaks down, because $b$ is no longer $O(\Omega^2 /\mathrm{Pr})$. For these modes, with ${\boldsymbol k}$ lined up with gravity, and $a$ negative, the fastest growing modes will be small $k^2$ adiabatic modes with $\sigma = \sqrt{-a}$, i.e. on the fast rotational timescale.  

 \subsubsection{The non-shellular case when ${\boldsymbol{e}_g}$  lies outside the wedge between $\boldsymbol{\Omega}^{\perp}$ and ${\nabla} \ell$ }
 \label{appendixsub}

This is the case where \ref{eq:2.14} gives a physically satisfactory maximum growth rate with ${\boldsymbol k}$ in the unstable wedge. Expanding the expression for \ref{eq:1.6}a using sine and cosine rules for sums, and using \ref{eq:2.13}, the condition for $s$ to be a local maximum, we may obtain, after some simplifications, 
\begin{equation}
a =  - \frac{ \Omega | \nabla \ell |}{2 \varpi} \sin^2 {\tilde \theta_k}  \frac{ \sin^2 ( {\tilde \gamma}-{\tilde \Lambda})}{\sin  {\tilde \Lambda} \sin {\tilde \gamma}}.
\label{eq:2.19}
\end{equation}
Now we have an expression for R/Ri in order to express $b$ in  terms of Ri rather than R. To do this we use the definition of $\nabla \ell$,
to deduce
\begin{equation}
\frac{ 2 \Omega \varpi}{| \nabla \ell |}   = \frac{\sin \gamma}{\sin \Lambda}, \quad \frac{\varpi {\cal S}}{ | \nabla \ell | } = \frac{\cos \Lambda \sin \gamma}{\sin \Lambda} - \cos \gamma. \;\;
%\eqno{(\theequation \textit{a,b})}
\label{2.21}
\end{equation}
Now $\mathrm{R} = {\cal N}^2 \varpi / 2 \Omega | \nabla \ell |$ and $\mathrm{Ri} = {\cal N}^2 / {\cal S}^2$ by definition, so $\mathrm{R} / \mathrm{Ri} = {\cal S}^2 \varpi / 2 \Omega |\nabla \ell |$. So the square of \ref{2.21}b divided by \ref{2.21}a gives
\begin{equation}
\frac{\mathrm{R}}{\mathrm{Ri}} = \left( \frac{\cos \Lambda \sin \gamma}{\sin \Lambda} - \cos \gamma \right)^2 \frac{\sin \Lambda}{\sin \gamma} = \frac {\sin^2 ( \gamma - \Lambda)}{\sin \gamma \sin \Lambda}.
\label{2.22}
\end{equation}
Now from \ref{eq:1.6}b 
\begin{equation}  b =  {\cal N}^2 \sin^2  {\tilde \theta_k} = \mathrm{R} \frac{ 2 \Omega | \nabla \ell |}{ \varpi} \sin^2 {\tilde \theta_k} = \mathrm{Ri} \frac{ 2 \Omega | \nabla \ell |}{ \varpi} \sin^2 {\tilde \theta_k} \frac {\sin^2 ( \gamma - \Lambda)}{\sin \gamma \sin \Lambda}.
\label{2.23}
\end{equation}
Now in the $\lambda \to 0$ limit, \ref{eq:2.1} tells us that $a = -\mathrm{Pr} b$. Putting together our expressions for $a$ and $b$,
\begin{align}
\nonumber
\mathrm{Pr} b &= \mathrm{RiPr}  \frac{ 2 \Omega | \nabla \ell |}{ \varpi} \sin^2 {\tilde \theta_k} \frac {\sin^2 ( \gamma - \Lambda)}{\sin \gamma \sin \Lambda} \\
&= -a =  \frac{ \Omega | \nabla \ell |}{2 \varpi} \sin^2 {\tilde \theta_k}  \frac{ \sin^2 ( {\tilde \gamma}-{\tilde \Lambda})}{\sin  {\tilde \Lambda} \sin {\tilde \gamma}}.
\label{2.23}
\end{align}
Looking at these expressions, $\sin^2 (\gamma - \Lambda) = \sin^2 ( {\tilde \gamma} - {\tilde \Lambda})$, so these terms cancel. We thus end up with
\begin{equation}
 \mathrm{RiPr} = \frac{\sin \gamma \sin \Lambda}{4 \sin (\gamma + \phi) \sin(\Lambda + \phi)},
\label{2.24}
\end{equation}
which is different from the shellular $\phi = 0$ case where the limit $\lambda \to 0$ corresponded to the simpler $\mathrm{Ri Pr} = 1/4$. 

If $\Lambda$ and $\gamma$ are both positive (they must have the same sign) and $\phi$ is positive, then the limit is $\mathrm{Ri Pr}<1/4$. This is stabilising, because it means that $Ri$ has to be smaller for instability, and since $\mathrm{Ri}= {\cal N}^2 / {\cal S}^2$ this means the shear ${\cal S}$ has to be larger for instability. Example: $\Lambda = 30^{\circ}$, $\gamma= 60^{\circ}$, $\phi = 30^{\circ}$ gives the limit as $Ri Pr = 1/8$, so the range of instability is
reduced from $0 < \mathrm{Ri Pr}  < 0.25$ down to $0 < \mathrm{Ri Pr} < 0.125$ confirming positive $\phi$ is stabilising if $\gamma$ and $\Lambda$ are positive. 

However, if $\phi$ is negative when $\gamma$ and $\Lambda$ are positive, $\phi$ is destabilising. Example: $\Lambda = 30^{\circ}$, $\gamma= 60^{\circ}$, $\phi = - 15^{\circ}$, then the upper limit of Ri Pr is increased to 0.5915 so a smaller shear ${\cal S}$ will still be unstable.

If $\phi$ is negative and $|\phi|$ approaches the smaller of $\Lambda$ or $\gamma$ then $\sin (\Lambda + \phi)$ or $\sin ( \gamma + \phi)$ will become small so that  \ref{2.24} will diverge to infinity. This is correct, because as $\phi$ approaches the wedge of instability, we expect the system to become adiabatically unstable, i.e. unstable whatever Ri is. If ${\boldsymbol{e}_g}$ is inside the wedge, fluid
motion perpendicular to gravity cannot be stabilised by the stratification, and since it is inside the wedge it is driven by the shear, so it is very unstable.  

 \subsubsection{The non-shellular case close to $\lambda = 0$}
\label{appendixsub1}
Now suppose that $\lambda$ is small but not quite zero (i.e.~the limit $\mathrm{RiPr}=O(1)$ as $\mathrm{Pr}\to 0$), so that squares and higher powers of $\lambda$ can be neglected. Then \ref{eq:1.14} gives
\begin{equation}
\lambda = \frac{a+\mathrm{Pr} b}{3a}.
\label{2.26}
\end{equation}
Now $a + \mathrm{Pr} b$ is small and negative, but not quite zero. If we put in the expressions \ref{eq:2.19} for $a$ and \ref{2.23} for $b$, \ref{2.26} becomes
\begin{equation}
\lambda = \frac{1}{3} \left( 1 - 4\mathrm{RiPr} \frac{\sin {\tilde \gamma} \sin {\tilde \Lambda}} {\sin  \gamma \sin \Lambda} \right) = \frac{\kappa \nu k^4}{b},
\label{2.27}
\end{equation}
so
\begin{equation}
k^4 = \frac{{\cal N}^2 \sin^2 {\tilde \theta_k}}{3}  \left( 1 - 4\mathrm{RiPr} \frac{\sin {\tilde \gamma} \sin {\tilde \Lambda}} {\sin  \gamma \sin \Lambda} \right) .
\label{2.28}
\end{equation}
This means that if RiPr is just a little less than the limiting value given by \ref{2.24} there is an unstable solution with a long wavelength, because $k$ is small from \ref{2.28}, and the growth rate
is also small from \ref{eq:1.10}, giving
\begin{equation}
s  = \frac{2 \sqrt{\mathrm{Pr}}{\cal N} \sin {\tilde \theta_k}}{\sqrt{3}}  \left( 1 - 4\mathrm{RiPr} \frac{\sin {\tilde \gamma} \sin {\tilde \Lambda}} {\sin  \gamma \sin \Lambda} \right)^{1/2}.
\label{2.29}
\end{equation}
This result has been confirmed numerically, and describes the slow growth which occurs when the strength of the differential rotation is only just above the minimum value required for instability.

\section{Tables of simulations}
\label{Tables}

\begin{table*}
\begin{center}
\begin{tabular}{cccccccccccc}
\hline
$S$ & $\phi$ & $\Lambda$ & $\Gamma$ & $\gamma$ & RiPr & $k_x$ & $k_z$ & $|\boldsymbol{k}|$ & $\sigma$ & $\theta_k$ & Adiabatically stable? \\
\hline
& & & & \multicolumn{4}{c}{$\mathrm{Latitude} = 0^{\circ}$} \\

\hline
$ 2 $& $ -90 ^{\circ}$ & $ 90 ^{\circ}$ & $-66. 42^{\circ}$& $135 ^{\circ}$  & 0.025 & -0.17 & 0.51 & 0.54 & 0.70 & 71.08 $^\circ$ & $\checkmark $ \\ 
$ 2 $& $ -60 ^{\circ}$ & $ 60 ^{\circ}$ &$ -39.73 ^{\circ}$  & $120 ^{\circ}$   & 0.025 & -0.053 & 0.58 & 0.58 & 0.76 & 84.81 $^\circ$ & $\checkmark $ \\ 
$ 2 $& $ -30 ^{\circ}$ & $ 30 ^{\circ}$ & $-18.46 ^{\circ}$ & $105 ^{\circ}$ &   0.025 & -0.30 & 0.51 & 0.59 & 0.61 & 59.21 $^\circ$ & $\checkmark $ \\ 
$ 2 $& $ 0 ^{\circ}$ & $ 0 ^{\circ}$ & NA & NA  & 0.025 & NA & NA & NA & NA & NA & $\checkmark $ \\ 

$ 2 $& $ 30 ^{\circ}$ & $ -30 ^{\circ}$ & $18.46 ^{\circ}$ & $ -105 ^{\circ}$  & 0.025 & -0.30 & 0.51 & 0.59 & 0.61 & $59.21^\circ$ & $\checkmark $ \\ 

$ 2 $& $ 60 ^{\circ}$ & $ -60^{\circ}$ & $ 39.73 ^{\circ}$ & $ -120 ^{\circ}$  & 0.025 & -0.05 & 0.58 & 0.58 & 0.76 & $ 84.81 ^\circ$ & $\checkmark $ \\ 

$ 2 $& $ 90 ^{\circ}$ & $ -90 ^{\circ}$ & $ 66.42 ^{\circ}$ & $ -135 ^{\circ}$  & 0.025 & -0.17 & 0.51 & 0.54 & 0.70 & $ 71.08 ^\circ$ & $\checkmark $ \\

\hline
 & & & & \multicolumn{4}{c}{$\mathrm{Latitude} = 30^{\circ}$}  \\
\hline

$ 2 $& $ -90 ^{\circ}$ & $ 120 ^{\circ}$ & $ -69.73 ^{\circ}$ & $ 150 ^{\circ}$ & 0.025 & 0.47 & 0.49 & 0.68 & 0.36 & $ 46.28 ^\circ$ & $\checkmark $ \\ 

$ 2 $& $ -60 ^{\circ}$ & $ 90 ^{\circ}$ & $ -36.42 ^{\circ}$ & $ 135 ^{\circ}$ & 0.025 & 0.26 & 0.67 & 0.72 & 0.55 & $ 69.19 ^\circ$ & $\checkmark $ \\ 

$ 2 $& $ -30 ^{\circ}$ & $ 60 ^{\circ}$ & $ -9.73 ^{\circ}$ & $ 120 ^{\circ}$ & 0.025 & -0.03 & 0.74 & 0.74 & 0.62 & $ 87.72 ^\circ$ & $\checkmark $ \\ 

$ 2 $& $ 0 ^{\circ}$ & $ 30 ^{\circ}$ & $ 11.54^{\circ}$ & $ 105 ^{\circ}$ & 0.025 & 0.32 & -0.67 & 0.74 & 0.49 & $ 64.14 ^\circ$ & $\checkmark $ \\ 

$ 2 $& $ 30 ^{\circ}$ & $ -30 ^{\circ}$ & NA & NA  & 0.025 & NA & NA & NA & NA & NA & $\checkmark $ \\ 

$ 2 $& $ 60 ^{\circ}$ & $ -30 ^{\circ}$ & $ 48.46 ^{\circ}$ & $ -105 ^{\circ}$ & 0.025 & -0.01 & 0.015 & 0.018 & 0.88 & $ 57.09 ^\circ$ & $\times $ \\ 

$ 2 $& $ 90 ^{\circ}$ & $ -60 ^{\circ}$ & $ 69.73 ^{\circ}$ & $ -120 ^{\circ}$ & 0.025 & -0.001 & 0.01 & 0.01 & 1.10 & $ 81.87 ^\circ$ & $\times $ \\

\hline
& & & & \multicolumn{4}{c}{$\mathrm{Latitude} = 60^{\circ}$} \\
\hline

$ 2 $& $ -90 ^{\circ}$ & $ 150^{\circ}$ & $ -78.46 ^{\circ}$ & $ 165 ^{\circ}$ & 0.025 & 0.52 & 0.22 & 0.56 & 0.03 & $ 22.77 ^\circ$ & $\checkmark $ \\ 

$ 2 $& $ -60 ^{\circ}$ & $ 120 ^{\circ}$ & $ -39.73 ^{\circ}$ & $ 150 ^{\circ}$ & 0.025 & 0.52 & 0.52 & 0.73 & 0.29 & $ 45.20 ^\circ$ & $\checkmark $ \\ 

$ 2 $& $ -30 ^{\circ}$ & $ 90 ^{\circ}$ & $ -6.42 ^{\circ}$ & $ 135 ^{\circ}$ & 0.025 & 0.29 & 0.70 & 0.76 & 0.51 & $ 67.21 ^\circ$ & $\checkmark $ \\ 

$ 2 $& $ 0.0 ^{\circ}$ & $ 60 ^{\circ}$ & $ 20.27 ^{\circ}$ & $ 120 ^{\circ}$ & 0.025 & 0.015 & 0.76 & 0.76 & 0.60 & $ 88.86 ^\circ$ & $\checkmark $ \\ 

$ 2 $& $ 30 ^{\circ}$ & $ 30 ^{\circ}$ & $ 41.54 ^{\circ}$ & $ 105 ^{\circ}$ & 0.025 & -0.25 & 0.70 & 0.75 & 0.48 & $ 70.19 ^\circ$ & $\checkmark $ \\

$ 2 $& $ 60 ^{\circ}$ & $ 0 ^{\circ}$ & NA & NA  & 0.025 & NA & NA & NA & NA & NA & $\checkmark $ \\ 

$ 2 $& $ 90 ^{\circ}$ & $ -30 ^{\circ}$ & $ 78.46 ^{\circ}$ & $ -105 ^{\circ}$ & 0.025 & 0.001 & 0.01 & 0.01 & 0.85 & $ 81.87 ^\circ$ & $\times $ \\

\hline
& & & & \multicolumn{4}{c}{$\mathrm{Latitude} = 90^{\circ}$} \\
\hline

$ 2 $& $ -90 ^{\circ}$ & $ 180 ^{\circ}$ & NA & NA  & 0.025 & NA & NA & NA & NA & NA & $\checkmark $ \\ 

$ 2 $& $ -60 ^{\circ}$ & $ 150 ^{\circ}$ & $ -48.46 ^{\circ}$ & $ 165 ^{\circ}$ & 0.025 & 0.51 & 0.21 & 0.55 & 0.024 & $ 22.30 ^\circ$ & $\checkmark $ \\ 

$ 2 $& $ -30 ^{\circ}$ & $ 120 ^{\circ}$ & $ -9.73 ^{\circ}$ & $ 150 ^{\circ}$ & 0.025 & 0.51 & 0.49 & 0.71 & 0.32 & $ 43.98 ^\circ$ & $\checkmark $ \\ 

$ 2 $& $ 0 ^{\circ}$ & $ 90 ^{\circ}$ & $ 23.58 ^{\circ}$ & $ 135 ^{\circ}$ & 0.025 & 0.29 & 0.62 & 0.69 & 0.58 & $ 65.20 ^\circ$ & $\checkmark $ \\ 

$ 2 $& $ 30 ^{\circ}$ & $ 60 ^{\circ}$ & $ 50.27 ^{\circ}$ & $ 120 ^{\circ}$ & 0.025 & 0.049 & 0.65 & 0.65 & 0.70 & $ 85.69 ^\circ$ & $\checkmark $ \\

$ 2 $& $ 60 ^{\circ}$ & $ 30 ^{\circ}$ & $ 71.54 ^{\circ}$ & $ 105 ^{\circ}$ & 0.025 & -0.16 & 0.60 & 0.62 & 0.59 & $ 75.34 ^\circ$ & $\checkmark $ \\ 

$ 2 $& $ 90 ^{\circ}$ & $ 0 ^{\circ}$ & NA & NA  & 0.025 & NA & NA & NA & NA & NA & $\checkmark $ \\

\hline
& & & & \multicolumn{4}{c}{Variations in shear (GSF instability at $S=2$ in Fig.~\ref{ShearComp} panels (a) and (b))}  \\

\hline

$ 0.5 $& $ -30 ^{\circ}$ & $ 60 ^{\circ}$ & $ -25.037 ^{\circ}$ & $ 73.90 ^{\circ}$ & 0.40 & 0.18 & -0.39 & 0.43 & 0.013 & $ 66.03 ^\circ$ & $\checkmark $ \\ 

$ 1.0 $& $ -30 ^{\circ}$ & $ 60 ^{\circ}$ & $ -20.035 ^{\circ}$ & $ 90 ^{\circ}$ & 0.10 & -0.18 & 0.60 & 0.63 & 0.18 & $ 72.95 ^\circ$ & $\checkmark $ \\ 

$ 1.5 $& $ -30 ^{\circ}$ & $ 60 ^{\circ}$ & $ -14.94 ^{\circ}$ & $ 106.1 ^{\circ}$ & 0.044 & -0.11 & 0.69 & 0.70 & 0.40 & $ 80.55 ^\circ$ & $\checkmark $ \\ 

$ 2.0 $& $ -30 ^{\circ}$ & $ 60 ^{\circ}$ & $ -9.73 ^{\circ}$ & $ 120 ^{\circ}$ & 0.025 & -0.029 & 0.74 & 0.74 & 0.62 & $ 87.72 ^\circ$ & $\checkmark $ \\

$ 2.5 $& $ -30 ^{\circ}$ & $ 60 ^{\circ}$ & $ -4.34 ^{\circ}$ & $ 130.89 ^{\circ}$ & 0.016 & 0.053 & 0.76 & 0.76 & 0.84 & $ 86.04 ^\circ$ & $\checkmark $ \\ 

$ 3.0 $& $ -30 ^{\circ}$ & $ 60 ^{\circ}$ & $ 1.31 ^{\circ}$ & $ 139.11 ^{\circ}$ & 0.011 & -0.12 & -0.76 & 0.77 & 1.06 & $ -81.12 ^\circ$ & $\checkmark $ \\

\hline
& & & & \multicolumn{4}{c}{Variations in shear (adiabatic instability at $S=2$ in Fig.~\ref{ShearComp} panels (c) and (d))} \\

\hline
$ 0.5 $& $ 90 ^{\circ}$ & $ -60 ^{\circ}$ & $ -25.03 ^{\circ}$ & $ -73.90 ^{\circ}$ & 0.40 & 0.18 & -0.39 & 0.43 & 0.013 & $ 66.03 ^\circ$ & $\checkmark $ \\ 

$ 1.0 $& $ 90 ^{\circ}$ & $ -60 ^{\circ}$ & $ 99.97 ^{\circ}$ & $ -90 ^{\circ}$ & 0.10 & 0 & 0 & 0 & 0.48 & NA & $\times $ \\ 

$ 1.5 $& $ 90 ^{\circ}$ & $ -60 ^{\circ}$ & $ 105.06 ^{\circ}$ & $ -106.1 ^{\circ}$ & 0.044 & 0 & 0 & 0 & 0.85 & NA & $\times $ \\ 

$ 2.0 $& $ 90 ^{\circ}$ & $ -60 ^{\circ}$ & $ 110.27 ^{\circ}$ & $ -120 ^{\circ}$ & 0.025 & 0 & 0 & 0 & 1.10 & NA & $\times $ \\

$ 2.5 $& $ 90 ^{\circ}$ & $ -60 ^{\circ}$ & $ 115.66 ^{\circ}$ & $ -130.89 ^{\circ}$ & 0.016 & 0 & 0 & 0 & 1.31 & NA & $\times $ \\ 

$ 3.0 $& $ 90 ^{\circ}$ & $ -60 ^{\circ}$ & $ 121.31 ^{\circ}$ & $ -139.11 ^{\circ}$ & 0.011 & 0 & 0 & 0 & 1.49 & NA & $\times $ \\ 

\hline
\label{Table1}
\end{tabular}
\caption{Table of linear properties for simulation parameters. For all of these we fix $\mathrm{Pr}=10^{-2}$ and $N^2 = 10$. $k_x$ and $k_z$ are wavevector components of the fastest growing linear mode, $\sigma$ is the corresponding growth rate. The cases which investigated variations in shear $S$ were considered at $\Lambda+\phi=30^\circ$ latitude with $\Lambda = 60^\circ, \phi = -30^\circ$ in the case that was GSF-unstable at $S=2$, and $\Lambda = -60^\circ, \phi = 90^\circ$ in the case that is adiabatically unstable if $S=2$. Our simulation units are determined by setting $\Omega=d=1$.}
\end{center}
\end{table*}

\begin{table*}
\begin{center}
\begin{tabular}{ccccc|cccccc}
\hline
$S$ & $\phi$ & $\Lambda$ & $L_x $ & $L_z$  & $\langle u_xu_y\rangle$ & $\langle u_xu_z\rangle$ & $\langle u_yu_z\rangle$ & $\sqrt{\langle u_x^2\rangle}$ & $\sqrt{\langle u_y^2\rangle}$ & $\sqrt{\langle u_z^2\rangle}$
 \\
\hline
& & & & & \multicolumn{4}{c}{$ \text{Latitude} = 0^{\circ} $} \\

\hline
2 & $-90^{\circ}$ & $90^{\circ}$ & 100 & 100  & $2883.03 \pm490.52$ & $-13.37\pm69.90$ &  $92.94\pm 147.36$ & $34.14\pm 260.66$ & $32.20\pm 89.30$ & $27.15\pm 260.66$ \\

2 & $-60^{\circ}$ & $60^{\circ}$ & 100 & 100  & $510.02\pm41.53$ & $73.13\pm9.30$ &  $195.36\pm20.84$ & $15.08\pm0.58$ & $31.56\pm 1.28$ & $13.00\pm 0.56$ \\

2 & $-30^{\circ}$ & $30^{\circ}$ & 100 & 100  & $135.81\pm 16.78$ & $37.73\pm5.74$ &  $112.04\pm13.98$ & $7.82\pm0.43$ & $15.19\pm 0.89$ & $7.69\pm0.39$ \\

2 & $0^{\circ}$ & $0^{\circ}$ & 100 & 100  & NA & NA &  NA & NA & NA & NA \\

2 & $30^{\circ}$ & $-30^{\circ}$ & 100 & 100  & $138.81\pm 20.62$ & $-38.93\pm6.97$ &  $-115.60\pm17.19$ & $7.88\pm0.52$ & $15.36\pm 1.06$ & $7.97\pm0.50$ \\

2 & $60^{\circ}$ & $-60^{\circ}$ & 100 & 100  & $134.84\pm16.54$ & $-36.92\pm5.59$ &  $-111.77\pm14.02$ & $7.79 \pm0.41$ & $15.14\pm0.82$ & $7.98 \pm0.41 $ \\

2 & $90^{\circ}$ & $-90^{\circ}$ & 100 & 100  & $2838.20\pm 578.87$ & $14.92\pm76.82$ &  $-128.26\pm130.56$ & $34.64\pm3.97$ & $88.33\pm7.04$ & $28.10\pm3.75$   \\

\hline

& & & & & \multicolumn{4}{c}{$ \text{Latitude} = 30^{\circ} $} \\

\hline
2 & $-90^{\circ}$ & $120^{\circ}$ & 100 & 100  & $25.71\pm 8.29$ & $-1.55\pm1.13$ &  $-20.68\pm7.95$ & $2.93 \pm0.61$ & $13.81\pm 2.30$ & $3.21 \pm0.69$ \\

2 & $-60^{\circ}$ & $90^{\circ}$ & 100 & 100  & $25.86\pm 7.13$ & $-2.68\pm0.51$ &  $-7.49\pm2.07$ & $3.29 \pm0.29$ & $8.14 \pm2.67 $ & $3.46 \pm 0.30$ \\

2 & $-30^{\circ}$ & $60^{\circ}$ & 100 & 100  & $24.08\pm 1.56$ & $-1.77\pm0.21$ &  $0.72\pm0.61$ & $3.22 \pm 0.09$ & $7.85 \pm0.26$ & $3.27 \pm0.10$ \\

2 & $0^{\circ}$ & $30^{\circ}$ & 100 & 100  & $11.29\pm1.48$ & $-1.14\pm0.37$ &  $5.17\pm1.15$ & $2.36 \pm0.16 $ & $8.08 \pm1.51 $ & $2.55 \pm0.22$ \\

2 & $30^{\circ}$ & $0^{\circ}$ & 100 & 100  & $0$ & $0$ &  $0$ & NA & NA & NA \\

2 & $60^{\circ}$ & $-30^{\circ}$ & 100 & 100  & $958.54\pm509.53$ & $-334.17\pm216.16$ &  $-516.59\pm311.61$ & $28.37\pm6.34$ & $29.85\pm6.63$ & $23.80\pm5.12$ \\

2 & $90^{\circ}$ & $-60^{\circ}$ & 100 & 100  & $9219.30\pm4186.12$ & $-633.00\pm529.89$ &  $-499.52\pm483.91$ & $83.77\pm15.41$ & $86.67\pm 16.71$ & $53.12\pm11.22$   \\

\hline

& & & & & \multicolumn{4}{c}{$ \text{Latitude} = 90^{\circ} $} \\

\hline
2 & $-90^{\circ}$ & $180^{\circ}$ & 100 & 100  & $0$ & $0$ &  $0$ & NA & NA & NA \\

2 & $-60^{\circ}$ & $150^{\circ}$ & 100 & 100  & $3.54\pm 0.77$ & $-0.88\pm0.18$ &  $-9.52\pm2.07$ & $0.52 \pm0.08$ & $4.55 \pm0.26$ & $1.24 \pm 0.14$ \\

2 & $-30^{\circ}$ & $120^{\circ}$ & 100 & 100  & $17.54\pm 2.13$ & $-5.37\pm0.31$ &  $-22.90\pm2.23$ & $2.22 \pm 0.17$ & $7.08 \pm 0.52$ & $3.23 \pm0.17$ \\

2 & $0^{\circ}$ & $90^{\circ}$ & 100 & 100  & $36.99\pm1.97$ & $-20.74\pm1.00$ &  $-25.59\pm1.72$ & $4.45 \pm 0.09$ & $6.61 \pm0.13$ & $5.09 \pm 0.12$ \\

2 & $30^{\circ}$ & $60^{\circ}$ & 100 & 100  & $162.84 \pm22.49$ & $-113.43 \pm11.59$ &  $-77.34 \pm12.22$ & $10.59 \pm0.52$ & $11.13 \pm0.69$ & $7.96 \pm0.47$ \\

2 & $60^{\circ}$ & $30^{\circ}$ & 100 & 100  & $138.36\pm19.83$ & $-110.84\pm25.12$ &  $-31.54\pm10.89$ & $15.67\pm1.26$ & $8.15\pm0.52$ & $8.35\pm0.59$ \\

2 & $90^{\circ}$ & $0^{\circ}$ & 100 & 100  & $0$ & $0$ &  $0$ & NA & NA & NA \\
\hline

& & & & & \multicolumn{4}{c}{Variations in shear (GSF instability at $S=2$ in Fig.~\ref{ShearComp} panels (a) and (b))} \\

\hline
1 & $-30^{\circ}$ & $60^{\circ}$ & 100 & 100  & $15.46\pm 1.28$ & $0.99 \pm0.13$ &  $2.88\pm 0.30$ & $2.16 \pm 0.12$ & $5.67 \pm0.45$ & $1.24 \pm0.13$ \\

1.5 & $-30^{\circ}$ & $60^{\circ}$ & 100 & 100  & $17.82\pm 1.52$ & $-0.65\pm0.16$ &  $0.98\pm0.31$ & $2.60 \pm 0.09$ & $4.72\pm 0.20$ & $2.06\pm 0.10$ \\

2 & $-30^{\circ}$ & $60^{\circ}$ & 100 & 100  & $24.08\pm 1.39$ & $-1.77\pm0.21$ &  $0.75\pm0.59$ & $3.22 \pm 0.09$ & $7.84 \pm 0.28$ & $3.26 \pm 0.09$ \\

2.5 & $-30^{\circ}$ & $60^{\circ}$ & 100 & 100  & $29.48\pm2.06$ & $-3.65\pm 0.36$ &  $-0.12\pm0.95$ & $3.95 \pm0.13$ & $10.99 \pm 1.22$  & $4.23 \pm0.14$ \\

\hline

& & & & & \multicolumn{4}{c}{Variations in shear (adiabatic instability at $S=2$ in Fig.~\ref{ShearComp} panels (c) and (d))} \\

\hline
1 & $90^{\circ}$ & $-60^{\circ}$ & 100 & 100  & $160.95\pm 69.64$ & $-64.15\pm71.27$ &  $-18.56\pm24.81$ & $24.52\pm 7.61$ & $11.14\pm 3.35$ & $12.61\pm 6.14$ \\

1.5 & $90^{\circ}$ & $-30^{\circ}$ & 100 & 100  & $646.72\pm 306.23$ & $-790.23\pm531.47$ &  $-365.74\pm 313.57$ & $76.15\pm 14.92$ & $65.94\pm 12.66$ & $42.89\pm 9.95$ \\

2 & $90^{\circ}$ & $-60^{\circ}$ & 100 & 100  & $916.75\pm 459.88$ & $-639.07\pm563.43$ &  $-493.17\pm481.91$ & $83.52\pm17.09$ & $86.28\pm 18.35$ & $52.87\pm12.45$   \\

2.5 & $90^{\circ}$ & $30^{\circ}$ & 100 & 100  & $1082.42\pm758.08$ & $-567.85\pm768.16$ &  $-580.33\pm842.64$ & $87.96\pm 24.89$ &$99.80\pm 29.88$  & $60.97\pm 19.59$ \\

\hline
\label{Table2}
\end{tabular}
\caption{Table of simulation parameters and nonlinear outcomes. All simulations have $\mathrm{Pr}=10^{-2}$ and $N^2=10$. Time-averages are based on the entire simulation after the initial linear growth phase. 
Simulation parameters not listed in this table are given in \S~\ref{sec:maths}. Our simulation units are determined by setting $\Omega=d=1$.}
\end{center}
\end{table*}

\section{Linear growth rates with fixed $\Lambda$}
\label{LobesAppendix}
\begin{figure*}
  \subfigure[]{\includegraphics[
    trim=0cm 0cm 0cm 0cm,clip=true,
    width=0.4\textwidth]{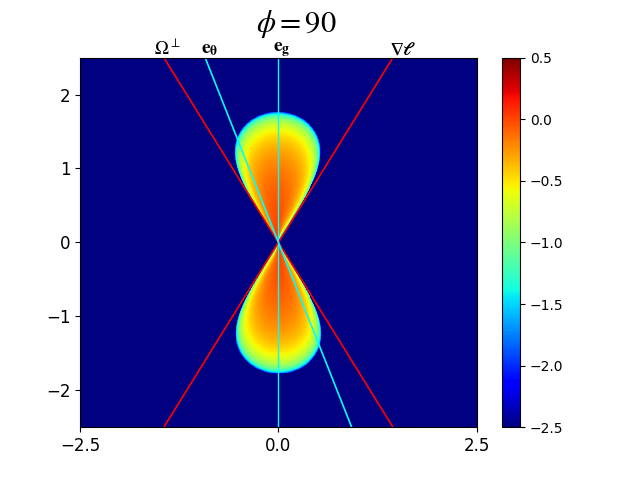}}
    \subfigure[]{\includegraphics[
    trim=0cm 0cm 0cm 0cm,clip=true,
    width=0.4\textwidth]{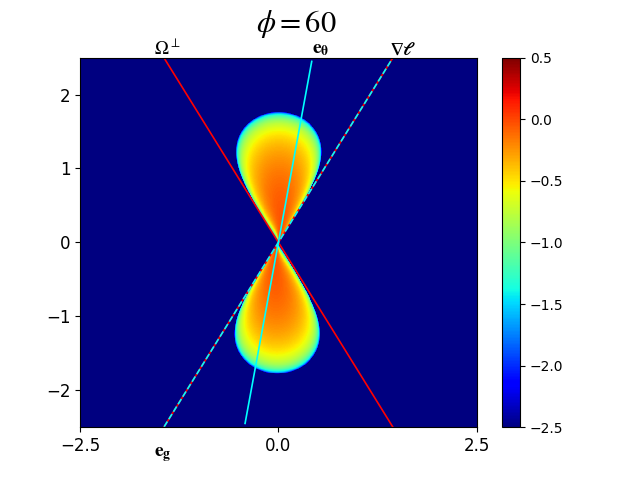}}
    \subfigure[]{\includegraphics[
    trim=0cm 0cm 0cm 0cm,clip=true,
    width=0.4\textwidth]{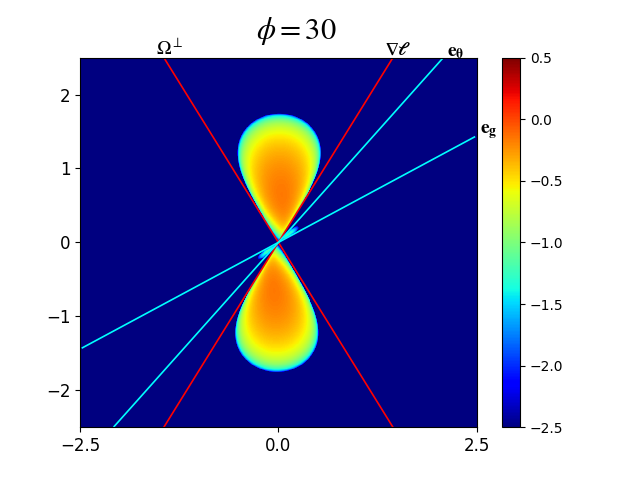}}
     \subfigure[]{\includegraphics[
    trim=0cm 0cm 0cm 0cm,clip=true,
    width=0.4\textwidth]{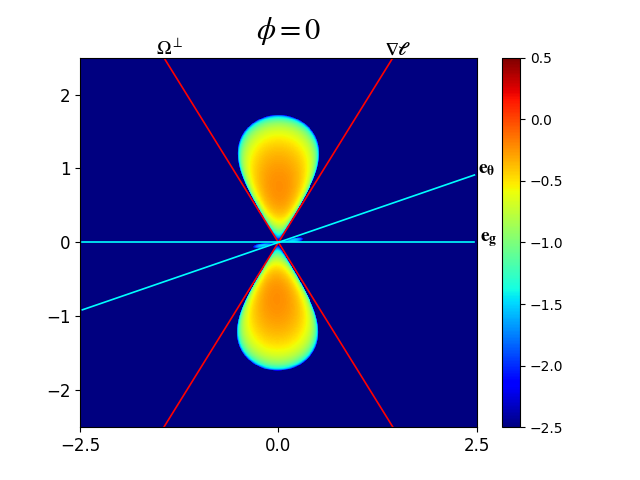}}
    \subfigure[]{\includegraphics[
    trim=0cm 0cm 0cm 0cm,clip=true,
    width=0.4\textwidth]{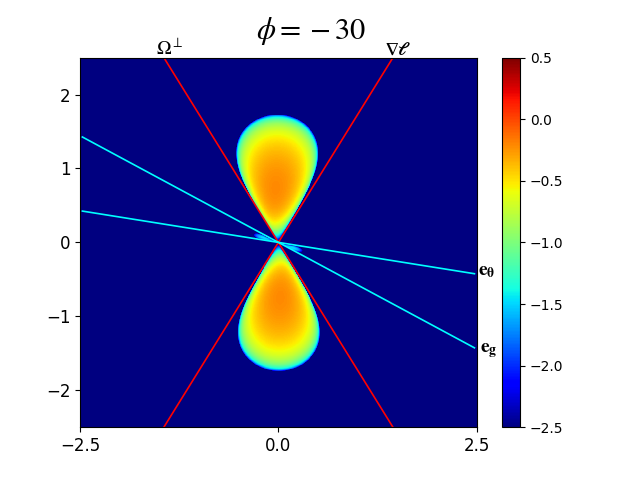}}
    \subfigure[]{\includegraphics[
    trim=0cm 0cm 0cm 0cm,clip=true,
    width=0.4\textwidth]{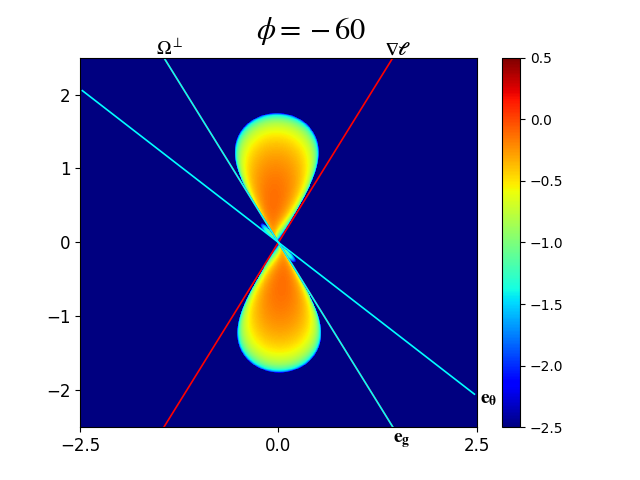}}
  \caption{Figures of linear growth rate $\log_{10} (\sigma/\Omega)$ for the axisymmetric GSF (or adiabatic) instability for various $\phi$ on the $(k_x, k_z)$-plane for $\mathcal{N}^2/\Omega^2 = 10$, Pr$=10^{-2}$, $\mathcal{S}/\Omega=2$, with $\Lambda $ = $60^\circ$ fixed. Here we vary $\phi$ in multiples of $30^\circ$ from $-90^\circ$ to $60^\circ$. GSF (or adiabatically) unstable modes are contained within the wedge bounded by the two vectors $\hat{\boldsymbol{\Omega}}^\perp$ and $\nabla \ell$. We also observe two secondary lobes outside the primary one. While the primary lobe is fixed in orientation, the strength and orientation of secondary modes depends on $\phi$ and are only present when we are adiabatically stable.}
  \label{Lobes2}
\end{figure*}

\bsp
\label{lastpage}
\end{document}